\documentclass[iop]{emulateapj}
\usepackage{graphicx}
\usepackage{amsmath}
\usepackage{afterpage}
\usepackage[]{natbib}
\usepackage[usenames]{color} 
\usepackage{ulem}
\newcommand{\kmss}{\>{\rm km}\,{\rm s}^{-2}}

\newcommand{\Msun}{\mbox{$\rm M_{\odot}$}}

\newcommand\degrees{^\circ}
\newcommand{\etal}{{et al.~}}
\newcommand\eg{{\it e.g.~}}

\newcommand\ie{{\it i.e.~}}

\def\apj{{ApJ}}
\def\apjs{{ApJS}}

\def\aj{{AJ}}
\def\mnras{{MNRAS}}
\def\nar{{NewAR}}   
\def\prd{{PhRvD}}   
\def\lrr{{LRR}}     
\def\ban{{BAN}}     
\def\jphg{{JPhG}}   
\def\phrvl{{PhRvL}} 
\def\newa{{NewA}}   

\shorttitle{Mapping the Milky Way Dark Matter Halo with SDSS}

\shortauthors{Loebman \etal}
\begin{document}

\title{The Milky Way Tomography with SDSS. V. Mapping the Dark Matter Halo}

\author{Sarah R. Loebman\altaffilmark{1,2}}
\author{\v{Z}eljko Ivezi\'{c}\altaffilmark{3}}
\author{Thomas R. Quinn\altaffilmark{3}}
\author{Jo Bovy\altaffilmark{4,5}}
\author{Charlotte R. Christensen\altaffilmark{6}}
\author{Mario Juri\'{c}\altaffilmark{3}}
\author{Rok Ro\v{s}kar\altaffilmark{7}}
\author{Alyson M. Brooks\altaffilmark{8}}
\author{Fabio Governato\altaffilmark{3}}

\altaffiltext{1}{{Department of Astronomy, University of Michigan, 
                  500 Church Street, Ann Arbor, MI 48109};       
		 {\tt sloebman@umich.edu}}
\altaffiltext{2}{Michigan Society of Fellows}
\altaffiltext{3}{Astronomy  Department, University of  Washington, 
                 Seattle, WA}
\altaffiltext{4}{Institute for Advanced Study, Einstein Drive, Princeton, NJ}
\altaffiltext{5}{Hubble Fellow}
\altaffiltext{6}{Physics Department, Grinnell College Grinnell, IA}
\altaffiltext{7}{Institute for Computational Science, University of Z\"urich, 
                 Z\"urich, Switzerland}
\altaffiltext{8}{Department of Physics \& Astronomy, 
                 Rutgers University, NJ}

\begin{abstract} 
We present robust constraints from the Sloan Digital Sky Survey (SDSS)
on the shape and distribution of the dark matter halo within the Milky
Way  (MW).  Using the  number density  distribution and  kinematics of
SDSS halo stars, we probe the dark matter distribution to heliocentric
distances   exceeding   $\sim$10~kpc   and  galactocentric   distances
exceeding  $\sim$20~kpc.   Our analysis  utilizes  Jeans equations  to
generate two-dimensional acceleration maps throughout the volume; this
approach   is   thoroughly   tested   on  a   cosmologically   derived
$N$--body+SPH simulation of a MW-like  galaxy.  We show that the known
accelerations  (gradients  of  the  gravitational  potential)  can  be
successfully  recovered in  such a  realistic system.   Leveraging the
baryonic  gravitational potential derived  by Bovy  \& Rix  (2013), we
show that the gravitational potential implied by the SDSS observations
cannot  be explained,  assuming Newtonian  gravity, by  visible matter
alone: the gravitational force  experienced by stars at galactocentric
distances of $\sim$20~kpc is as much as three times stronger than what
can be attributed to purely visible matter. We also show that the SDSS
data provide a strong constraint on  the shape of the dark matter halo
potential.  Within galactocentric  distances of $\sim$20~kpc, the dark
matter halo  potential is well described  as an oblate  halo with axis
ratio $q^\Phi_{DM}=0.7$$\pm$$0.1$;  this corresponds to  an axis ratio
$q^\rho_{DM}\sim0.4$$\pm$$0.1$    for   the   dark    matter   density
distribution.  Because of  our precise two-dimensional measurements of
the acceleration of the halo  stars, we can reject several  MOND models as an
explanation  of the  observed  behavior.
\end{abstract}


\section{Introduction}
\label{s:intro.tex}

This  paper  is  the  last  in  a  series  of  papers  utilizing  SDSS
observations   of   stars    to   map   their   spatial   distribution
\citep[][]{Juric2008},             metallicity            distribution
\citep[][]{Ivezic2008},   kinematics   \citep[][]{Bond2010}  and   the
distribution of interstellar dust \citep[][]{Berry2012}. Here we focus
on  observations  of  distant halo  stars  and  use  them to  map  the
distribution of dark matter in the Milky Way halo.

The nature of dark matter is  one of the most fundamental questions in
the physical  sciences today: determining  the make-up of  dark matter
and  its spatial  distribution has  important implications  for fields
ranging from  theories of galaxy  formation and evolution  to particle
physics  and cosmology.   While  the gravitational  arguments for  the
existence     of     dark      matter     are     well     established
\citep{Rubin1980,Spergel2003,Markevitch2004},     its    most    basic
properties   are  still  disturbingly   ambiguous \citep{Read2014}.   
We   can  address fundamental questions about dark  matter's properties 
by examining the distribution and shape of dark  matter structure within and around our
Galaxy \citep{Tremaine1979,HD2000}.

A     myriad     of    techniques     --     from    tidal     streams
\citep[e.g.,][]{Johnston1999,  Ibata2001,   Law2010,  Koposov2010}  to
Jeans equations \citep[e.g.,][]{Loebman2012, BovyVc} -- have been used
to  explore  the  Milky  Way's  (MW)  dark  matter  distribution.   In
particular,  applying  Jeans  equations  to  MW  stars  to  infer  the
underlying  mass  distribution  has   a  long  and  solid  theoretical
foundation \citep{Jeans1915, Oort1932}.

\subsection{Jeans Equations as a Tool for Estimating Stellar Acceleration}

While it is hard to measure stellar acceleration for individual stars,
which  would directly  constrain  the gravitational  potential, it  is
possible to  estimate it  statistically from stellar  kinematics using
Jeans  equations.   Jeans  equations  follow  from  the  collisionless
Boltzmann  (or  Vlasov)  equation;   for  a  detailed  derivation  see
\citet{Binney1987}.   Using cylindrical  coordinates  and assuming  an
axisymmetric  (motivated  by  SDSS  results, discussed  in  detail  in
\S\ref{sec:galfast})  and  steady-state system,  the  gradient of  the
potential in  the radial  ($R$) and vertical  ($Z$) directions  can be
expressed  in  terms  of  observable quantities:  the  stellar  number
density distribution, $\nu$,  the mean azimuthal (rotational) velocity
$\overline{v_{\phi}}$,     and      four     velocity     dispersions,
$\sigma_{\phi\phi}$,  $\sigma_{RR}$, $\sigma_{ZZ}$,  and $\sigma_{RZ}$
(all six quantities as functions of $R$ and $Z$), as

\begin{eqnarray}
\label{eq:eq1}
-\frac{\partial \Phi}{\partial R} = a_{R} =
\sigma_{RR}^2 \frac{1}{\nu} \frac{\partial \nu}{\partial R} +
\frac{\partial \sigma_{RR}^2}{\partial R} + \\
\sigma_{RZ}^2 \frac{1}{\nu} \frac{\partial \nu}{\partial Z} + 
\frac{\partial \sigma_{RZ}^2}{\partial Z} +
\frac{\sigma_{RR}^2 }{R} - \frac{\sigma_{\phi\phi}^2 }{R} -
\frac{{ \overline{v_\phi} }^2 }{R}, \nonumber
\end{eqnarray}
and
\begin{eqnarray}
\label{eq:eq2}
-\frac{\partial \Phi}{\partial Z} = a_{Z} =
\sigma_{RZ}^2 \frac{1}{\nu} \frac{\partial \nu}{\partial R} +
\frac{\partial \sigma_{RZ}^2}{\partial R} + \\
\sigma_{ZZ}^2 \frac{1}{\nu} \frac{\partial \nu}{\partial Z} + 
\frac{\partial \sigma_{ZZ}^2}{\partial Z} +
\frac{\sigma_{RZ}^2 }{ R} \nonumber. 
\end{eqnarray}

Given accelerations $a_{R}(R,Z)$ and $a_{Z}(R,Z)$, \ie the gradient of
the  gravitational   potential,  dark  matter   contributions  to  the
potential  can be  estimated after  accounting for  contributions from
visible  matter.    For  notational  simplicity,  we   call  the  term
$-\partial \Phi  / \partial R$ the  ``acceleration'' $a_{R}$; however,
it is  only one component of  the true $R$ acceleration  (that is, the
time derivative  of the velocity  component in the $R$  direction): $d
v_R    /d    t    =    -    \partial   \Phi    /    \partial    R    +
(\sigma_{\phi\phi}^2+\overline{v_\phi}^2)/R$.  Of  course, in case the
of $Z$ component, $d v_Z /d t = - \partial \Phi / \partial Z$.
 
Traditionally, Galactic studies utilizing Jeans equations were limited
by  data  to  the   solar  neighborhood  \citep[within  $\sim$150  pc,
  $\eg$][]{Kapteyn1922,  Oort1960, Bahcall1984}.  The  main conclusion
drawn from local  studies is that dark matter  contributes a small (of
the  order  10$\%$)  fraction  of  gravitational  mass  in  the  solar
neighborhood \citep[corresponding to  about 0.01 $\Msun$ pc$^{-3}$, or
  $0.38$        GeV       cm$^{-3}$,][]{Kuijken1989c,       Creze1998,
  Holmberg2000}. None  of the  local studies produced  a statistically
significant detection of dark matter.

Several groups have extended these studies to a few kpc from the plane
of the disk  \citep{Kuijken1991, Siebert2003, Holmberg2004, Smith2012,
  Bovy2012a}.  Recently,  \citet{Garbari2012} used a sample  of 2000 K
dwarf  stars that extend  to 1  kpc above  the plane  of the  disk and
estimated the  local dark  matter density distribution  ${\rho}_{DM} =
(0.022 \pm  0.015) \, \Msun$  pc$^{-3}$, and \citet{Zhang2013}  used a
sample of 9000  K dwarfs with spectra from  SDSS/SEGUE that extends to
$\sim$2  kpc from  the plane  to estimate  ${\rho}_{DM} =  (0.0065 \pm
0.0023) \, \Msun$ pc$^{-3}$.  Using kinematic data for $\sim$400 thick
disk stars  at distances  of a  few kpc from  the Galactic  plane from
\citet{MoniBidin2012},   \citet{Bovy2012}  estimated   ${\rho}_{DM}  =
(0.008 \pm 0.003)  \, \Msun$ pc$^{-3}$ ($0.3 \pm  0.1$ GeV cm$^{-3}$).
Note that  the remarkably small quoted errors  by \citet{Bovy2012} and
\citet{Zhang2013}   imply   a   statistically  significant   dynamical
detection of dark matter in the solar neighborhood.

It has been difficult to extend these measurements to distances beyond
a few kpc from  the solar neighborhood.  \citet{Loebman2012} presented
a brief research  note which applied the Jeans  equations technique to
SDSS  observations of  Galactic halo  stars;  here we  present a  more
detailed analysis  and discussion of  this technique and  motivate its
future application in the era of Gaia and LSST.

This  paper  consists  of  two   logical  parts:  we  first  test  the
performance  of Jeans equations  when applied  to a  realistic stellar
system,  and  then  we  apply   Jeans  equations  to  SDSS  data.   In
\S\ref{s:background} we describe the $N$--body+SPH simulation employed
in this work  to test the Jeans equations approach, as  well as a code
for generating  mock samples of  Galactic populations trained  on SDSS
data. The main  purpose of this analysis is to  estimate the errors in
our acceleration  estimates when  using Jeans equations.  These errors
include  contributions  from   both  the  unsatisfied  assumptions  of
steady-state and smoothness, and from the shot noise that results from
analyzing  finite-sized stellar  samples. The  simulation-based tests,
presented   in   \S\ref{s:results},   demonstrate   that   the   known
accelerations  (gradients  of  the  gravitational  potential)  can  be
successfully  recovered   in  such  a  realistic   system.   Then,  in
\S\ref{Sec:DMpot},  we leverage  the baryonic  gravitational potential
recently derived from disk stars by  Bovy \& Rix (2013), and show that
the accelerations of  SDSS halo stars provide strong  evidence for the
existence of an  extended dark matter halo. We  also test whether MOND
can provide  an alternative explanation for  the observed acceleration
in  \S\ref{Sec:MOND}.  We summarize  and discuss  the validity  of our
results in \S\ref{s:conclusion}.

\section{Background}
\label{s:background}

Here we utilize  a novel application of Jeans  equations made possible
by   the   Sloan   Digital  Sky   Survey\footnote{www.sdss.org}   data
\citep[][hereafter  SDSS]{SDSS2000}.  Recently,  a  series of  studies
\citep[][hereafter,          J08,         I08          and         B10
  respectively]{Juric2008,Ivezic2008,Bond2010}     leveraged    SDSS's
substantial sky  coverage and  accurate multi-color photometry  to map
the   Galactic  stellar  number   density  distribution   and  stellar
kinematics  out to  galactocentric distances  of $\sim$20  kpc.  Using
numerous main  sequence stars, these distributions  are extremely well
sampled  and span a  sufficiently large  physical space  to investigate
stellar acceleration  via Jeans equations.  The key  issue in applying
this form of Jeans equations is determining the spatial derivatives of
the    velocity    dispersions    (see   Equations~\ref{eq:eq1}    and
\ref{eq:eq2}); they are hard, if not impossible, to reliably constrain
using only the local  Solar neighborhood data.  However, these spatial
derivatives can be directly measured using SDSS data.

We address here the following main questions:
\begin{itemize}
\item  Given  that both  observations  of  the  MW  and  modern
       $N$--body  simulations  do not  support  a simple  steady-state
       picture  (e.g.  due  to  mergers), nor  a  perfect  cylindrical
       symmetry,  is   it  indeed   possible  to  recover   the  known
       gravitational  potential in  a $N$--body  simulation  by simply
       applying  Jeans equations to  simulated stellar  number density
       distribution and kinematic data?
\item  If so,  are the stellar  acceleration maps  derived from  SDSS data
       consistent with expectations based only on visible matter?
\item  If not,  what are the differences in  the morphology of stellar
       acceleration maps  due to the  inclusion of a dark  matter component
       and what can be inferred about its distribution?
\end{itemize}

In  this section  we  describe the  background  information and  tools
needed to investigate these questions,  and we then provide answers in
the following section.

\subsection{The $N$--body+SPH Simulation}
\label{s:simulation}

To test the Jeans equations approach, we apply our analysis tools to a
simulation  with  {\it known}  stellar  accelerations, velocities  and
stellar  spatial (number density)  distribution.  This  simulation has
been previously studied in \citet{Zolotov2012} and \citet{Munshi2013}.
It  is a  cosmologically  derived \citep[WMAP3,][]{Spergel2003}  Milky
Way--mass   galaxy  evolved   for  $13.7$   Gyr  using   the  parallel
$N$--body$+$SPH{\footnote{For  notational  simplicity,  hereafter,  we
    refer to this galaxy  as ``the $N$--body simulation;'' however, it
    is   truly   a   $N$--body$+$SPH  simulation.}}    code   GASOLINE
\citep{Wadsley2004}, which contains realistic gas, cooling and stellar
feedback \citep{Stinson2006, Shen2009, Christensen2012}.  We track the
galaxy's   formation  and   evolution  using   the   zoomed-in  volume
renormalization  technique\footnote{The simulation  in this  paper was
  initially selected  from a uniform resolution,  DM-only, 50 comoving
  Mpc box.  The galaxy was  then resimulated at higher resolution (and
  with gas particles).  The volume renormalization technique simulates
  only the region within a few virial radii of the primary halo at the
  highest resolution, while still  maintaining the large 50 Mpc volume
  at low  resolution.  This accounts  for the large scale  tidal field
  that    builds   angular   momentum    in   tidal    torque   theory
  \citep{Peebles1969,Barnes1987}.}     \citep{Katz1993,    Brooks2011,
  Pontzen2008,  Governato2012}.   Our   simulated  galaxy  includes  a
stellar halo, which  is built up primarily during  the merging process
in a $\Lambda$CDM cosmology \citep[\eg][]{Bullock2005,Zolotov2009}.

GASOLINE simultaneously calculates  the potential and the acceleration
that  particles feel;  force  calculations are  consistent with  other
state-of-the-art  cosmological  gas-dynamical codes  \citep{Power2003,
Scannapieco2012}.  The typical RMS  acceleration error is $\sim$ 0.2\%
\citep{Wadsley2004}.   Full  6D phase  space  ($x$,  $y$, $z$,  $v_x$,
$v_y$, $v_z$) and mass information is also tracked.  

At  the end  of  the simulation,  the  average star  particle mass  is
$\sim5800$   $\Msun$   and   the   dark  matter   particle   mass   is
$1.3\times10^5$ \Msun,  with the minimum dark  matter spline softening
length of $173$  pc.  At redshift of zero, the  simulated galaxy has a
virial radius  of $227$  kpc and a  virial mass  of $6.8\times10^{11}$
$\Msun$\footnote{The  virial mass  and  virial radius  is measured  at
  $100$*$\rho_{critical}$}, of this final mass,  7\% is in gas, 6\% is
in stars, and  87\% is in dark matter. The dark  matter to baryon mass
ratio  in a  region corresponding  to the  solar neighborhood  ($7 \le
R/kpc  \le  9$,  ($0  \le  |Z|/kpc  \le  1$)  is  36\%.   A  total  of
$4.6\times10^{6}$    dark    matter,    $2.1\times10^{6}$   gas    and
$7.4\times10^{6}$  star  particles are  within  the  virial radius  at
redshift  of  zero.  The  simulated  galaxy is  \textit{approximately}
rotationally symmetric (a total enclosed matter axis ratio $b:a > 0.9$
within 100 kpc, and a stellar matter axis ratio $b:a > 0.95$ at $R$=10
kpc; see \S~\ref{measuring_axial_symmetry} for details), has a Johnson
system $R$-band  disk scale length of $\sim$3.1  kpc and corresponding
bulge to  disk ratio of 0.33 \citep{Brooks2011},  and maximum circular
velocity of  $\sim$235 km/s.   These structural parameters  are within
10\%  of those measured  for the  Milky Way  \citep[for example,][find
  that the virial  mass of the Milky Way's dark matter  halo is in the
  range $8$--$13$  $\times$ $10^{11}$ $\Msun$,  see Table~\ref{t:comp}
  for details]{Xue2008}.

\begin{table*}
\begin{center}
\begin{tabular}{lcc}
\hline
                                          & Milky Way                               & $N$--body Simulation \\
Virial Radius ($kpc$)                     & $200$\tablenotemark{$a$}                  &  $227$               \\ 
\hline
Virial Mass ($\Msun$)                     & $1.0\times10^{12}$\tablenotemark{$b$,$c$}   &  $6.8\times10^{11}$  \\ 
\hline
Johnson R-band  Disk Scale Length ($kpc$) & $3.6$\tablenotemark{$d$}                  &   $3.1$              \\  
\hline   
Maximum  Circular  Velocity ($km$/$s$)    & $220$\tablenotemark{$b$}                  & $235$                \\
\hline
\end{tabular} 
\end{center}
\center{\tablerefs{$^a$\cite{Boylan2011}, $^b$\cite{Xue2008}, $^c$\cite{Klypin2002}, $^d$\cite{Juric2008}}}
\caption{A  comparison of  various structural  parameters  between the
  Milky Way and the adopted $N$--body Simulation}
\label{t:comp}
\end{table*}

\begin{figure*}[!h]
\epsscale{1}
\hskip -.15 in
  \plotone{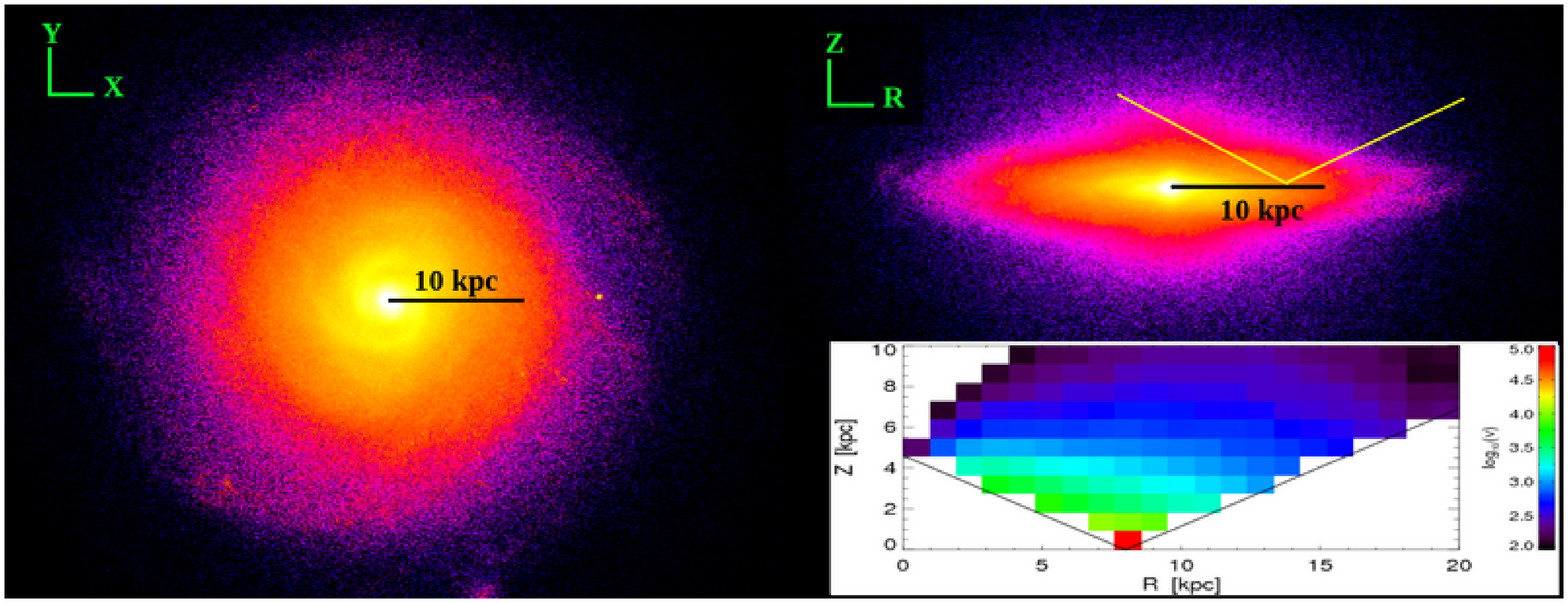} 
  \caption{(Left) top  down view of the  stellar particle distribution
  (shown on  a logarithmic  scale) at $Z=0$  of the  adopted simulated
  MW--like  galaxy.  (Top  right)  edge-on view  of  the same  stellar
  particle  distribution.   The   yellow  lines  indicate  the  region
  selected in our  analysis to mimic the SDSS  volume.  (Bottom right)
  the number of stellar particles within the selected SDSS volume when
  binned  in 1.0  kpc x  1.0 kpc  $R$-$Z$ bins;  this  high resolution
  simulation has  enough stellar particles  (at least 100 per  bin) to
  conduct a statistical analysis in the volume probed by SDSS.  }
\label{f:tipsy_view}
\end{figure*}

For reference, Figure~\ref{f:tipsy_view} gives a visual perspective of
the $N$-body      simulation      used      throughout     this  paper.
Figure~\ref{f:tipsy_view}  shows a  top-down and  edge-on view  of the
stellar  particle   distribution  at   $Z=0$  when  visualized   on  a
logarithmic scale.   The edge-on view has yellow  lines overplotted to
indicate the region selected in our analysis to mimic the SDSS volume.
Also plotted  is the number  of stellar particles within  the selected
SDSS volume when binned in 1.0 kpc x 1.0 kpc $R$-$Z$ bins.  This panel
illustrates  that  {\it  our  high resolution  simulation  has  enough
stellar  particles (at  least 100  per bin)  to conduct  a statistical
analysis in a synthetic SDSS volume.}

\subsubsection{The Spatial Distribution of Mass in the Simulated Galaxy}

\begin{figure*}[!h]
\epsscale{.49}
\hskip -.15 in
  \plotone{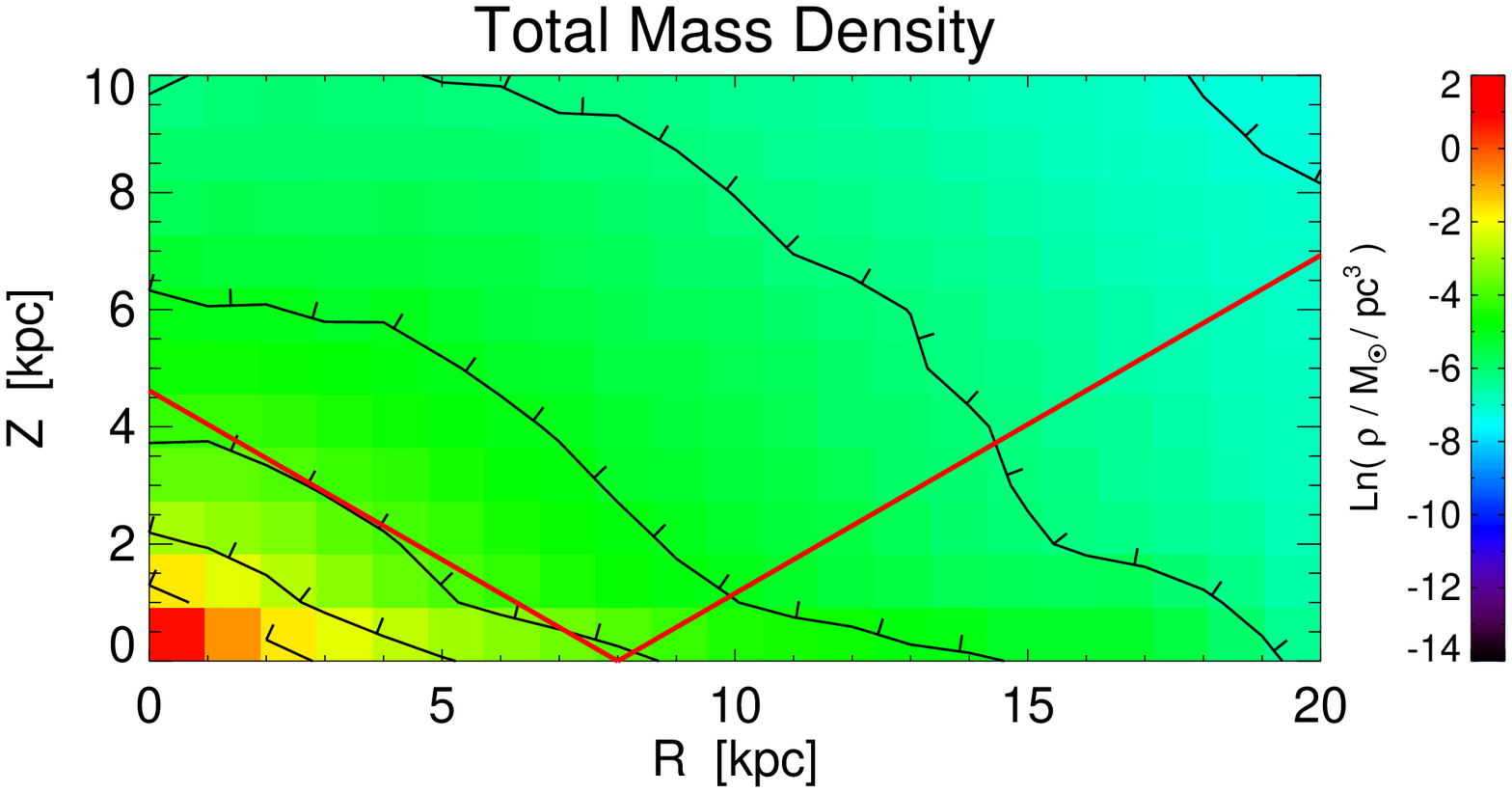}
  \plotone{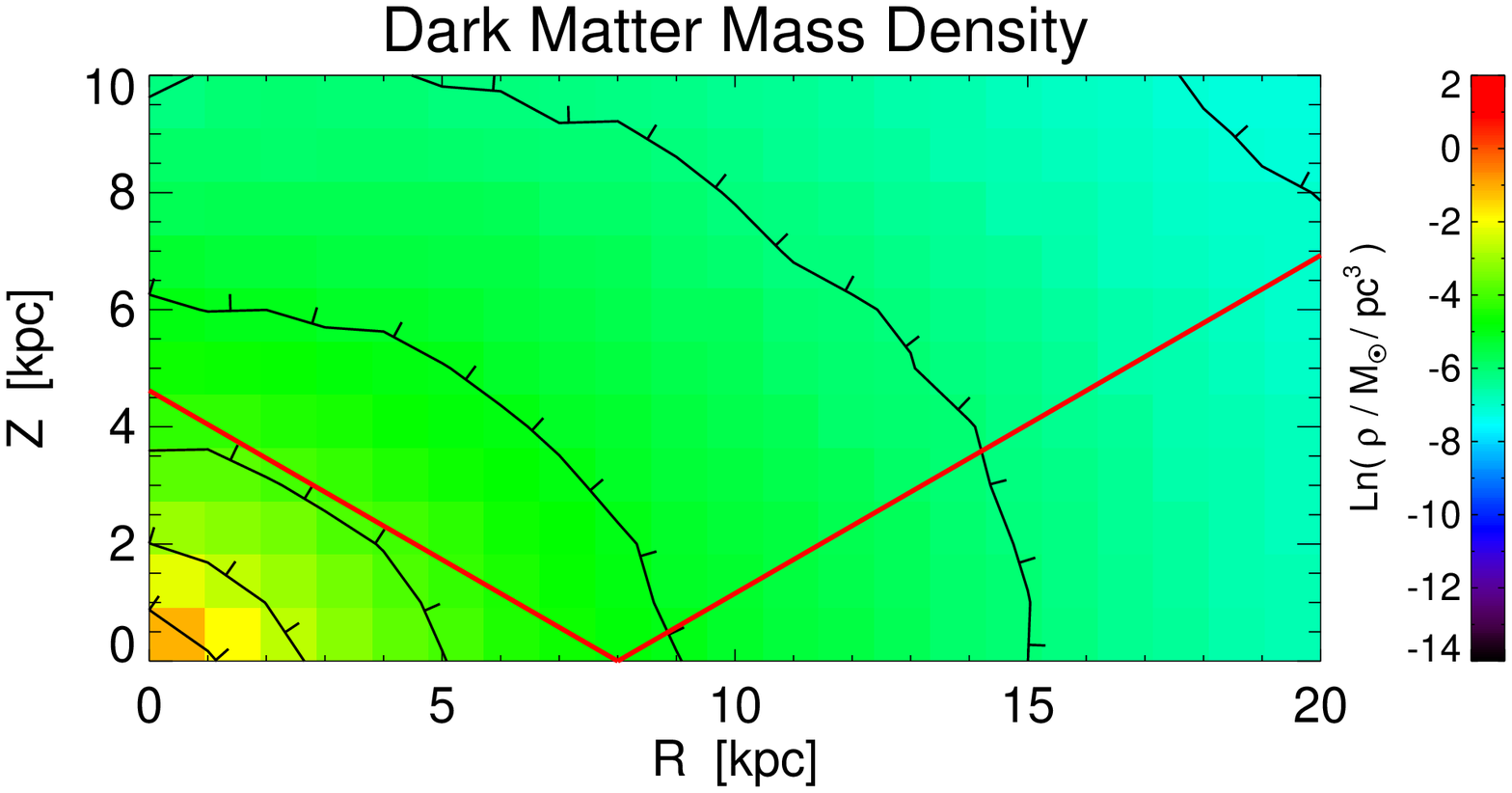}
  \plotone{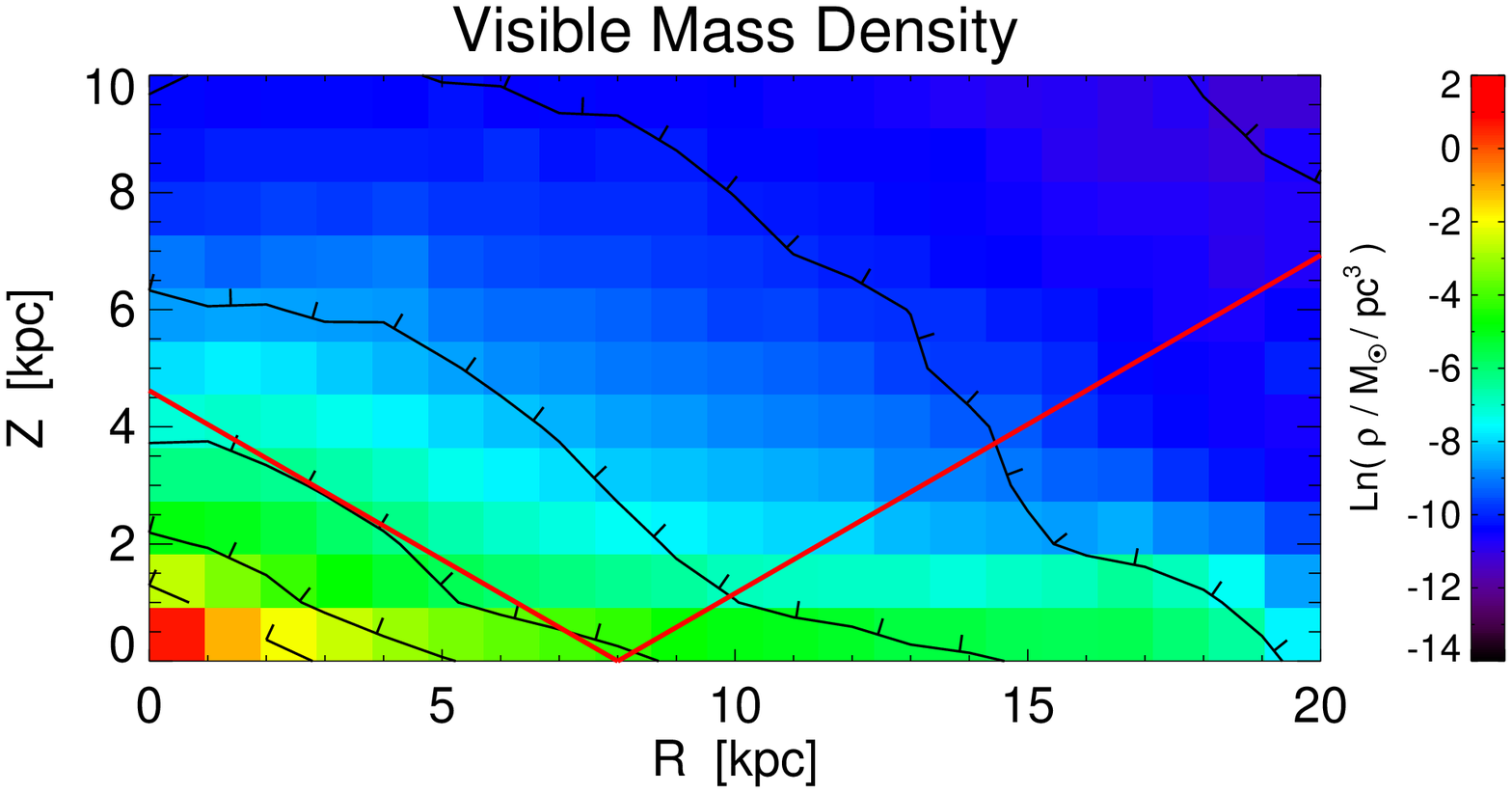}
  \plotone{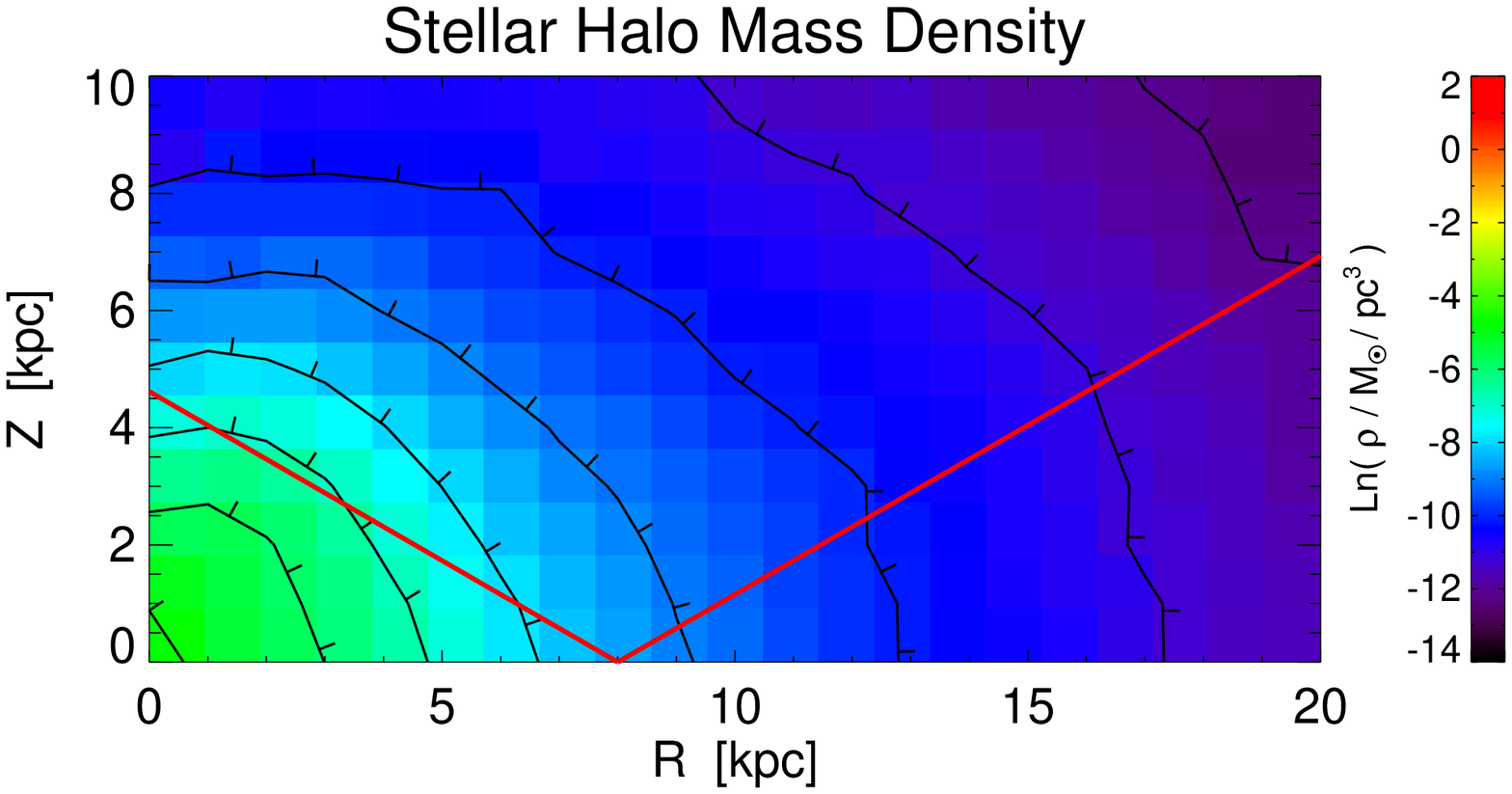}
  \caption{The azimuthally averaged mass density maps of four relevant
    quantities  within the $N$--body  simulation: total,  dark matter,
    visible, and  stellar halo mass.   The displayed dynamic  range is
    the  same in  all  panels for  easy  comparison.  Overplotted  are
    logarithmically  spaced isodensity  contours;  contour tick  marks
    correspond   to  the  direction   of  decreasing   density.   Also
    overplotted in  red is the  SDSS footprint within  the simulation.
    (Top left)  The total  mass density (gas,  dark matter  and stars)
    within R  $\le 20$ kpc  and Z  $\le 10$ kpc  of the center  of the
    $N$--body simulated  galaxy.  (Top right) The  dark matter density
    within  the simulation.  The  majority of  the total  mass density
    within the SDSS footprint is  from the dark matter.  (Bottom left)
    The mass density of all  visible matter (gas and stars) within the
    $N$--body simulation.   The bulge (R $\le  5$ kpc, Z  $\le 4$ kpc)
    and disk ($5$  kpc $\le$ R $\le 20$ kpc, Z  $\le 2$ kpc) structure
    are evident within this  distribution.  (Bottom right) The stellar
    halo  mass density  within the  simulation.  The  majority  of the
    visible mass within the SDSS footprint is from the stellar halo.}
\label{f:sim_mass_maps}
\end{figure*}

Many of the plots throughout this  paper show a total or mean quantity
mapped into rectilinear  bins in $R$-$Z$ space within  $0$ $\le$ R/kpc
$\le 20$ and $0$ $\le$ Z/kpc $\le 10$.  This perspective gives a sense
of the two dimensional distribution  of a quantity throughout the SDSS
volume.  Figure~\ref{f:sim_mass_maps} provides  an example of this for
four relevant quantities within  the $N$--body simulation: total, dark
matter, visible,  and stellar halo  mass density.  The  SDSS footprint
within the  simulation (shown here  in red) is always  overplotted for
reference.  The  top left panel  of Figure~\ref{f:sim_mass_maps} shows
the total mass density, including  gas, dark matter and stars.  To the
right  of this  panel is  the dark  matter density  distribution.  The
significance  of dark  matter relative  to the  gas and  stars  is not
constant, yet the  majority of the total mass  density within the SDSS
footprint  is  clearly  from  dark  matter.   The  bottom  two  panels
illustrate the  distribution of visible  matter.  The bottom  panel on
the left of Figure~\ref{f:sim_mass_maps} shows the mass density of all
gas  and   stars  within  the  $N$--body   simulation.   Two  striking
structural features stand out within this total visible matter density
map: the bulge (R $\le 5$ kpc,  Z $\le 4$ kpc) and disk ($5$ kpc $\le$
R  $\le  20$   kpc,  Z  $\le  2$  kpc).    These  structures  are  not
significantly sampled  by the SDSS volume within  the simulation.  The
bottom  right-hand  panel  of Figure~\ref{f:sim_mass_maps}  shows  the
stellar  halo  mass density  within  the  simulation.   Note that  the
majority of  the visible  mass within the  SDSS footprint is  from the
stellar halo.

\subsubsection{Tests of Axial Symmetry}\label{measuring_axial_symmetry}

\begin{figure*}[!h]
\epsscale{.6}
\hskip -.15 in
   \plotone{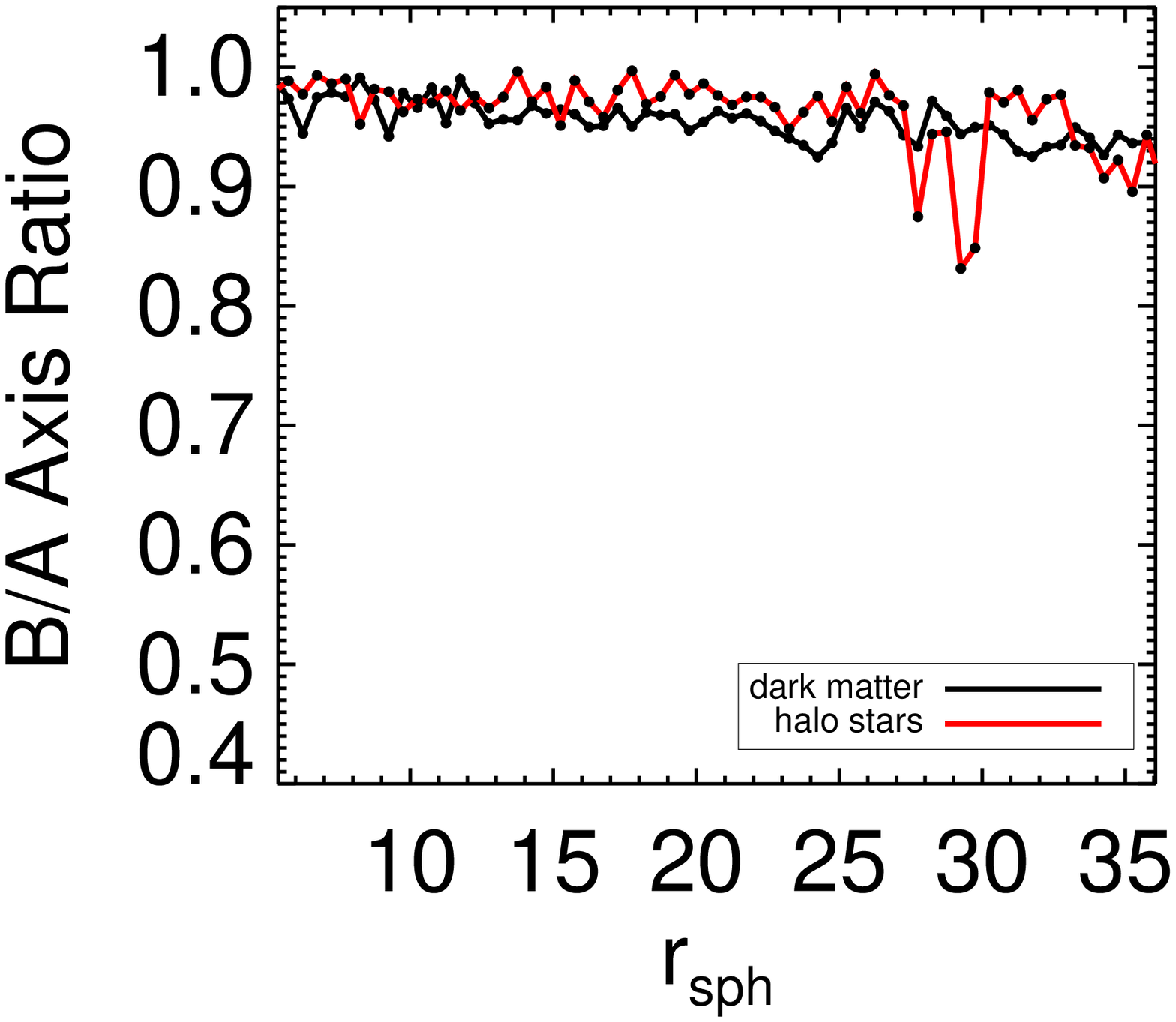}
   \plotone{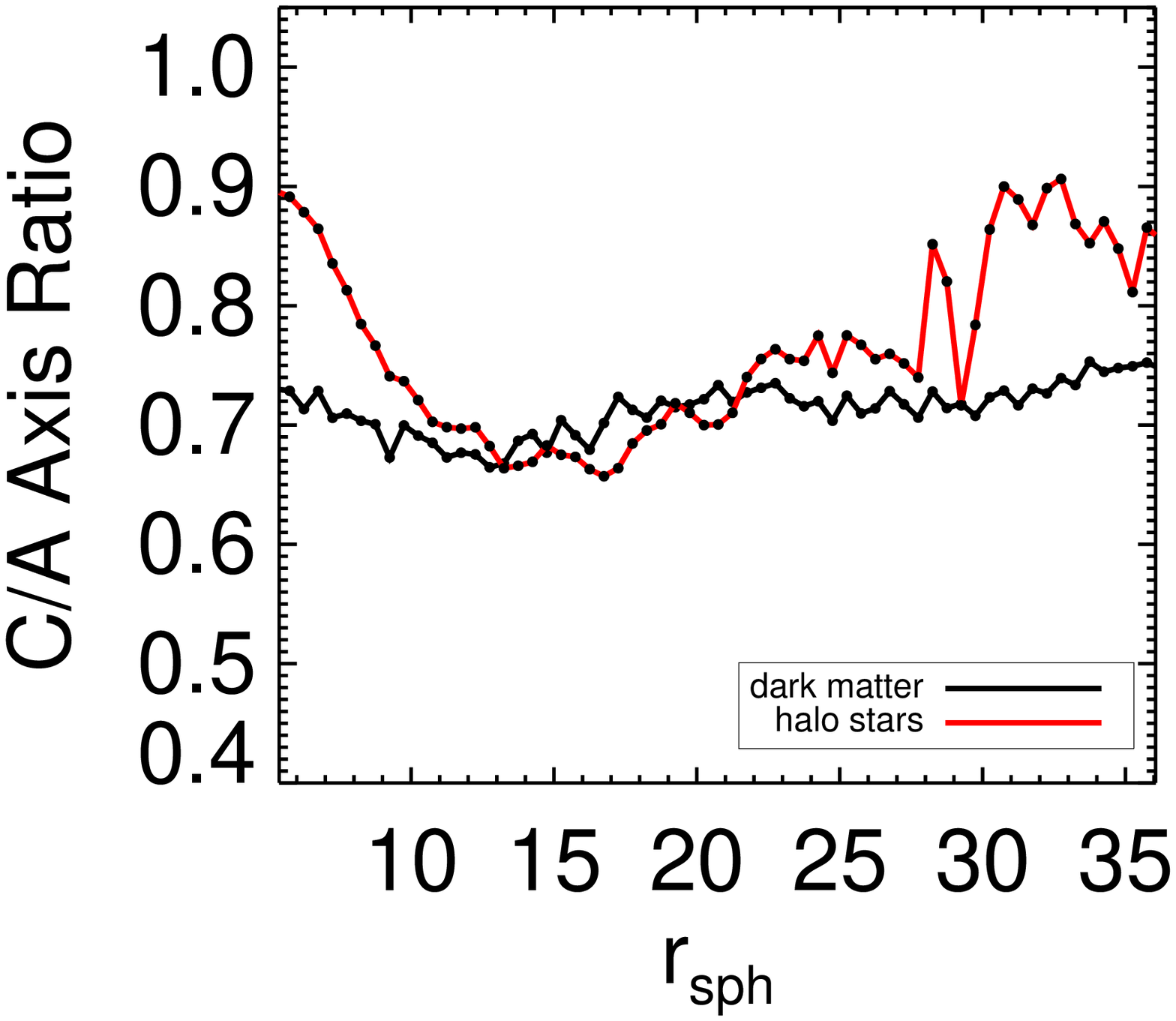}
   \caption{(Top)  The  semi-minor to  semi-major  axis  ratio in  the
   equatorial  plane ($b/a$) of  dark matter  and halo  star particles
   across the  SDSS volume within  the simulation. In both  cases, the
   $b/a$ axis ratio is always greater  than or equal to $0.8$ and less
   than  $1.0$,  indicating  the  distributions  are  nearly  but  not
   completely  axisymmetric  in  the  $\phi$ direction.   (Bottom)  An
   analogous  figure  to  the top  panel  but  for  the ratio  of  the
   semi-minor  axis  perpendicular to  the  equatorial  plane and  the
   semi-major  axis   ($c/a$).  Both  the  dark   matter  and  stellar
   distributions are oblate, and the dark matter $c/a$ axis ratio does
   not vary significantly within the SDSS volume.}
\label{f:axis_ratios}
\end{figure*}

Before we project  mean quantities in the $R$-$Z$  spatial grid or use
the axisymmetric form of  Jeans equations, we motivate the application
of these techniques by  illustrating the simulation's $\phi$ symmetry.
The  top  panel  of  Figure~\ref{f:axis_ratios}  shows  the  major  to
semi-major axis ratio  ($b/a$) of dark matter and  halo star particles
across the SDSS  footprint within the simulation.  Axis  ratios of the
particle distribution are determined following the iterative technique
outlined in \S 4.2  of \citet{Roskar2010}, which identifies isodensity
contours.    This   procedure   is   analogous   to   that   used   in
\citet{Katz1991}, though it uses differential shells (in increments of
0.5     kpc)    instead     of     cumulative    shells,     following
\citet{Debattista2008}.   For both dark  matter particles  and stellar
halo particles, the  $b/a$ axis ratio is always  greater than or equal
to $0.8$ and less than  $1.0$, indicating the distributions are nearly
but  not completely  axisymmetric  in the  $\phi$  direction.  At  the
virial radius, the $b/a$ axis  ratio for all particles is $0.91$.  The
bottom  panel of  Figure~\ref{f:axis_ratios} is  analogous to  the top
panel but for  the major to minor axis ($c/a$).   The $c/a$ axis ratio
is a measure of the departure from spherical symmetry for axisymmetric
shells.   A  $c/a$ axis  ratio  of 1  is  a  perfectly spherical  mass
distribution; $c/a$ $< 1$ indicates that the distribution is flattened
(oblate) in the same sense as the stellar disk.  At the virial radius,
the $c/a$ axis ratio for all particles is $0.74$.  As the bottom panel
of  Figure~\ref{f:axis_ratios} shows,  in  this $N$--body  simulation,
both the  dark matter  and stellar distributions  are oblate,  and the
dark  matter $c/a$  axis ratio  does not  vary significantly  over the
entire SDSS volume.

\subsubsection{The Acceleration Maps for a Simulated Galaxy}

One  final thing  to consider  before we apply  Jeans equations  to the
simulation:  what  do  the   true  accelerations  look  like  for  the
simulation?  The  top panel of Figure~\ref{f:ratio_az}  shows the mean
component of the acceleration in  the $Z$ direction projected into the
$R$-$Z$ grid;  here, the acceleration of each  particle was calculated
using the force from all  the particles in the entire simulation.  For
comparison, the  middle panel  shows an analogous  map, but  here, the
acceleration of each  star and gas particle was  calculated using only
the  contributions from  other star  and gas  particles (that  is, the
contribution  from the  dark matter  was not  included).   As evident,
there are substantial  differences in the morphology of  the two maps;
the bottom  panel shows the ratio  of the top and  middle panel.  This
panel  demonstrates  that  {\it  the  effect of  dark  matter  on  the
acceleration  in the  $Z$ direction  increases quickly  away  from the
plane of  the disk  and towards  the outer parts  of the  galaxy}; for
example, the ratio of accelerations  is doubled by $R$=8 kpc and $Z$=6
kpc. These distances are probed by SDSS -- hence these results suggest
that  the  effect  of  dark  matter on  stellar  acceleration  may  be
uncovered in SDSS  data, and that stellar populations  in the halo are
more sensitive to the existence of dark matter than disk stars.

Along  the  same  lines,  the  top  panel  of  Figure~\ref{f:ratio_ar}
illustrates  the  mean  component  of  the  acceleration  in  the  $R$
direction when  the force of all  the particles (gas,  dark matter and
stars) in the  simulation is considered, while the  middle panel shows
the mean component  when the force of just gas  and star particles are
considered.  The  bottom panel shows a  ratio of the top  panel to the
middle panel; the effects  of dark matter  are easily discernible; for
example, the ratio of accelerations  is doubled by $R$=8 kpc and $Z$=4
kpc. 

\subsection{SDSS-based Mock Catalogs: \textit{galfast}}\label{sec:galfast}

When constraining the Galactic potential via Jeans equations with SDSS
(or  any other survey)  data, several  preliminary analysis  steps are
required:
\begin{enumerate}

\item In order to quantify  the stellar number density distribution as
  a    function   of    coordinates    $R$   and    $Z$   ($\nu$    in
  Equations~\ref{eq:eq1}  and \ref{eq:eq2}),  the  appropriate stellar
  population needs to be selected (e.g.  halo stars), the distances to
  the  stars need  to be  estimated, and  the  observational selection
  function accounted for.  In  addition, the assumption of cylindrical
  symmetry  must  be tested,  and  the  impact  of local  substructure
  (e.g. stellar streams) quantified.

\item In order to quantify  the four velocity dispersions and the mean
azimuthal velocity  as functions of  coordinates $R$ and  $Z$, complex
kinematics (proper  motion and  radial velocity  measurements) are
needed  and  require substantial  analysis.  For  example, the  error
dependence for the radial velocity components and the error dependence
for the tangential velocity  components are fundamentally different as
a function  of distance.  Notably, the  tangential velocity components
are   computed  as  the   product  of   distance  and   proper  motion
measurements, and  these errors carry  their own hidden  dependence on
distance.  Proper motion  errors increase  for faint  stars,  and more
distant stars are generally fainter than closer ones.

\end{enumerate} 
These tasks  are far from  trivial, but fortunately they  have already
been undertaken and published.

\subsubsection{The Stellar Number Density Distribution for Halo Stars} 

J08  accomplished the  first task  of quantifying  the  stellar number
density distribution  for both disk and halo  components.  They showed
that the stellar number  density distribution, $\nu(R,Z,\phi)$, can be
well  described (apart  from  local  overdensities) as  a  sum of  two
cylindrically symmetric components

\begin{equation} 
\label{eq:nuDH}
           \nu(R,Z,\phi) = \nu_D(R,Z) + \nu_H(R,Z).
\end{equation}

The disk  component can be modeled  as a sum of  two exponential disks
(see their Equations~22 and 23),  while the halo component requires an
oblate bi-axial (cylindrically symmetric) power-law model

\begin{equation} 
\label{eq:nuDH2}
 \nu_H(R,Z)= \nu_D(R_\odot,0) \,  \epsilon_H \, \left({R_\odot^2 \over
 R^2 + (Z/q_H)^2}\right)^{n_H/2},
\end{equation} 

Here $\nu_D(R_\odot,0)$  is the  local solar neighborhood  density of
tracer   stars,  and  $\epsilon_H$   measures  the   local  fractional
contribution   of  halo  stars.    The  number   count  normalization,
$\nu_D(R_\odot,0)$,  reflects how  tracer stars  are selected,  and is
related  to   the  local  luminosity  function.    Since  the  overall
normalization of $\nu(R,Z)$ in Equations~\ref{eq:eq1} and
\ref{eq:eq2}  cancels  out,   $\nu_D(R_\odot,0)$  is  not  of  further
interest in this context.

J08 obtained best-fit MW  parameters using SDSS data, after accounting
for selection effects and masking regions with prominent substructure;
their results  are listed for both  the stellar disk  and stellar halo
components in their Table  10 (second column).  For completeness, they
obtained   $\epsilon_H=0.0051$,  $q_H=0.64$,   and   $n_H=2.77$,  with
estimated  uncertainties of  25\%,  $\lesssim$0.1, and  $\lesssim$0.2,
respectively.  We  note that the  best-fit values for $q_H$  and $n_H$
are covariant -- the more  symmetric halos correspond to larger $n_H$,
see their Figure  22.  They also tested for  cylindrical symmetry (see
their  Figure 11)  and could  not reject  their  best-fit axisymmetric
number counts model.

I08 studied the  metallicity distribution of disk and  halo stars and,
of   direct   relevance   to   this  work,   demonstrated   that   the
multi-component (i.e.  disk and halo) decomposition of $\nu(R,Z)$ from
Equation~\ref{eq:nuDH}  is  not a  case  of  over-fitting; instead,  a
fairly  simple selection, $[Fe/H]  = -1$,  clearly separates  disk and
halo  components  (see  their  Figures  5 and  9)  and  justifies  the
decomposition model from Equation~\ref{eq:nuDH}.

\subsubsection{The Kinematic Behavior of Halo Stars} 

B10 performed a detailed analysis  of available kinematic data for the
SDSS  stellar sample:  radial velocities  were derived  from  the SDSS
spectroscopic  survey and  proper motions  were obtained  by comparing
SDSS  astrometry   and  Palomar  Observatory   Sky  Survey  astrometry
\citep{Munn2004} from  $\sim$50 years  earlier.  Their main  result of
interest to this  work is a clear demonstration  (see their Figures~12
and 13)  that the  velocity ellipsoid for  halo stars is  invariant in
spherical  coordinates   within  the   volume  probed  by   SDSS  data
(galactocentric  distances  of $\lesssim$20  kpc).   The very  complex
behavior of  measured proper motions (see their  Figure~14) and radial
velocities (see  their Figure~15) on the  sky can be  explained with a
simple  triaxial velocity  ellipsoid  that is  invariant in  spherical
coordinates, $\sigma_{rr}$=141  km s$^{-1}$, $\sigma_{\phi\phi}$=85 km
s$^{-1}$,    and   $\sigma_{\theta\theta}$=75   km    s$^{-1}$,   with
uncertainties  of  about 5  km  s$^{-1}$.   Their  leading sources  of
uncertainty are distance scale  errors, local standard of rest errors,
and   systematic   errors   in   radial-velocity   and   proper-motion
measurements; see  their section 5.3  for details. Given  the velocity
ellipsoid  in   spherical  coordinates,  it  can   be  transformed  to
cylindrical coordinates as
\begin{equation} 
\label{eq:sigRR}
   \sigma_{RR}^2 = \sigma_{rr}^2 \,\cos(\alpha)^2  + \sigma_{\theta\theta}^2\,\sin(\alpha)^2,
\end{equation} 
\begin{equation} 
\label{eq:sigZZ}
   \sigma_{ZZ}^2 = \sigma_{rr}^2 \,\sin(\alpha)^2 + \sigma_{\theta\theta}^2\,\cos(\alpha)^2,
\end{equation} 
and
\begin{equation} 
 \label{eq:sigRZ}
  \sigma_{RZ}^2 = (\sigma_{rr}^2 - \sigma_{\theta\theta}^2)\,\sin(\alpha)\,\cos(\alpha),
\end{equation} 
where $\alpha = \tan^{-1}(Z/R)$.

Together  with  the  spatial  distribution  of  halo  stars  given  by
Equation~\ref{eq:nuDH2},  these equations  are sufficient  to evaluate
all  terms listed  in Equations~\ref{eq:eq1}  and  \ref{eq:eq2}. These
``direct''  analytic  acceleration maps  are  discussed  in detail  in
\S\ref{Sec:DMpot}. 

\subsection{The \textit{galfast} Code} 

The best-fit $\nu_H(R,Z)$ from J08 and the best-fit velocity ellipsoid
for   halo  stars  from   B10  can   be  inserted   analytically  into
Equations~\ref{eq:eq1}  and   \ref{eq:eq2}  to  compute   $a_{R}$  and
$a_{Z}$.   Such  analytic  results   properly  account  for  the  SDSS
selection function and  Galactic substructure.  However, this approach
does not include the effects of finite stellar counts, counting noise,
and volume edges.  Such sampling effects play an important role in the
analysis  of  the $N$--body  simulation,  where  we utilize  numerical
derivatives of  the ``observed'' velocity ellipsoid  and impose strict
stellar count criteria.

To  leverage the  computational methods  developed and  tested  in the
$N$--body framework, we instead generate  a mock catalog of SDSS stars
generated by the code \textit{galfast} \cite[][]{JuricGalfast} .  This
public\footnote{See   https://github.com/mjuric/galfast}  Monte  Carlo
code is  based on the best-fit parameterizations  of the distributions
of stellar  number density, metallicity and  kinematics constrained by
the SDSS  data mentioned  above.  It produces  catalogs with  the same
behavior of  observables (such  as counts, magnitudes,  colors, proper
motions, radial velocity) as seen in SDSS data, {\it except that there
  are no  effects of substructure},  and selection effects  are easily
accounted for  (e.g., one  can generate a  mock catalog for  the whole
Galaxy, and  then apply  exactly the same  selection criteria  to this
mock  catalog  and  to  the  $N$--body  simulation).   The  code  also
generates appropriate error distributions of all measured quantities.

We  note that  there  are  no hidden  inputs,  such as  star-formation
history, age-metallicity relation,  etc., included in \textit{galfast}
-- it  is simply  a sophisticated  Monte Carlo  generator  designed to
produce  a  snapshot of  the  current  sky  with the  stellar  content
consistent with SDSS observations.

Using \textit{galfast}, we generate  a flux-limited catalog with $14 <
r <  21$ and  mimic the  SDSS sky footprint  by only  considering high
Galactic   latitudes  ($|b|>30^\circ$).    The   catalog  lists   true
positions, absolute magnitudes, velocities and metallicity, as well as
corresponding simulated  SDSS observations convolved  with measurement
errors.

We treat  this mock catalog as  we would treat  any catalog downloaded
from the  SDSS Data  Release site. We  correct the magnitudes  in each
filter for interstellar dust  extinction and select a halo-like sample
using a color cut $0.25 < g-r < 0.35$.  The only instance where we use
the ``truth''  provided in  the mock catalog  is when  rejecting stars
with $M_r < 4$ to minimize  contamination by giants (in a real sample,
one could  envision obtaining a  spectrum for each star  to accomplish
the  same  step).  The  resulting  sample of  0.61  million  stars  is
dominated by  low-metallicity main  sequence F stars,  with kinematics
commensurate with a halo-dominated sample.

\subsection{Numerical Procedures}\label{sec:numprocs}

We process our mock catalog from \textit{galfast} and our mock catalog
from our  adopted $N$--body simulation  in {\it exactly the  same way,
using the same code}: for  a set of stars with given three-dimensional
positions  and three-dimensional  velocities, we  first  determine the
density,  $\nu(R,Z)$, and  the five  kinematic quantities  utilized in
Equations~\ref{eq:eq1} and \ref{eq:eq2},  and then compute $a_{R}$ and
$a_{Z}$.

The  computation of the  number density,  mean azimuthal  velocity and
velocity  dispersions is  done  for  each bin  in  the $R$-$Z$  plane.
{\footnote{In the case of the $N$--body simulation, we also calculated
    these  quantities in  45$\degrees$  slices in  $\phi$, rotated  in
    increments  of 90$\degrees$  from  0  to 360;  we  found that  our
    results varied no more that 10$\%$.}}   We set the bin width to be
1  kpc, and  we require  at  least 100  stellar particles  per bin  to
minimize the  shot noise.  All  quantities are computed  using weights
proportional to the mass of each stellar particle (assumed constant in
the \textit{galfast} catalogs).

To estimate  the  gradients  required in Equations~\ref{eq:eq1}
and \ref{eq:eq2}  (\ie the  spatial gradients of  the velocity dispersions and 
the stellar  number density) we  use  a  parametric  technique: we  fit  a
second-order polynomial  in $R$ and $Z$  to values from  the bin being
processed  and its  8 adjacent  neighbors (using  IDL  fitting routine
MPFIT2DFUN), and  determine $R$ and  $Z$ gradients by  taking the analytic
derivative of  the best fit. This method  filters numerical 
noise (due to counting noise and polynomial fitting) to some extent and 
produces smoother maps (with values closer to
the truth  in the \textit{galfast} catalogs,  where we  know that
velocity dispersion  gradients in spherical  coordinates are vanishing
by construction). We exclude  edge pixels (bins) from further analysis
because the parametric results are not as robust due to the smaller number
of adjacent pixels.

\subsubsection{Tests of Numerical Procedures} \label{sec:numprocsTests}

The \textit{galfast} catalogs provide a strong test of our algorithms;
we have verified that we can recover the number density and kinematics
used  as  input  to  \textit{galfast}.  Furthermore, we  can  test  the
resulting  acceleration maps  by directly  taking  appropriate spatial
derivatives of  the analytic expressions for  the spatial distribution
from  J08  and  kinematics  from  B10  (that is,  we  can  bypass  the
\textit{galfast}  step). Since  these derivatives  (``analytic'' maps)
can be  evaluated with negligible numerical  noise, unlike derivatives
based on a  mock sample (``numerical'' maps), we  can measure the bias
and scatter due  to a finite-size sample (to  the extent that analytic
expressions  from  J08  and  B10  are  correct,  these  analytic  maps
represent ground truth; for  their illustration and further discussion
see \S\ref{Sec:DMpot}).

A comparison of the analytic  and numerical maps reveals that they are
morphologically very similar;  we find that the latter  are biased low
by  3\% for  the $a_R$  maps and  by 14\%  for the  $a_Z$ map,  with a
root-mean-square  scatter  of 25\%  (over  all  the  pixels) for  both
maps. This  performance is satisfactory for  testing the applicability
of Jeans equations to a realistic $N$--body simulated galaxy. However,
at the smallest  Z ($\sim$2.5 kpc), the $a_Z$ map is  biased low by as
much as a  factor of 1.5 at $R=8.5$ kpc. This  biasing is probably due
to  edge effects  when  fitting polynomials,  or  due to  insufficient
curvature  in the fitting  functions.  When  comparing our  results to
related  published work  (see \S\ref{Sec:DMpot})  we use  the analytic
maps, and when  comparing mock stellar samples from  the $N$--body and
\textit{galfast} simulations we use the numerical maps.

In the case of the  $N$--body simulation, we have an additional test:
if  all  algorithms  are  correctly  implemented,  {\it  and}  if  all
assumptions that go into the derivation of Jeans equations are not too
incorrect, then we ought to be  able to reproduce the true $a_{R}$ and
$a_{Z}$ that are known  from direct force calculations. This analysis
is described in the following section.

\section{Validation of the Jeans Equations Method}
\label{s:results}

In this  section we  first test the  Jeans equations approach  using a
realistic  MW-like simulated galaxy  with known  stellar accelerations
from  force  computations.   The  simulated galaxy  is  not  perfectly
cylindrically symmetric, nor is  it in a steady-state.  The comparison
of known accelerations and  those computed by Jeans equations provides
a  quantitative  assessment  of  both  systematic  and  random  errors
inherent in this method.  After quantifying these errors, we apply the
same methodology  to the \textit{galfast} catalog  and demonstrate the
signature of dark matter in the Milky Way halo.

\subsection{Tests of the Jeans Equations Method Using Simulations}
\label{s:results_simulation}

To quantify acceleration errors in  the Jeans equations method, we use
an $N$--body simulation, with positions, velocities, and accelerations
for  $7.3$ million  stellar  particles within  the  virial radius.  To
maintain identical  selection effects as  with the SDSS data,  we only
use  simulation  data  within  the SDSS footprint;  this  region  contains
220,000   stellar   particles;   their   distribution  is   shown   in
Figure~\ref{f:sim_sdss_nstarmap}.  We  include all the  star particles
from  this region  in  our analysis  (that  is, there  is no  specific
selection of ``halo stars''); however, we exclude results within 1 kpc
of the plane  of the disk to minimize the influence  of disk stars and
their strong gradients in all relevant quantities.

Our  data  is  binned  in   1  kpc  square  $R$-$Z$  pixels;  we  also
investigated  smaller bin  sizes, down  to twice  the  force softening
length  (346  parsec).   Because  the  star  particle  number  density
decreases  quickly with increased  galactocentric radius,  the adopted
size of 1 kpc is a ``sweet spot'' that allowed us to spatially resolve
gradients in the acceleration  map, while simultaneously having enough
stellar  particles  per  bin  for  counting  errors  to  remain  small
($\sim$10\%).

\begin{figure*}[!h]
\epsscale{.65}
\hskip -.15 in
  \plotone{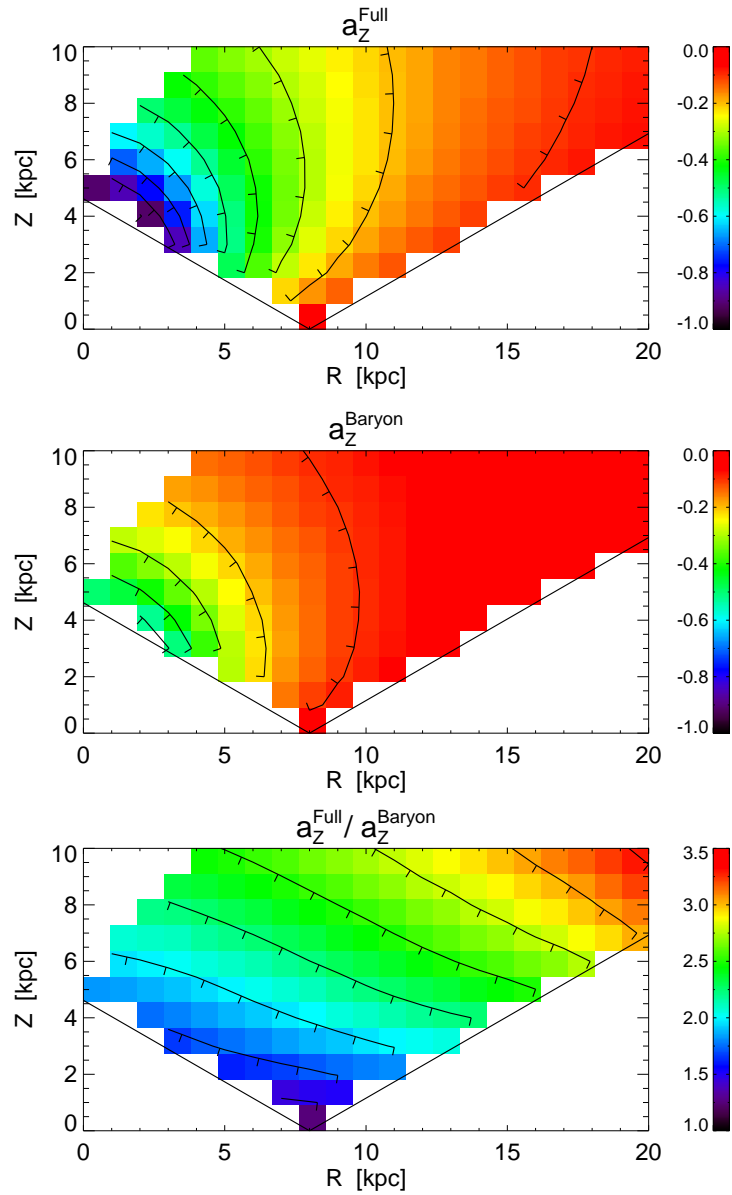}     
  \caption{ A comparison of the acceleration in the $Z$ direction from
  the $N$--body simulation when  all contributions are included (star,
  gas, and  dark matter  particles; top panel)  to the  result without
  dark  matter (middle  panel).  The  maps are  limited to  the volume
  explored by SDSS data, and the acceleration is expressed in units of
  $2.9\times10^{-13}$ $\kmss$. The  ratio of the two maps  is shown in
  the bottom panel.  The importance  of the dark matter increases with
  the distance  from the origin; at  the edge of the  volume probed by
  SDSS  ($R\sim20$  kpc, $Z\sim10$  kpc),  the  total acceleration  in
  the analyzed simulation  is about 3 times larger  than contribution from
  the visible matter.}
\label{f:ratio_az}
\end{figure*}

\begin{figure*}[!h]
\epsscale{.65}
\hskip -.15 in
  \plotone{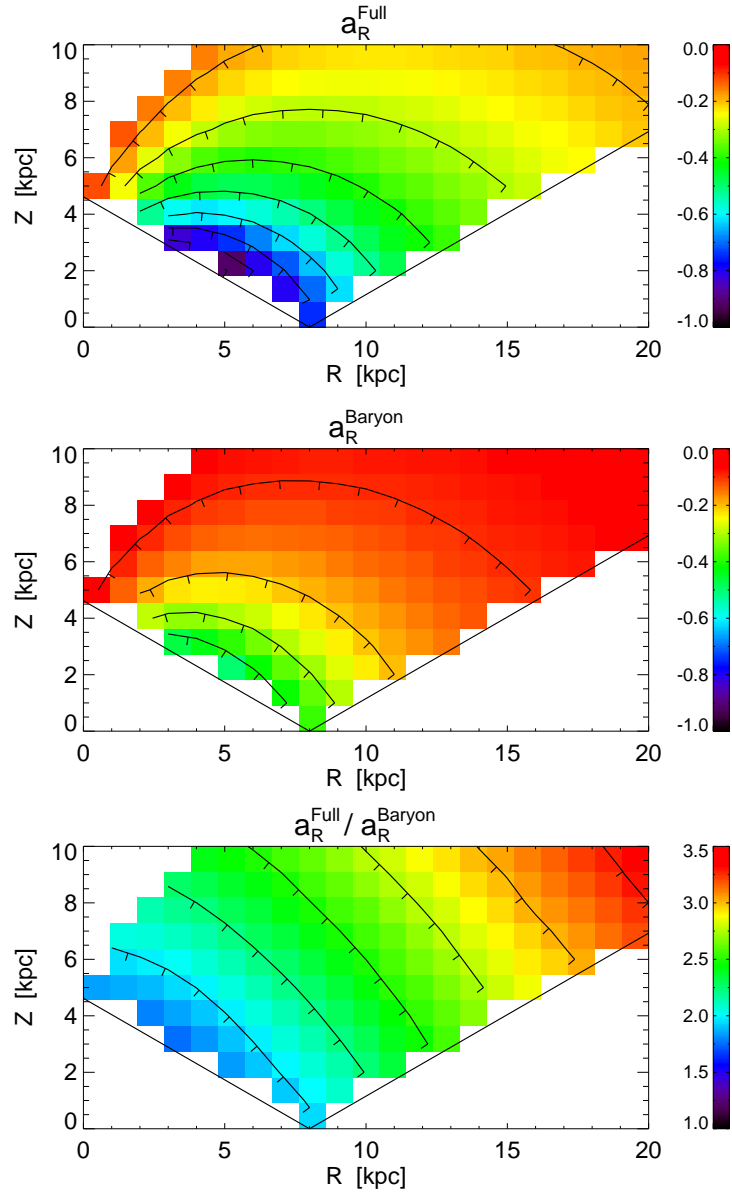}  
  \caption{  An analogous  figure  to Figure~\ref{f:ratio_az},  except
  that  the component  of the  acceleration  in the  $R$ direction  is
  shown.     The    acceleration   is    expressed    in   units    of
  $2.3\times10^{-13}$ $\kmss$.   Similar to the  acceleration map in
  the $Z$  direction shown in  Figure~\ref{f:ratio_az}, the importance
  of the dark matter increases with increased galactocentric distance.}
\label{f:ratio_ar}
\end{figure*}

\begin{figure*}[h]
\epsscale{.9}
\hskip -.15 in
  \plotone{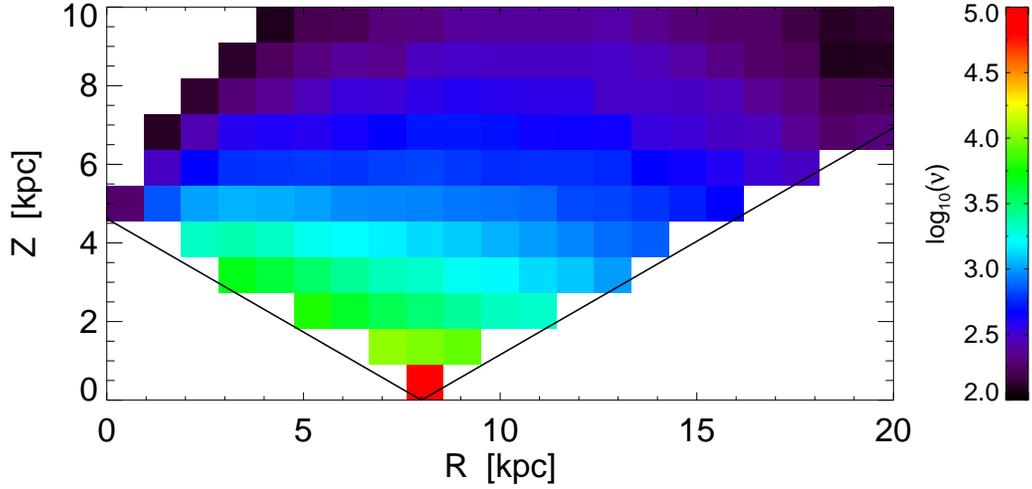}
  \caption{ The number counts  of stellar particles from the $N$--body
  simulation  restricted  to  the  volume  probed  by  SDSS.   Stellar
  particles have  been binned in  1 kpc x  1 kpc $R$-$Z$ bins  and only
  bins with at least 100 particles  are shown and used in our analysis.  In
  addition,  edge  pixels are  subsequently  excluded  from the  Jeans
  equations analysis due to less reliable count gradient estimation.}
\label{f:sim_sdss_nstarmap}
\end{figure*}

\begin{figure*}[h]
\epsscale{.7}
\hskip -.15 in
  \plotone{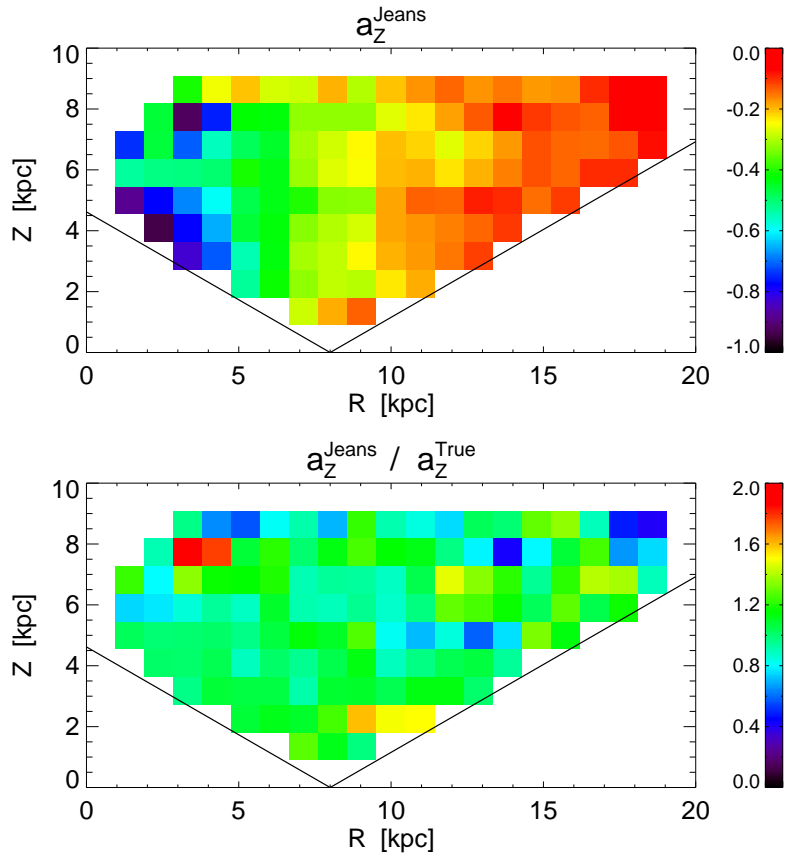}    
  \caption{   (Top)   The
  acceleration in the $Z$  direction for stellar particles from $N$--body
  simulation,  derived using  Equation~\ref{eq:eq2}  and expressed  in
  units of $2.9\times10^{-13}$ $\kmss$.  (Bottom) The ratio of the top
  panel and  the true acceleration  map known from  force computations
  (top panel of Figure~\ref{f:ratio_az}). The Jeans equations approach
  successfully reproduces the true  acceleration map with a bias below
  $\sim$10\%.  The  maps are spatially limited to  the volume explored
  by SDSS data.}
\label{f:jeans_works_az}
\end{figure*}

The  top  panel  of  Figure~\ref{f:jeans_works_az} shows  the  $a_{Z}$
acceleration  map  generated  by   applying  Jeans  equations  to  the
particles from the $N$--body simulation  in the region that mimics the
SDSS volume.  An overall gradient is easy to see; the magnitude of the
acceleration  decreases   with  increased  radius   ($R$).   The  true
acceleration map  (shown in the top  panel of Figure~\ref{f:ratio_az})
displays     similar     behavior;     the     bottom     panel     of
Figure~\ref{f:jeans_works_az}  shows  a  ratio  of the  top  panel  of
Figure~\ref{f:jeans_works_az} and the mean true accelerations from the
top  panel of  Figure~\ref{f:ratio_az}.  We find that Jeans equations
reproduce the true $a_{Z}$ map quite well: for the entire SDSS volume,
the  mean value  of $a^{Jeans}_{Z}  / a^{True}_{Z}$  is $1.05$  with a
dispersion\footnote{ Instead of using the classically defined standard
  deviation, which  is sensitive to non-Gaussian outliers,  we use the
  interquartile range of the  distribution to estimate the dispersion.
  The interquartile range is normalized to obtain a standard deviation
  in     case    of     Gaussian    distribution,     $\sigma_{G}    =
  0.7413\,(q_{75}-q_{25})$, where  $q_{25}$ and $q_{75}$  are the 25\%
  and  75\%  quartiles. For more details see \citet{DMbook}.} of  
$\sigma_{G}=0.18$.   When  we consider  a
column of data  that is unaffected by the bulge  in the simulation ($7
\leq $  R$/$kpc $\leq 9$), we  find that $\sigma_{G}$  drops to $0.15$,
with a mean of $1.08$.

\begin{figure*}[h]
\epsscale{.7}
\hskip -.15 in
  \plotone{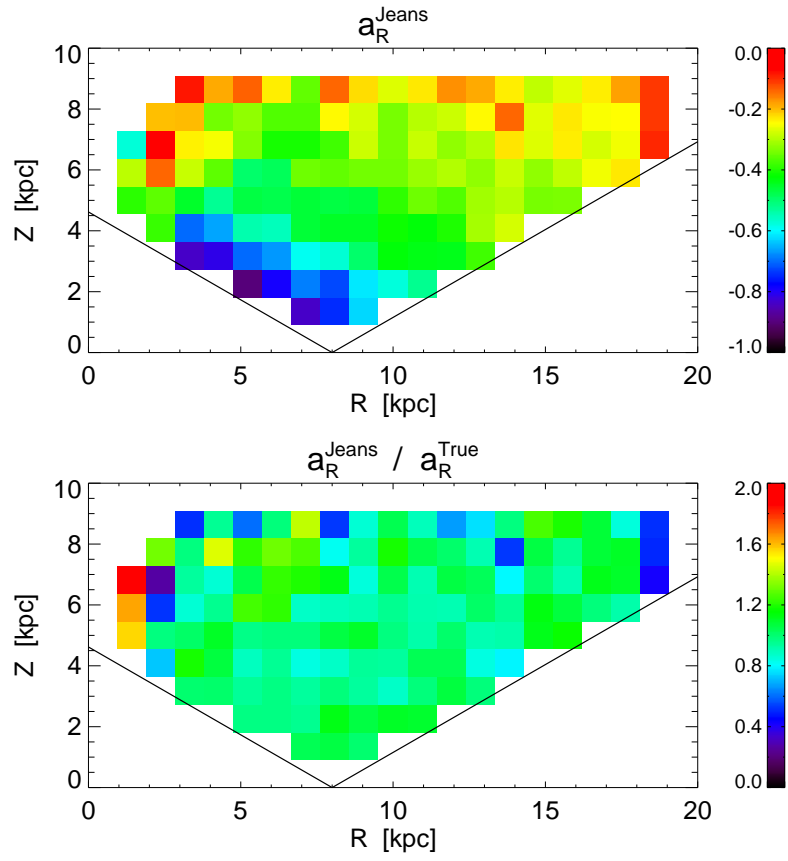} 
  \caption{Analogous to Figure~\ref{f:jeans_works_az}, except that the
  acceleration    in     the    $R$    direction,     derived    using
  Equation~\ref{eq:eq1},   is  shown   and  expressed   in   units  of
  $2.3\times10^{-13}$ $\kmss$.}  
\label{f:jeans_works_ar}
\end{figure*}

Figure~\ref{f:jeans_works_ar}  shows  an  analogous  set of  maps  for
acceleration in the $R$  direction, $a_{R}$.  The mean value  of the ratio
$a^{Jeans}_{R} /  a^{True}_{R}$ for the  entire SDSS volume  is $1.02$
with $\sigma_{G}$ of  $0.13$.  When the map is  subselected to include
data within $7 \leq $ R$/$kpc $\leq 9$, the mean value drops to $0.99$
with $\sigma_{G}=0.12$.   We note  that we tested  for the  effects of
non-axisymmetry on  these results by making  8 slices in  $\phi$ of 90
degrees offset  by 45  degrees.  We found  that the  mean acceleration
within these slices varied by around $10\%$.

We conclude from  this analysis of the $N$--body  simulation that even
in a non-steady state system with deviations from axial symmetry, {\it
Jeans  equations can still  recover meaningful  average accelerations;
within  a given  bin, an  individual acceleration  value  has expected
random error  below 20\%, with  a bias below  10\%}. As we  show next,
this  performance is  sufficient to enable tests for the
existence of dark matter in the MW halo.

\subsection{Application of the Jeans Equations Method to SDSS Data}
\label{s:sdss_results}

In this  section we apply Jeans  equations to a catalog  of stars from
the SDSS volume generated using \textit{galfast}.  We first assess the
relative significance of each term in Jeans equations as a function of
$R$ and  $Z$ to understand  the global distribution  of $a^{SDSS}_{Z}$
and $a^{SDSS}_{R}$, the components of  the acceleration in the $Z$ and
$R$  directions  implied  by the SDSS  data.  We  compare  the  resulting
$a^{SDSS}_{Z}$ and $a^{SDSS}_{R}$ maps to the maps generated using the
$N$--body simulation;  we inspect  the morphology of  the acceleration
maps to draw conclusions about  the presence of dark matter within the
SDSS Galactic volume.

\subsubsection{The Construction of the Acceleration Maps} 

\begin{figure*}[!h]
\epsscale{1}
\hskip -.15 in
  \plotone{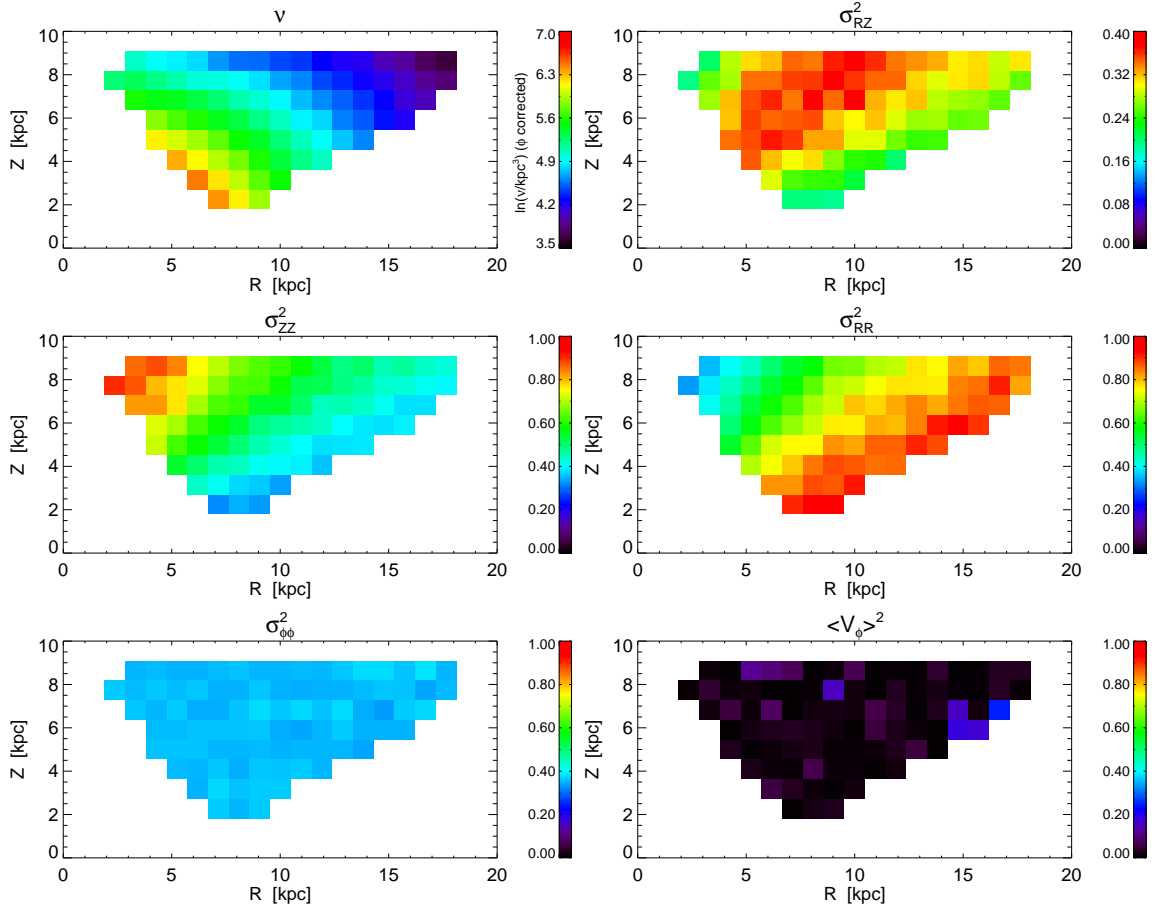}    
  \caption{(Top left) Stellar number density map for halo stars in the
  SDSS  footprint  generated using  \textit{galfast}.   This panel  is
  logarithmically scaled, while all other panels are shown on a linear
  scale and  re-normalized by $2\times10^4$ km$^2$  s$^{-2}$ to enable
  comparison of relative contributions of terms from Equations 1 and 2
  to accelerations $a_Z$ and $a_R$.  Each respective term is listed at
  the top of each panel.}
\label{f:moments}
\end{figure*}

We first  examine the spatial distribution  of stars with  $M_r \ge 4$
and $0.25 < g-r < 0.35$ (top left panel in Figure~\ref{f:moments}).  A
selection  function  correction has  been  applied  to compensate  for
the varying range  of the  axial ($\phi$) angle  sampled by the SDSS Galactic
data; the computed distribution is  a good match to the analytic model
used by  \textit{galfast} and verifies that the  binning algorithm and
the selection function correction are correctly implemented.

Figure~\ref{f:moments}   also  shows  velocity   distribution  moments
$\sigma^2_{RZ}$,        $\sigma^2_{RR}$,       $<$$V_{\phi}$$>$$^{2}$,
$\sigma^2_{\phi\phi}$, and $\sigma^2_{ZZ}$.  The strong variation with
$R$   and   $Z$  seen   for   $\sigma^2_{RZ}$,  $\sigma^2_{RR}$,   and
$\sigma^2_{ZZ}$ is due to the use of the cylindrical coordinate system. We
have  verified   that  analogous  estimates   performed  in  the spherical
coordinate system reproduce the spatially invariant velocity ellipsoid
used by \textit{galfast} to within numerical noise.

\begin{figure*}[h!]
\epsscale{1}
\hskip -.15 in
  \plotone{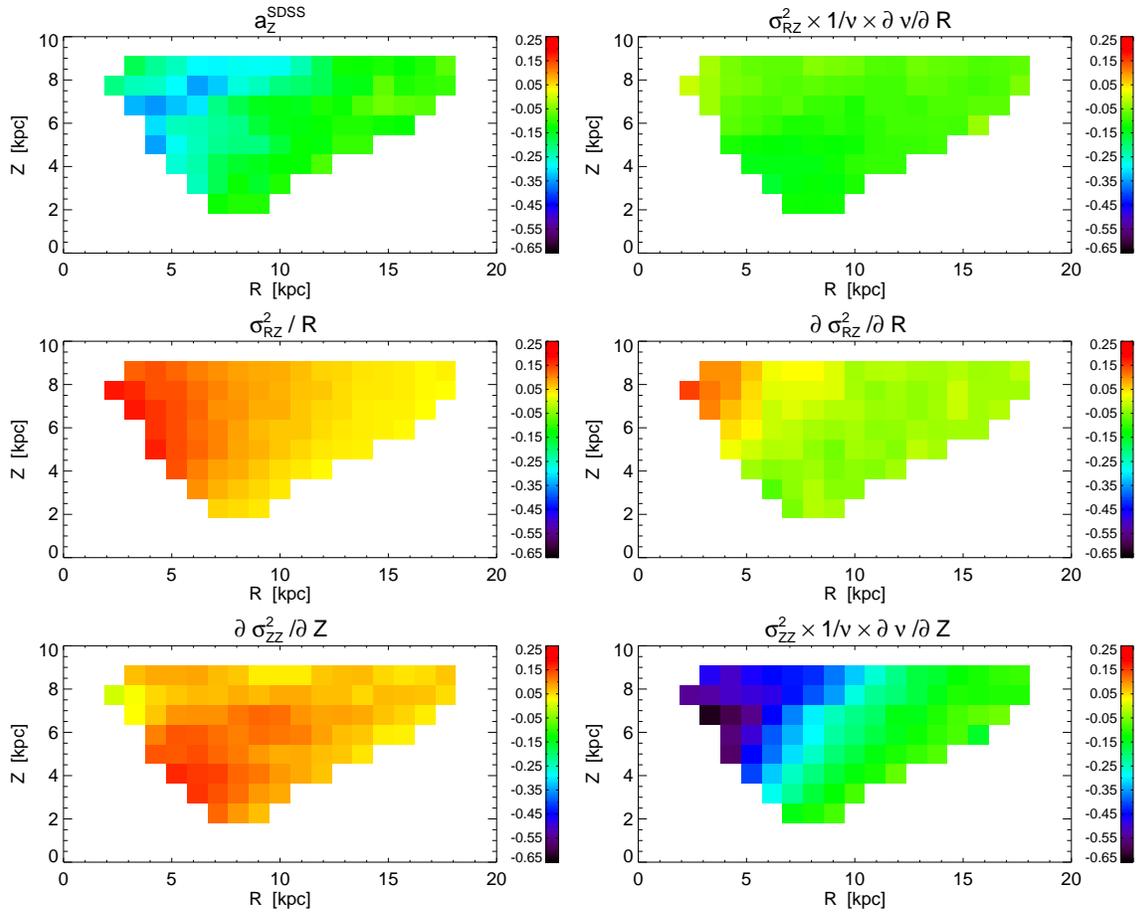}           
  \caption{ \textit{Galfast} $a^{SDSS}_{Z}$ map, expressed in units of
  $2.9\times10^{-13}$  km  s$^{-2}$, and  its  constituent terms  from
  Equation~\ref{eq:eq2}.   Note that  the  scale is  the  same in  all
  panels for  easy comparison.  Terms  are ordered clockwise  from top
  right to middle  left and add to equal  $a^{SDSS}_{Z}$, shown in the
  top left panel.  Each panel  is labeled with the term it corresponds
  to.}
\label{f:galfast_az_terms}
\end{figure*}

\begin{figure*}[h!]
\epsscale{1}
\hskip -.15 in
  \plotone{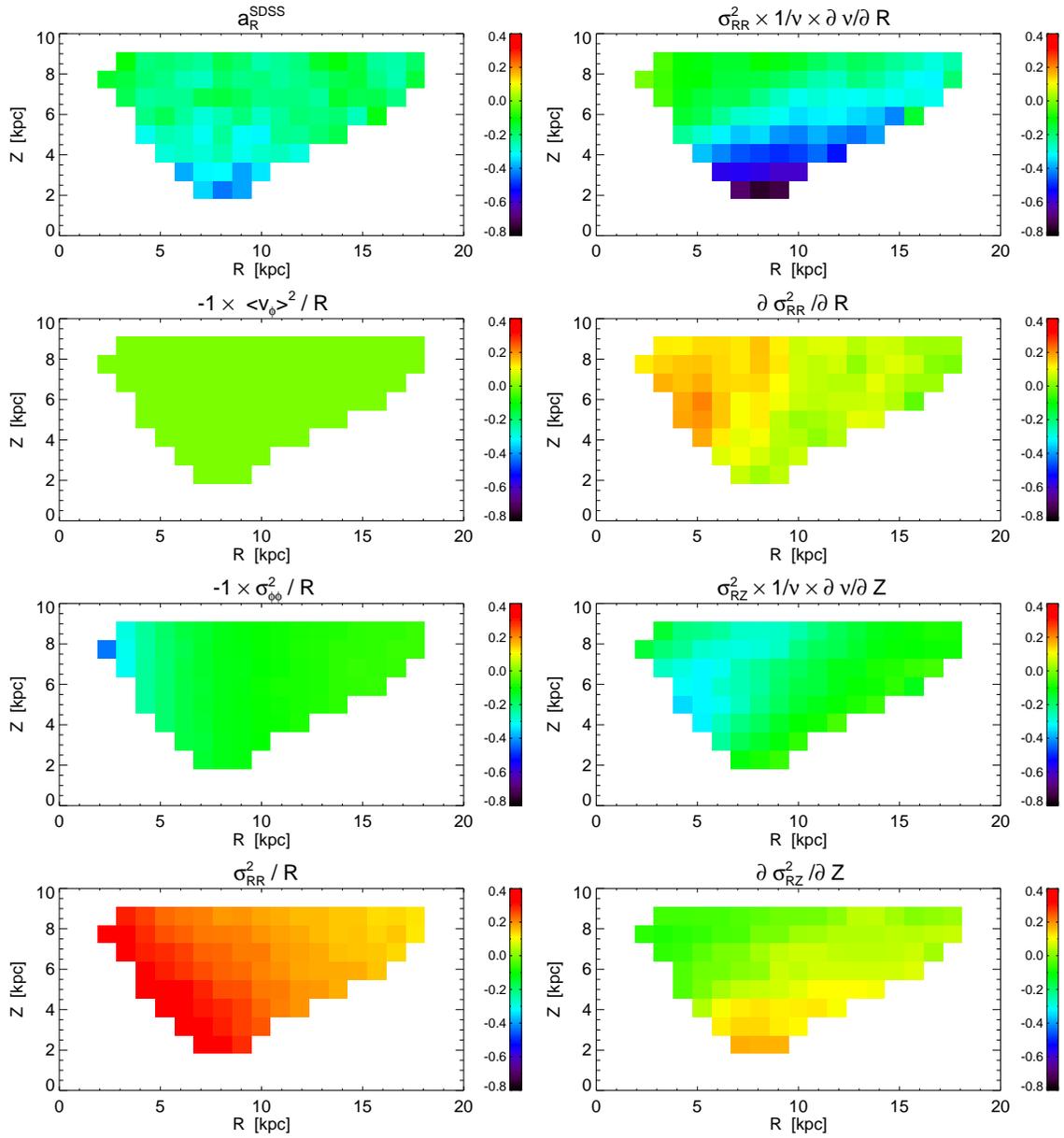}   
  \caption{Analogous   to   Figure~\ref{f:galfast_az_terms}  but   for
  $a^{SDSS}_{R}$ and  the constituent terms  in Equation~\ref{eq:eq1}.
  Each  panel  is  expressed  in  units  of  $2.3\times  10^{-13}$  km
  s$^{-2}$.  The  constituent terms are labeled  and ordered clockwise
  from top right to upper-middle  left and add to form $a^{SDSS}_{R}$,
  shown in the top left panel.}
\label{f:galfast_ar_terms}
\end{figure*}

The    spatial   derivatives    of   these    terms   are    used   in
Equations~\ref{eq:eq1} and \ref{eq:eq2}  to compute $a^{SDSS}_{Z}$ and
$a^{SDSS}_{R}$;         they         are        illustrated         in
Figures~\ref{f:galfast_az_terms} and Figures~\ref{f:galfast_ar_terms},
together  with the  main result  of our  analysis,  $a^{SDSS}_{Z}$ and
$a^{SDSS}_{R}$ maps  shown in  the top left  panel in each  figure. In
each  figure, the other panels show  all  the  additive   terms  from
Equations~\ref{eq:eq1}  and \ref{eq:eq2}.   Note that  different terms
have varying  contributions towards  the final acceleration  maps. All
terms contributing  to acceleration maps show  smooth global behavior,
with only a small number of pixels deviating from the overall trends.

\subsubsection{The Initial Interpretation of the Acceleration Maps} 

Now  that  we have  maps  for  $a^{SDSS}_{Z}$  and $a^{SDSS}_{R}$,  we
consider what these maps tell  us about the underlying distribution of
matter within the SDSS volume.  To assess this, we again draw upon our
$N$--body simulation to predict what behavior we would expect when the
dark matter contribution is and is not included (in \S\ref{Sec:DMpot}, 
we continue this discussion using a baryon potential derived from SDSS
measurements for disk stars). 

Recall    the    top    panels   of    Figures~\ref{f:ratio_az}    and
\ref{f:ratio_ar},   which  shows   the  map   of   $a^{Full}_{Z}$  and
$a^{Full}_{R}$   from  $N$--body  simulation.    In  this   case,  the
acceleration of each particle was  calculated using the force from all
the particles  in the entire  simulation.  For comparison,  the middle
panels  of  these two  figures  show  analogous  maps, but  there  the
acceleration   was   calculated    without   including  the dark   matter
contribution.

\begin{figure*}[h!]
\epsscale{.7}
\hskip -.15 in
  \plotone{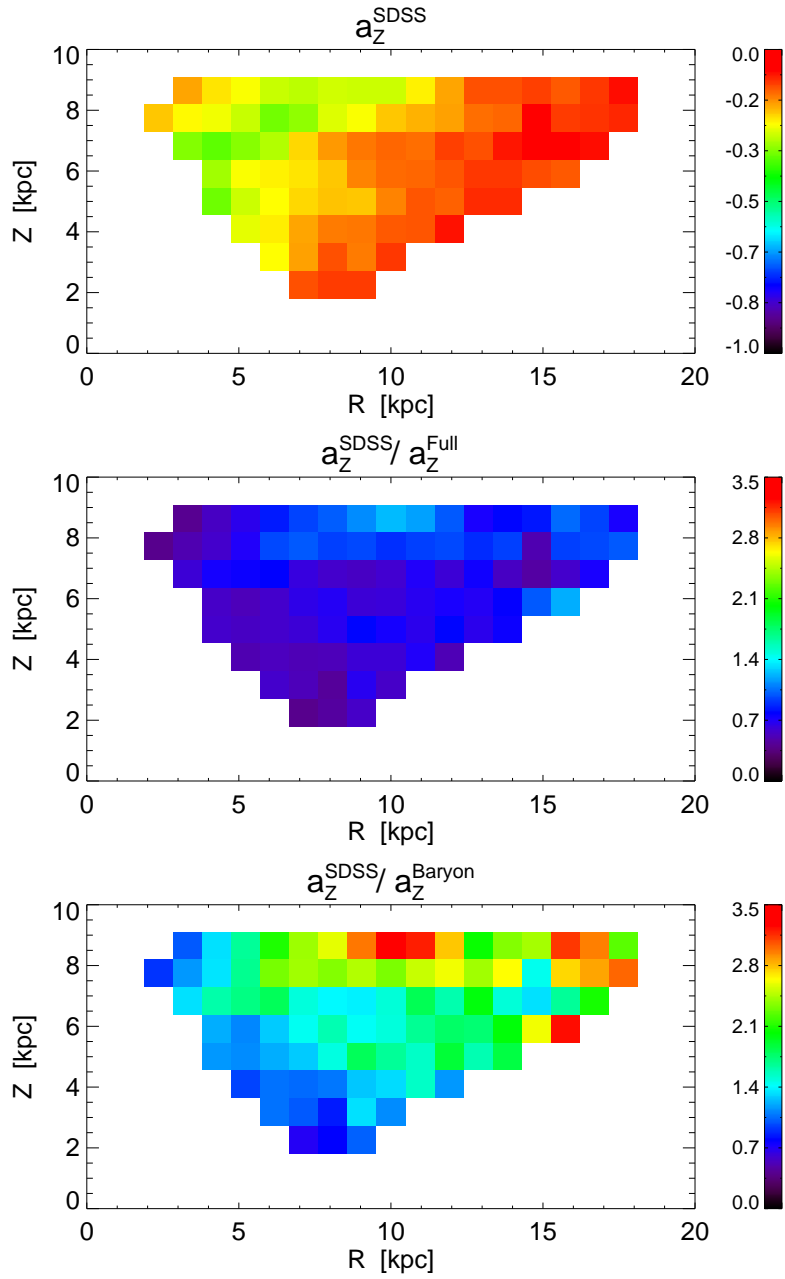}  
  \caption{The results   of   applying  Jeans   equations   to  the   SDSS
  observations simulated using  \textit{galfast}.  The top panel shows
  a map  of acceleration  in the $Z$  direction expressed in  units of
  $2.9\times  10^{-13}$ km  s$^{-2}$ (same  as the  top left  panel in
  Figure~\ref{f:galfast_az_terms}, except for different scaling).  The
  middle and  bottom panels  show the  ratio of the  map from  the top
  panel and  the two model-based maps  shown in the top  two panels in
  Figure~\ref{f:ratio_az}.}
\label{f:galfast_az}
\end{figure*}

\begin{figure*}[h!]
\epsscale{.7}
\hskip -.15 in
  \plotone{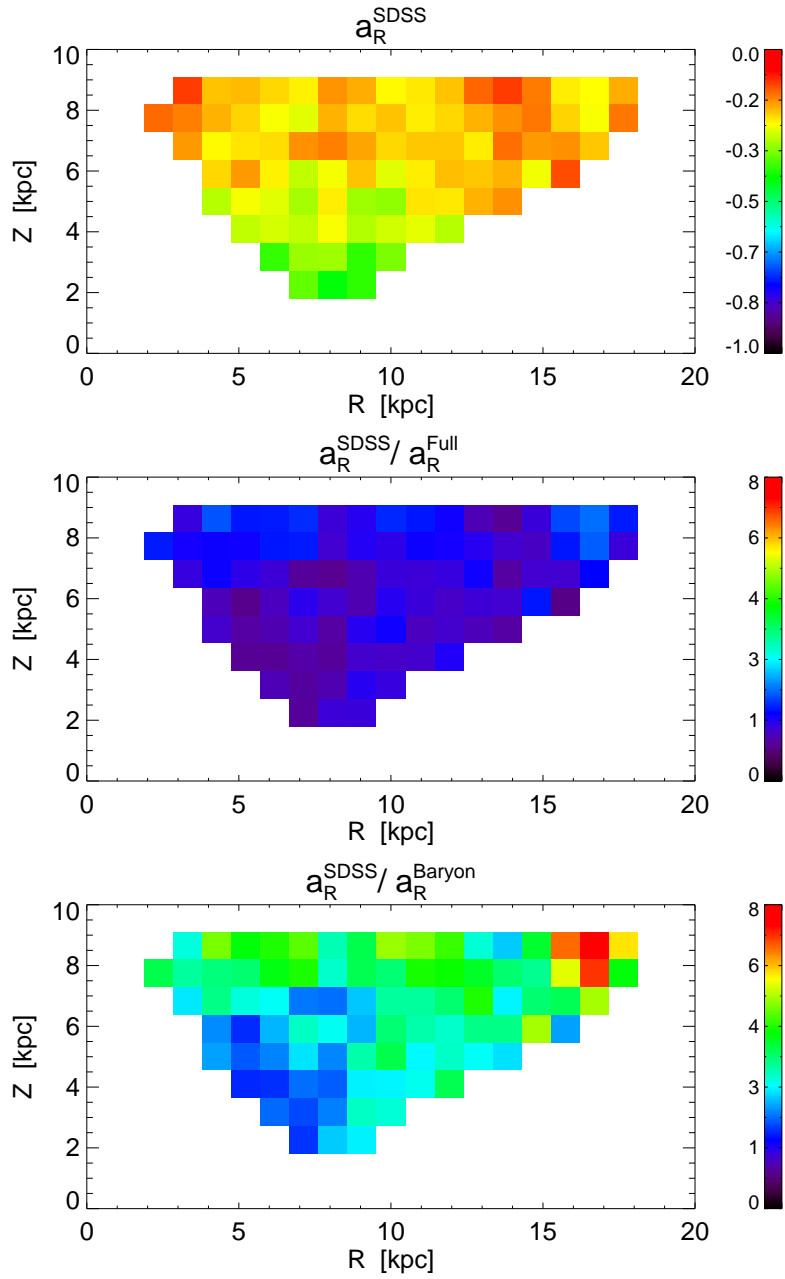}      
  \caption{Analogous   to  Figure~\ref{f:galfast_az},   but   for  the
  component  of the acceleration  in the  $R$ direction,  expressed in
  units of $2.3\times10^{-13}$ $\kmss$.}
\label{f:galfast_ar}
\end{figure*}

Similarly, the middle  and bottom panels of Figures~\ref{f:galfast_az}
and  \ref{f:galfast_ar}  show  the  ratio of  the  $a^{SDSS}_{Z}$  and
$a^{SDSS}_{R}$   map   to    the   simulation's   $a^{Full}_{Z}$   and
$a^{Baryon}_{Z}$   and   $a^{Full}_{R}$   and  $a^{Baryon}_{R}$   maps
respectively.  Clearly,  the acceleration  maps derived from  the SDSS
data   are  closer   to   the  model-based   acceleration  maps   that
\textit{include}  contributions  from both  baryons  and dark  matter.
       {\it At large  galactocentric distances, the SDSS accelerations
         are  as much  as  three  to four  times  stronger than  those
         predicted by a non-dark matter model!}

Therefore,  by generating acceleration  maps using  the SDSS  data and
comparing these maps to  expectations from an $N$--body simulation, we
have demonstrated that a model containing dark matter is a much better
fit  to observations  than  the model  that  contains baryonic  matter
alone. While it is encouraging  to see yet another aspect of $N$--body
simulation  that  at  least   qualitatively  agrees  with  data,  this
far-reaching  conclusion  can  be   derived  without  a  reference  to
simulation, as we show next.

\section{Constraints on the Dark Matter Gravitational Potential}
\label{Sec:DMpot}

The analysis  in the previous  section shows that our  Jeans equations
approach   can  successfully  recover   stellar  accelerations   in  a
non-steady-state and non-cylindrically symmetric $N$--body simulation.
A  comparison  between the  $N$--body  simulation  and the  SDSS-based
acceleration maps  strongly suggests that  a dark matter  component is
needed  to  account  for  the observed  accelerations.   However,  the
strength  of  this  conclusion  depends  on  how  well  the  $N$--body
simulation  matches  the  observed  MW.   To  supplement  our  earlier
argument,  in this  section we  perform  an analysis  of the  observed
acceleration  maps that  does not  require  the use  of the  $N$--body
simulation.   Instead, we  utilize a  new  observationally constrained
description  of  the  MW  gravitational potential;  we  quantitatively
compare  this potential to  our SDSS-based  acceleration maps  to draw
conclusions   about   the   dark   matter  potential.    Because   our
\textit{galfast}-based acceleration  maps suffer from  numerical noise
(recall   \S\ref{sec:numprocs}),   here   we  use   the   ``analytic''
acceleration maps computed directly using Equation~\ref{eq:nuDH2} from
J08  and  the  velocity  ellipsoid   for  halo  stars  from  B10  (see
Equations~\ref{eq:sigRR}--\ref{eq:sigRZ}).

\subsection{Analytic SDSS Acceleration Maps for Halo Stars}
\label{sec:anamaps}
 
\begin{figure*}[h!]
\epsscale{1}
\hskip -0.3in
  \plotone{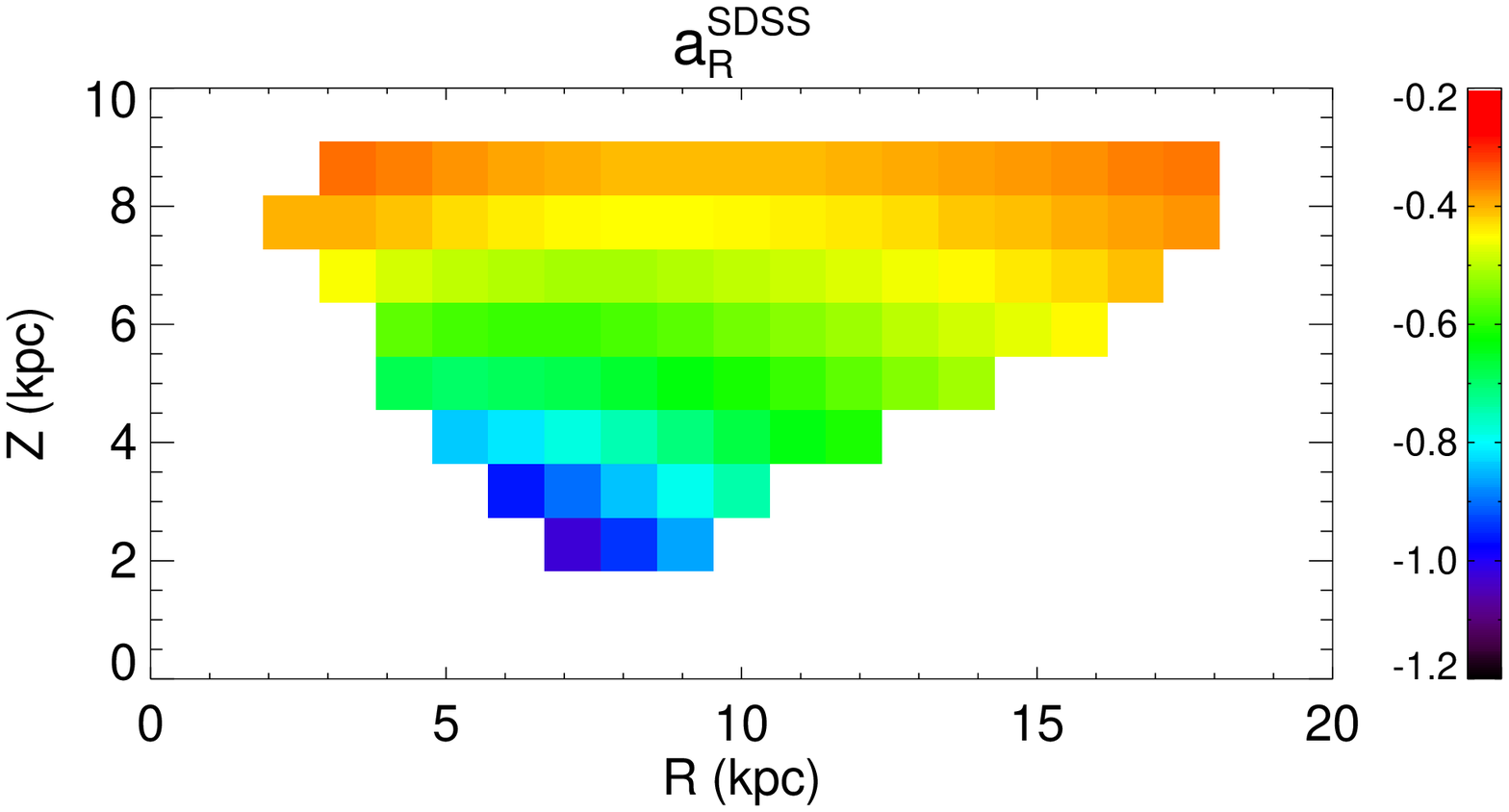}
  \plotone{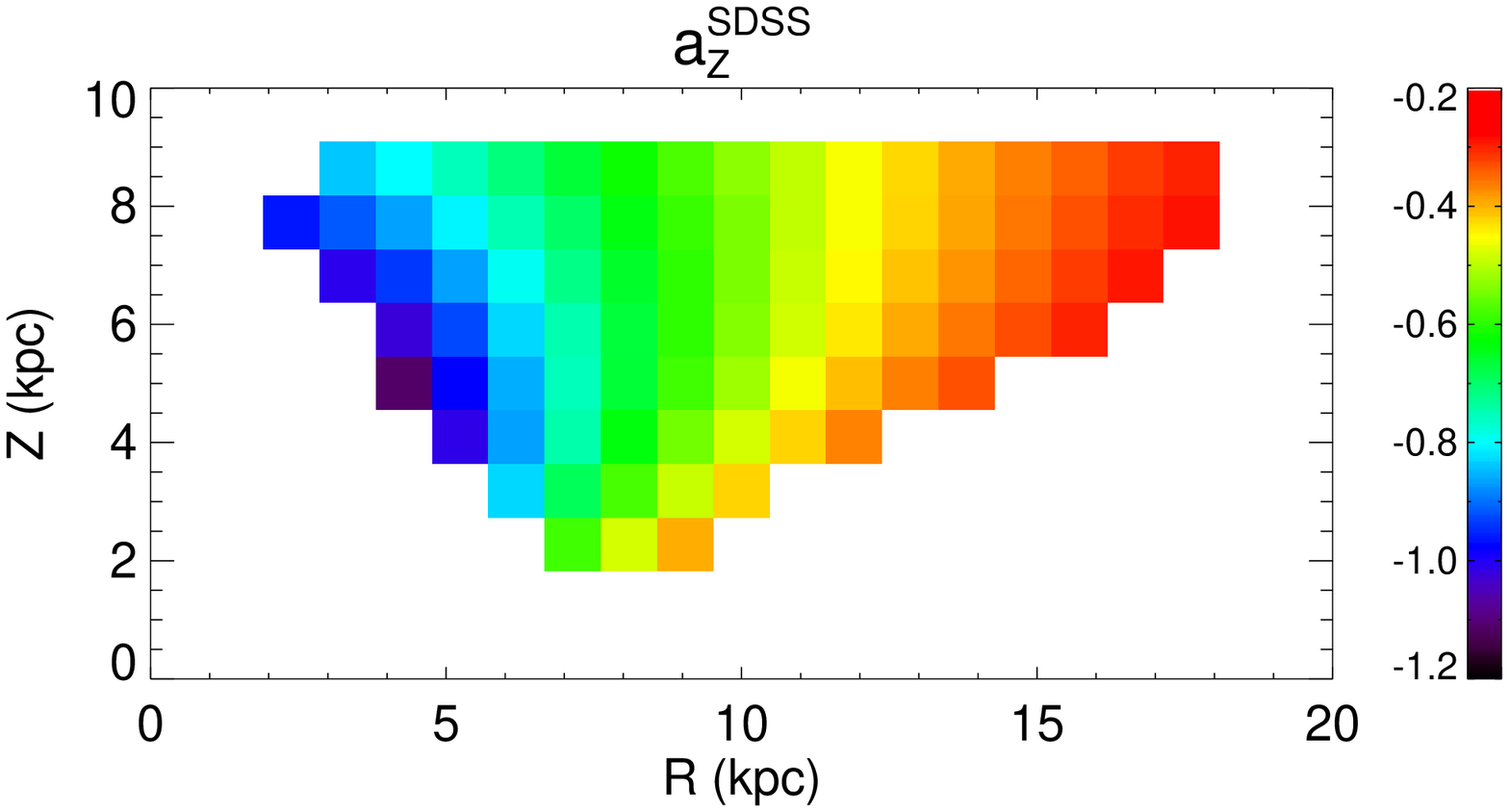}  
  \epsscale{1.0}
  \caption{Acceleration   maps    computed   using   Jeans   equations
  (Equations~\ref{eq:eq1}   and   \ref{eq:eq2}),   with  the   spatial
  distribution of halo  stars described by Equation~\ref{eq:nuDH2} and
  the         velocity          ellipsoid         described         by
  Equations~\ref{eq:sigRR}--\ref{eq:sigRZ}, as inputs
  (expressed in units of $10^{-13}$ $\kmss$).  These maps are
  morphologically very similar to the  maps shown in the top panels in
  Figures~\ref{f:galfast_az} and Figures~\ref{f:galfast_ar} (note that the
  stretch for color palette is different in this figure to emphasize spatial
  variation).}
\label{fig:analytics}
\end{figure*}

Our      analytic     acceleration      maps     are      shown     in
Figure~\ref{fig:analytics}.       As      already      implied      in
\S\ref{sec:numprocsTests},  they are  morphologically very  similar to
numerical maps  shown in the top  panels in Figures~\ref{f:galfast_az}
and  \ref{f:galfast_ar}  (for ease  of  comparison  we  used the  same
$R$-$Z$ grid although  the analytic maps can be evaluated  on an arbitrary
grid). Although these analytic maps are formally noise-free, as we demonstrated in the preceding section, we anticipate that the
random errors  due to deviations from cylindrical  symmetry and steady
state can be up to about 20\% (with a bias below 10\%).

We find that these maps cannot be described by a spherically symmetric
potential. Case  in point, there is a large  class of potentials of the functional form
\begin{equation}
\label{eq:xvar}
          x = \left( {R^2 + (Z/q)^2 + R^2_{core} \over R^2_\odot} \right)^{1/2}.
\end{equation} 
For this class, the isopotential surface axis ratio $q$ can be estimated as
\begin{equation}
\label{eq:qDMdirect}
          q = \left( {Z\, a_R(R,Z) \over R \,a_Z(R,Z)} \right)^{1/2}. 
\end{equation}

When  we apply the  maps shown  in Figure~\ref{fig:analytics}  to this
equation,  we  find  that  the  median  value of  $q$  is  0.80,  with
(inter-quartile based) scatter  of $\sigma_G=0.04$.  This evidence for
oblateness comes directly from  the fact that the spatial distribution
of      halo      stars      is      oblate      ($q_H=0.64$;      see
Equation~\ref{eq:nuDH2}). Nevertheless, it does not follow immediately
that the dark matter potential must be oblate because the contribution
of disk baryons  to the potential is non-negligible.   We now turn our
attention to a recent model where the disk baryons have been carefully
accounted for.

\subsection{SEGUE G Dwarfs and the Bovy-Rix Potential} 

Recently,  \citet[][henceforth, BR13]{BR2013}  studied  in detail  the
dynamics of $\sim$16,000 G dwarfs  drawn from the SDSS Sloan Extension
for   Galactic   Understanding   and  Exploration   \citep[][hereafter
  SEGUE]{Yanny2009}.  The  SEGUE G dwarf  sample is dominated  by disk
stars and  extends to 3  kpc from the  Galactic plane, with  a similar
extent    in    the    radial   direction    \citep{Bovy2012b}.     As
\citet{Bovy2012a}  show,  these  disk  stars  can  be  separated  into
sub-populations based upon chemical abundance parameters ($[Fe/H]$ and
$[\alpha/Fe]$).  \citet{Bovy2012a} find  that the spatial distribution
of each  sub-population is well  fit by a single  exponential profile,
both as  a function  of height above  the midplane  and galactocentric
radius.  Moreover,  the kinematic  behavior of each  sub-population is
relatively  simple  \citep{Bovy2012c}, making  it  possible  to fit  a
three-integral action-based distribution  function and parametrize the MW
potential to the SEGUE data \citep[][ BR13]{Ting2013}.

BR13's parametrization  of the  MW potential includes  a two-component
gravitational potential,  corresponding to the baryon  and dark matter
content.   The   former  is  likely   the  most  robust   and  precise
determination  of the MW  baryonic potential  to date.   Additionally, the
local normalization for the dark matter component is consistent with a
more  direct  measurement  from  \citet{Bovy2012} and  has  a  similar
precision.   However, due  to  the relatively  local  nature of  their
sample,  BR13   \textit{cannot  strongly  constrain   deviations  from
  spherical symmetry  for the  dark matter model},  and thus  for this
component they adopt a  spherically symmetric potential.  We note that
BR13's potential model  is publicly available via the  galactic and MW
dynamics        python       package       \textit{galpy}\footnote{See
  http://galpy.readthedocs.org/en/latest/}.

\begin{figure*}[h!]
\epsscale{1}
\hskip -.15 in
  \plotone{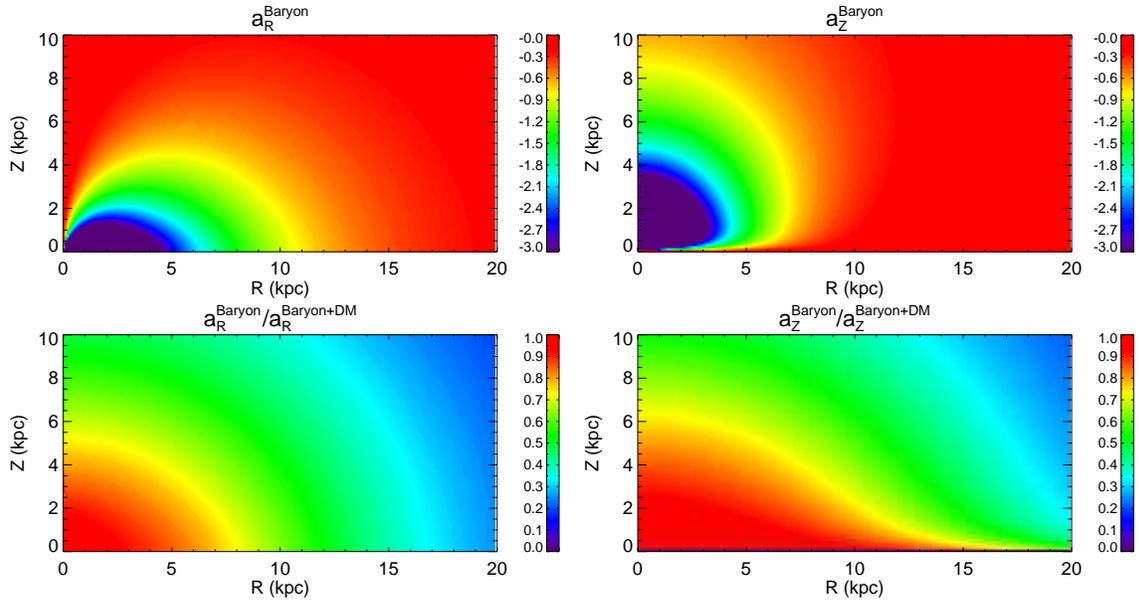}      
  \caption{Top: acceleration maps  predicted by the baryonic component
           of   the   Bovy-Rix   potential  (left:   $a_{R}$,   right:
           $a_{Z}$; expressed in units of $10^{-13}$ $\kmss$). 
           Bottom: predicted  fractional contribution of the
           accelerations from  the baryonic component  relative to the
           total potential model.  Note  that the contours of constant
           fraction at  $R=8$ kpc  are roughly horizontal  for $a_{Z}$
           and   relatively  more   perpendicular   for  $a_{R}$,   in
           qualitative  agreement  with  predictions from  the  N-body
           simulation  (see bottom panels  in Figures~\ref{f:ratio_az}
           and \ref{f:ratio_ar}).}
\label{fig:BRfractions} 
\end{figure*}

In Figure~\ref{fig:BRfractions} we explore the accelerations predicted
by  the  BR13 potential  model.   The top  left  and  right panels  of
Figure~\ref{fig:BRfractions} show the $a_{R}$ and $a_{Z}$ acceleration
maps generated from  the BR13 baryon potential.  We  also consider the
relative   significance  of   the   baryon  potential   to  the   dark
matter+baryon   model;   the  bottom   left   and   right  panels   of
Figure~\ref{fig:BRfractions} show  the fractional contribution  of the
baryons to the dark matter+baryon accelerations.  These panels include
contours of constant fraction.  In the case of $a_{Z}$ (bottom right),
at $R\sim8$ kpc  the contours are roughly horizontal  (parallel to the
$R$ axis), and  in the case of $a_{R}$ (bottom  left), at $R\sim8$ kpc
the contours  are relatively more perpendicular.  Encouragingly, these  
trends are in qualitative  agreement   with  the  predictions   from  the  
$N$--body simulation   (see  bottom   panels  in   Figures~\ref{f:ratio_az}  
and \ref{f:ratio_ar}).

We next compare the  accelerations predicted by the BR13 two-component
potential  model to  our SDSS-based  analytic acceleration  maps.  Our
goal is to  understand how well the two-component  model, containing a
spherically symmetric  dark matter halo, fits  our SDSS-based results.
We begin  by considering the  data/model ratio: for the model to be a
good  match to our data,  the median  value of  the ratio  should be
roughly 1.0  with a small  rms ($\la10\%$--$20\%$ on a  linear scale).
In  both  the  cases  ($a_{R}$  and  $a_{Z}$),  to  achieve  a  median
data/model ratio of  1.0, we must rescale the  model by multiplying by
0.66  and 0.57  respectively.  After  these renormalizations,  the rms
scatter is  fairly small (23\%  on a linear scale).   However, because
the  model  derived  $a_{R}$  and $a_{Z}$  require  \textit{different}
renormalizations, and there are  systematic deviations as functions of
$R$  and   $Z$,  we  conclude  that  \textit{the   data  versus  model
  discrepancy cannot be resolved by a simple rescaling alone.}

To re-emphasize this  point, we draw upon an  illustrative example.  At
$R=18$  kpc,   $Z=8$  kpc,   the  extrapolated  BR13   model  predicts
$a_R=-0.61$  and   $a_Z=-0.30$  (in  units   of  $10^{-13}$  km/s$^2$)
resulting  from  the  dark   matter  component,  and  $a_R=-0.18$  and
$a_Z=-0.10$  from baryon  component.  However,  at this  location the
measured  SDSS-based  accelerations are $a_R=-0.35$  and $a_Z=-0.29$.   
That is,  \textit{the  model dark  matter component  by itself exceeds the 
total measured acceleration}.

In contrast to this, we  consider $a_{Z}$ in the solar neighborhood at
small $Z$ ($R=8$ kpc, $Z=3$ kpc).  Here our SDSS-based acceleration is
$a_Z=   -0.56\times  10^{-13}$  km/s$^2$.    Converting  this   to  an
equivalent  surface  density   yields  65  $M_\odot$/pc$^2$,  with  an
uncertainty  of $\sim$10\%.   This value  agrees within  errors  to the
constraints   on   the  equivalent   surface   density  presented   in
\citet{Bovy2012}:  using Figure~1  from \citet{Bovy2012},  the surface
density correction  factor is 0.9  at $Z=3$ kpc, yielding  a predicted
surface density  of 77$\pm$9 $M_\odot$/pc$^2$.   Hence, the SDSS-based
$a_Z$  derived from  our halo  sample is  $\sim$16\% smaller  than the
acceleration based on disk sample at $R=8$ kpc, $Z=3$ kpc, but the two
values are  consistent within quoted statistical  errors.  

In summary,
our    acceleration   maps    are   statistically    consistent   with
\cite{Bovy2012} at  $Z$ as  small as 3  kpc; additionally,  we explore
much  larger galactocentric  distances  which allows  us  to draw  new
constraints on  the dark matter  potential. However, we note  that the
extrapolation of  our acceleration  maps for halo  stars to  $Z<3$ kpc
predicts weaker $a_Z$ acceleration  than experienced by disk stars (as
summarized  by  the BR13  model);  we  return  to this  discussion  in
\S\ref{sec:validity}.

We  now generate  model  maps in  better  agreement with  the data  by
including two  modifications to the  original BR13 model: 1)  we allow
for a renormalization of the  baryonic component (which is much better
constrained than  dark matter component, in both  shape and amplitude,
and thus we expect a renormalization factor close to unity), and 2) we
allow departures from spherical symmetry for the dark matter component
(as motivated by disagreements at large galactocentric radii).

\subsection{Modified BR13 Potential} 

Henceforth,  we  adopt the following description for the gravitational
potential used to generate the model acceleration maps
\begin{equation}
\label{eq:potTot}
   \Phi(R,Z) = f_{BR} \, \Phi_{bar}(R,Z) + \Phi_{DM}(R,Z),
\end{equation}
where  $\Phi_{bar}(R,Z)$ is the  component due  to baryons  (stars and
interstellar medium), $f_{BR}$  is a renormalization factor (discussed
in more  detail below)  and $\Phi_{DM}(R,Z)$ is  the component  due to
dark matter (e.g., see Equation~2-54a in \citealt{Binney1987}),
\begin{equation}
\label{eq:potDM}
\Phi_{DM}(R,Z) = {1 \over 2} v_{o}^2 \, \ln \left( {R^2 + (Z/q_{DM})^2
  + R^2_{core} \over R^2_\odot} \right).
\end{equation}
The  free model  parameters are  $f_{BR}$, the  dark  matter potential
scale ($v_{o}$),  the spatial scale ($R_{core}$), and  the dark matter
axis ratio ($q_{DM}$).  Given  $\Phi(R,Z)$, we compute $a_{R}(R,Z) = -
\partial  \Phi(R,Z)  /  \partial  R$  and  $a_{Z}(R,Z)  =  -  \partial
\Phi(R,Z) / \partial Z$. 

The   chosen   logarithmic  potential   is   convenient  because   its
corresponding  matter  density  can  be  expressed  analytically  (see
eq.~\ref{eq:rhoDM}). We  discuss the  uniqueness of this  potential in
more detail in \S\ref{sec:uniqDMpot}.

\subsection{The Best-fit Dark Matter Potential} 

Next we discuss our  procedure for identifying the best-fit parameters
in Equation~\ref{eq:potTot} and Equation~\ref{eq:potDM}.  We first fix
$f_{BR}=1$, and exhaustively explore the two-dimensional $R_{core}$ --
$q_{DM}$  parameter space.   For  a given  trial  pair of  $R_{core}$,
$q_{DM}$, we determine the best-fit value for $v_{o}$.  To do this, we
directly compute  the $a_{R}$ data/model  ratio for a list  of $v_{o}$
values.  The $v_{o}$ that corresponds  to a median data/model ratio of
1.0 is selected  as the best fit value of  $v_{o}$ for that particular
$R_{core}$, $q_{DM}$ pair.  There  is no a priori guarantee that
the  corresponding $a_{z}$ data/model  ratio will  equal 1.0  as well;
however, deviations are minor in practice ($\sim$1\%;  this agreement 
implies  that the  chosen  model form is satisfactory).

Adopting the best $v_{o}$ for  each $R_{core}$, $q_{DM}$ pair, we then
track the goodness  of the $R_{core}$, $q_{DM}$ fits  by measuring the
robust ``residual metric.''  This metric  is defined as the sum of the
two ($a_{R}$  and $a_{Z}$) median absolute  deviations; smaller values
correspond to better fits (that  is, we do not assume the model$-$data
differences    follow     a    Gaussian    distribution).\footnote{For
  completeness, we tried a  residual metric that only includes $a_{Z}$
  or $a_{R}$, to constrain the dark matter potential, but we find that
  these constraints are much weaker than when both datasets (maps) are
  considered    together.    }    We    define   the    deviation   as
$\delta$=log$_{10}$(data/model) for both maps.

\begin{figure*}[h!]
\epsscale{.7}
  \plotone{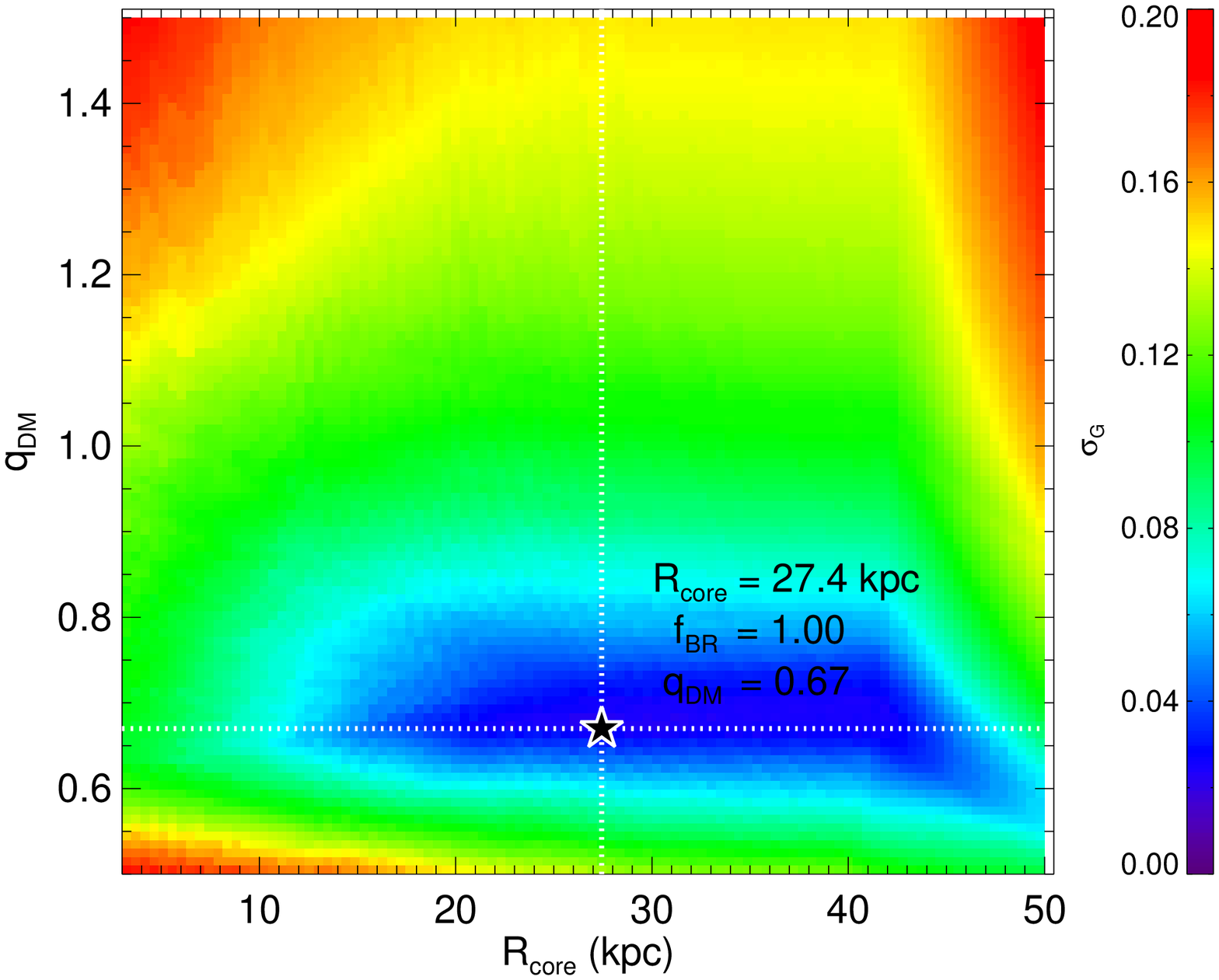}
  \vskip -1in  
  \plotone{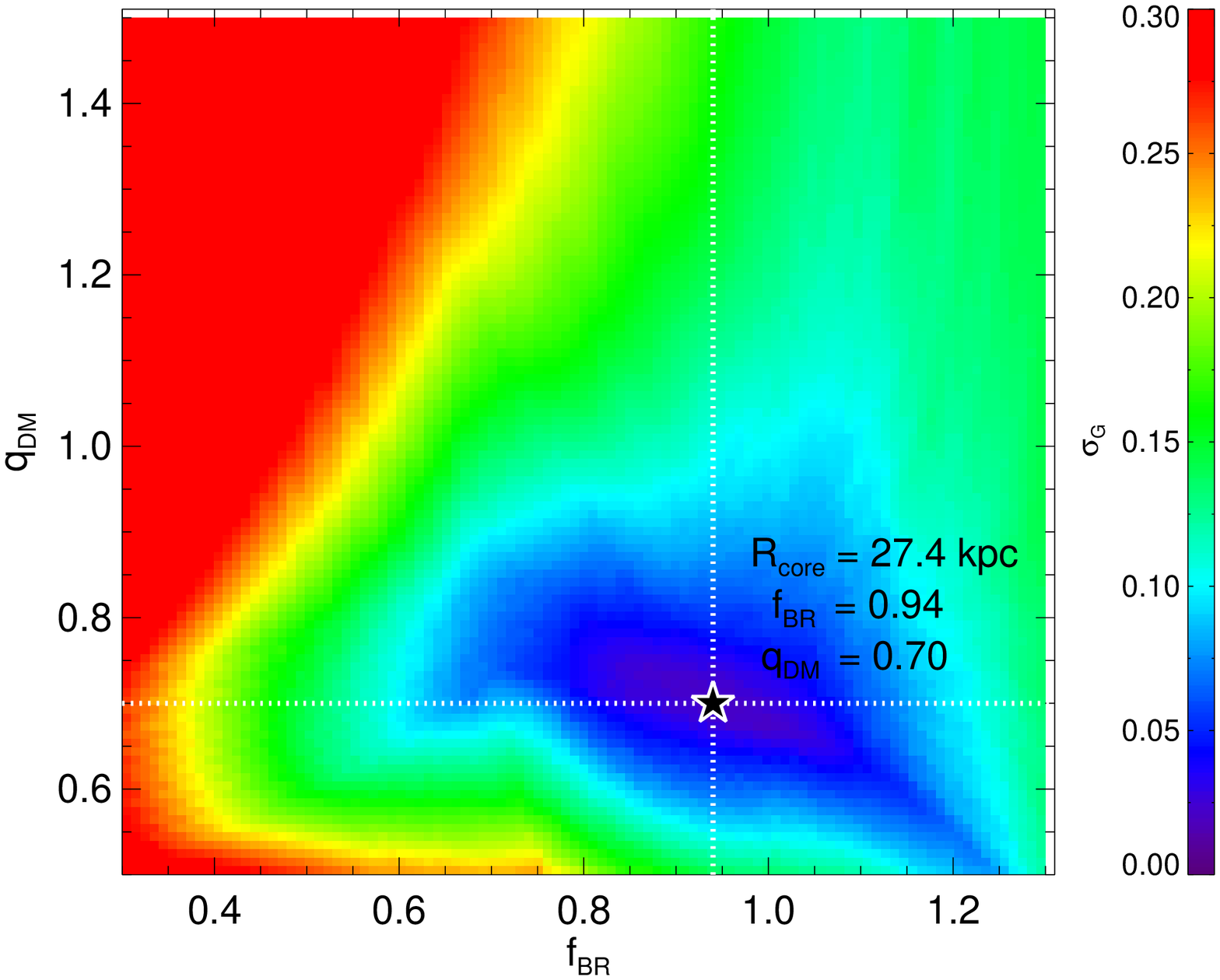}
  \vskip -.5in    
  \caption{The  variation  of  the  robust  residual  metric  for  the
  parameters  $R_{core}$  and  $q_{DM}$ from  Equation~\ref{eq:potDM}.
  Lower values (shown in blue)  correspond to better model fits to the
  SDSS-based  acceleration maps.   Top panel:  the metric  with baryon
  renormalization   factor   from   Equation~\ref{eq:potTot}  set   to
  $f_{BR}=1$.  Bottom panel: the metric  as a function of $f_{BR}$ and
  $q_{DM}$, with spatial scale set to $R_{core}=27.4$ kpc.}
\label{fig:deltaFit1}
\end{figure*}

The variation of $\delta(R_{core},q_{DM})$  is shown in the left panel
in  Figure~\ref{fig:deltaFit1}.   The  plausible  (\ie,  not  strongly
excluded)  range   for  the  spatial  scale,  $R_{core}$,   is  $22  <
R_{core}/{\rm  kpc} <  42$;  outside this  range  the residual  metric
rapidly  increases to  statistically implausible  values (given its
minimum attained value).   The formal
(but  very shallow)  local  minimum is  found  at $R_{core}=27.4$  and
$q_{DM}=0.68$,  corresponding to  $v_o=195$  km/s.  We  find that  the
best-fit  values of $R_{core}$  and $v_o$  are strongly  covariant and
related via  $v_o =  (55 + 5.1  \times R_{core}/kpc)$ km/s.   The axis
ratio for  the dark  matter potential is  strongly constrained  to the
range $0.65  < q_{DM} < 0.75$,  and is essentially  independent of the
choice of $R_{core}$.

\begin{figure*}[h!]
\epsscale{1}
  \plotone{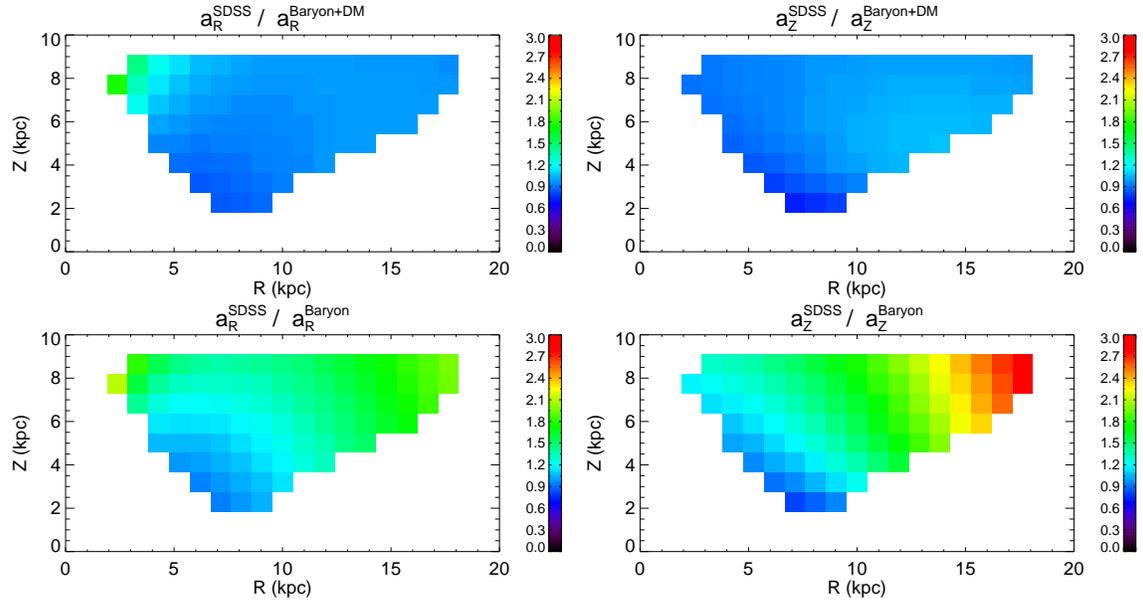}  
  \caption{The  top  two  panels  show  the ratio  of  the  SDSS-based
           acceleration   maps  (left:   $a_R$,   right:  $a_Z$;   see
           Figure~\ref{fig:analytics}) and  the best-fit two-component
           model  based on  baryon potential  from  \citet{BR2013} and
           dark matter  potential described by Equation~\ref{eq:potDM}
           (with $q_{DM}=0.7$).  The bottom  two panels show the ratio
           of  the SDSS-based  acceleration maps  and  the predictions
           based on baryon potential from \citet{BR2013}. } 
\label{fig:ACCdmRatios}
\end{figure*}

Now  that we  have  determined our  best-fit  $R_{core}$ and  $q_{DM}$
parameters,  we examine  the (data/model)$_{best-fit}$  residuals.  In
Figure~\ref{fig:ACCdmRatios},      we      show      the      residual
(data/model)$_{best-fit}$ maps for  $a_{R}$ and $a_{Z}$.  Allowing for
a non-spherical  dark matter potential greatly  improves the agreement
between the data  and the model in both cases: the  rms scatter is 5\%
for $a_R$ and 3\% for $a_Z$, whereas there was a 23\% scatter and need
for differing renormalization factors  for the original model.  As can
be  seen  in  the  right panel  of  Figure~\ref{fig:ACCdmRatios},  the
largest  discrepancy between  the $a_Z$  data and  the  best-fit $a_Z$
model (shown in  dark blue) is found at small $Z$,  where we know that
our map is  biased by 16\% relative to the BR13  results.  In the left
panel of Figure~\ref{fig:ACCdmRatios}, the largest discrepancy between
the  $a_{R}$ data and  the best-fit  $a_{R}$ model  (shown in  red) is
found at  the smallest $R$  and large $Z$, with  measured acceleration
1.8  times larger  than the  best-fit model  value.  We  conclude that
either  one  of the  two  adopted  SDSS results  from  J08  or B10  is
problematic in  this region,  or that the  adopted model  potential is
incapable   of  fully  explaining   observations.  We   continue  this
discussion in \S\ref{sec:validity}. 

\subsubsection{The Impact of Uncertainty in $R_{core}$ on Other Quantities} 

As noted above, our constraints on $R_{core}$ are weak (\eg, $R_{core}$
is  plausibly  within $22  <  R_{core}/{\rm  kpc}  < 42$).   Here,  we
investigate  if   relevant  local  measurements   can  strengthen  our
constraints on  $R_{core}$ and provide  a check for our  best-fit dark
matter potential.
 
Before we can utilize any local mass measurements, we must convert our
analytic  gravitational  potential  to  a mass  density  distribution.
Fortunately, \cite{Binney1987}  provides the following  expression for
converting  to  matter  density  (see  their  Equation~2-54b)  from  a
gravitational   potential  of   the  functional   form   described  by
Equation~\ref{eq:potDM},

\begin{eqnarray}
\label{eq:rhoDM}
\rho_{DM}(R,Z) = \left({v_{o}^2 \over 4 \pi G q_{DM}^2} \right) \times \\
{(2q_{DM}^2+1)R_{core}^2 + R^2 + 2(1-q_{DM}^{-2}/2)Z^2 \over 
(R_{core}^2 + R^2 + Z^2 q_{DM}^{-2})^2} \nonumber. 
\end{eqnarray}

Using  this expression  with our  best-fit values  of $R_{core}=27.4$,
$q_{DM}=0.67$,  and $v_o=195$ km/s,  we obtain  $\rho_{DM}(R=8,Z=0)$ =
0.004 $M_\odot/pc^3$, with an estimated uncertainty of only about 10\%
due   in  large   part  to   our  uncertainty   in   $R_{core}$.  This
$\rho_{DM}(R=8,Z=0)$ estimate is within statistical agreement with the
result   of  $\rho_{DM}(R=8,Z=0)=0.008\pm0.003$   $M_\odot/pc^3$  from
\citet{Bovy2012}.

As  discussed  in \cite{Binney1987},  when  $q_{DM}=0.7$, the  density
distribution predicted by Equation~\ref{eq:rhoDM} is negative near the
$Z$ axis  for $|Z|>7 \, R_{core}$.  However,  this unphysical behavior
is of  no concern here because  \textit{$7 \,R_{core} >  100$ kpc even
  for the  smallest allowed $R_{core}$},  which is clearly  far beyond
the probed volume.

For the small $Z$ relevant here, the isodensity contours 
are elliptical with the axis ratio given by 
\begin{equation}
\label{eq:qrhoDM}
          q^\rho_{DM} = { 1 + 4\,q_{DM}^2 \over 2 + 3/q_{DM}^2},
\end{equation}
yielding $q^\rho_{DM}$ = 0.36. This result is in good agreement with the
estimate $q^\rho_{DM} = 0.47 \pm 0.14$ from \cite{Loebman2012}, but
we note that the baryon contribution to the potential was not accounted
for in their study and thus it is superseded by the above result. 

We also  consider the dark  matter contribution to the  local circular
speed,   which   can  be   computed   from   (see  Equation~2-54c   in
\citealt{Binney1987})
\begin{equation}
\label{eq:vcDM}
v_c^{DM}(R_\odot,Z=0) = v_{o} \, {R_\odot \over \sqrt{R_\odot^2 + R^2_{core}}}, 
\end{equation}
Considering  the plausible  range of  $R_{core}$ ($22  < R_{core}/{\rm
  kpc}  <  42$), we  estimate  $v_c^{DM}$ to  be  $v_c^{DM}  = (63.8  -
0.33 \times R_{core}/kpc)$ km/s via a  linear  fit.   Our best-fit  value,
$v_c^{DM}=55$ km/s, is uncertain to within $\sim$3 km/s (again, due to
the weak  constraints on $R_{core}$).  Thus, we  find that considering the
local  circular  speed  does   not  provide  a  strong  constraint  on
$R_{core}$.

The best-fit  contribution of dark  matter halo to the  local circular
speed,  $v_c^{DM}=55$ km/s,  is  a  factor of  two  smaller than  that
estimated by BR13.  This discrepancy probably implies that the adopted
logarithmic  potential given by  Equation~\ref{eq:potDM} close  to the
Galactic plane does not have sufficient curvature in the $Z$ direction
to produce larger $v_c$ (the normalization $v_o$ cannot be responsible
because  it's  value  is  set  by  distant  halo  stars).   A  similar
``curvature problem''  close to the  plane (and close to  the symmetry
axis) with J08 and B10 results is discussed in \S\ref{sec:validity}.

\subsection{The Uniqueness of Adopted Dark Matter Potential}\label{sec:uniqDMpot} 

At  large galactocentric  radii the  dark matter  contribution  to the
force    felt   by    halo   stars    dominates   over    the   baryon
contribution.  Measurements   are  precise  enough   to  test  whether
functional forms other than the adopted logarithmic potential given by
Equation~\ref{eq:potDM} would fit the data. 

We first test whether a  single value of $q_{DM}$ is sufficient: using
Equation~\ref{eq:qDMdirect},  but  this  time  with  the  BR13  baryon
contribution  subtracted from  the  measured maps,  we  find that  the
median  value (per  bin)  of $q_{DM}$  is  0.67, with  (inter-quartile
based) scatter  of $\sigma_G=0.05$.  The  small width of  the $q_{DM}$
distribution indicates that the  spatial variation of the acceleration
maps   is    well   captured   by   the   $x$    variable   given   by
Equation~\ref{eq:xvar}.

It is straightforward to show that for a generalized potential 
\begin{equation}
\label{eq:potDM2}
              \Phi_{DM}(R,Z) = {1 \over 2} v_{o}^2 \, f(x),
\end{equation}
where $x$ is given by Equation~\ref{eq:xvar}, 
\begin{equation}
       q_{DM}^2 \, {a_Z \over Z}  =  {a_R \over R} = {v_o^2 \over R_\odot} \, {df \over dx}.
\end{equation}
For  the  logarithmic   potential,  $f(x)=ln(x)$,  deviations  from  a
logarithmic  potential will  be seen  as deviations  of the  first two
terms from the expected $1/x$  behavior. For the region with $Z>5$ kpc
and  $R>8$  kpc,  we  find  no  evidence  of  the  departures  from  a
logarithmic potential.

Nevertheless, the  dynamic range of $x$  is fairly small,  from 3.5 to
4.5  for   $R_{core}=27.4$  kpc,   and  for  a   power-law  potential,
$x^\alpha$,  can  provide a  very  good  description  of the  best-fit
logarithmic  potential,  especially  if  $R_{core}$ is  re-fit.   With
$R_{core}=27.4$ kpc, $\alpha=0.73$  provides almost the same potential
(with per bin scatter of  0.2\%) as the best-fit logarithmic potential
(that is, 2\,ln(x)  is very close to 1.007\,x$^{0.73}$ for  $3.5 < x <
4.5$). When  $R_{core}$ is  changed to 22  kpc, the agreement  is even
better with  a best-fit  $\alpha=0.53$. Forcing $R_{core}=0$  does not
provide  a satisfactory  fit  for any  $\alpha$.   Therefore, while  a
logarithmic potential is  fully consistent with the data,  it is not a
unique  solution. The  main reason  for  this ambiguity  is the  small
dynamic range of  $x$ due to the finite  sampled volume.  We reiterate
that $q_{DM}=0.7$ is robustly  determined irrespective of the detailed
form for $f(x)$.

\subsection{Test of the BR13 Baryon Potential}

In our  analysis above,  we assumed that  the baryon component  in the
BR13 potential  is perfect ($f_{BR}=1$). We now  relax this assumption
and allow $f_{BR}$ and $q_{DM}$  to be free fitting parameters, with a
fixed $R_{core}=27.4$ kpc.

Our  resulting variation  in $\delta(f_{BR},q_{DM})$  is shown  in the
right panel in Figure~\ref{fig:deltaFit1}; we find the best-fit values
(shown in dark blue) to be $f_{BR}=0.94$, $q_{DM}=0.70$.  Although the
formal best fit is found  at $f_{BR}=0.94$, \textit{the data are fully
  consistent with  $f_{BR}=1$}.  Based  on the variations  of best-fit
$q_{DM}$ with  other fitting parameters,  we conclude that  its formal
uncertainty  is much  smaller  than 0.1.   However,  due to  plausible
deviations  between the  adopted analytic  potential and  reality, and
departures  from  steady state  and  cylindrical  symmetry, we  cannot
exclude the  possibility that its  systematic uncertainty could  be as
large as 0.1.  Hence, we adopt as our best-fit  model $f_{BR}=1$ and a
logarithmic   dark    matter   potential   with   $q_{DM}=0.70\pm0.1$,
$R_{core}=27.4$ kpc and $v_o=194$ km/s.

\subsection{The Impact of Uncertainty from the J08 Best-fit Parameters}

The upper  limits on the  uncertainty of the best-fit  halo parameters
quoted by J08 are 0.2 for $n_H$  and 0.1 for $q_H$ (see their Table 10
and  discussion in  \S~4.2.4).   The formal  uncertainties in  fitting
(based on  a $\chi^2$ analysis)  are several times smaller.  These two
parameters  are highly  covariant (see  figure 22  in J08);  values of
$n_H$ that  are larger than the  best-fit value (i.e.,  a steeper halo
stellar number density profile)  correspond to larger values of $q_H$,
and vice versa. 

When  the  parameter values  for  the  stellar  halo are  varied  from
($n_H$=2.57,  $q_H$=0.54)   to  ($n_H$=2.97,  $q_H$=0.74)   along  the
direction of covariance,  the resulting $q_{DM}$  (potential) varies from
0.55  to 0.82,  with the  implied $q^{\rho}_{DM}$  varying from  0.19 to
0.57.  Because the adopted variation in $n_H$ and $q_H$ represents
an upper  limit, we  conclude that our  final errors of  $\sim$0.1 for
$q_{DM}$ and $q^{\rho}_{DM}$ are {\it not} dominated by uncertainties in
the best-fit values for $n_H$ and $q_H$.

\subsection{Comparison with Other Results}

This is  the first study  where distant halo  stars have been  used to
constrain  the  shape  of  dark  matter potential  {\it  in  situ}  at
galactocentric distances of up to 20 kpc. However, related constraints
on $q_{DM}$  have been  obtained in studies  of stellar  streams, most
notably     using    the     Sagittarius     dwarf    tidal     stream
\citep[e.g.,][]{Law2010} and  the GD-1 stream (see  below). The former
provides constraints  at much larger galactocentric  radii (20-50 kpc)
than  considered here  and  thus we  focus  below on  the analysis  of
GD-1. We  note that \citet{CVH2013}  used the Sagittarius  dwarf tidal
stream to constrain  the dark halo potential and  found that within 10
kpc from the  Galactic center it is axisymmetric  and flattened toward
the disk plane with $q = 0.9$.

\subsubsection{Tidal Stream GD-1}\label{sec:K2010} 

GD-1 is a long, thin stellar stream discovered in SDSS DR 4 photometry
in  2006 \citep{Grillmair2006}.   GD-1 spans  $80^{\circ}$  across the
northern sky, passes within 30$^{\circ}$  of the Galactic pole, and is
at its midpoint  about  8  kpc   away  from  the  midplane  of  the  disk
\citep{Carlberg2013}.  Based upon  the SDSS photometry, USNO-B astrometry,
and SEGUE spectroscopy, \citet[][hereafter K10]{Koposov2010} were able
to  construct  an empirical  six-dimensional  phase-space  map of  the
stream.

From their analysis,  K10 conclude that GD-1 is  on an eccentric orbit
that is consistent with a single flattened isothermal potential.  That
is,  K10 fit  GD-1 using  a  model of  the same  functional form  as
Equation~\ref{eq:potDM}, but they suppose it accounts for both baryons
and dark matter.   In this limit, K10 adopt  $R_{core}=0$ and conclude
that  the total  axis  ratio, $q$,  is  equal to  $q=0.87\substack{+0.07
  \\ -0.04}$.

To emulate K10's results, we  have repeated our fitting procedure with
a fixed $f_{BR}=0$ (that is, supposing no separate baryon component to
the  potential). We  find that  in this  case, $R_{core}=0$  is indeed
strongly preferred,  and we obtain $q=0.80$, which  is consistent with
the K10 results.

K10 go on to estimate $q_{DM}$ by including a simple baryon bulge+disk
model  in  their analysis;  from  this, they  find  a  lower limit  on
$q_{DM}$  ($q_{DM}>0.89$ at  the 90\%  confidence level).   This lower
limit appears  to exclude our  best-fit value; however,  the differing
results are  not surprising given  the fact that  \textit{their baryon
  model is significantly  different than ours}. It is  likely that the
baryon component  determined by  BR13 is much  more robust  than the model
used by K10.  \textit{We reiterate that when using the same functional
  form for the potential, we get the same best-fit model parameters as
  K10.}   Thus, in an  indirect sense,  the accelerations  measured by
halo  stars  are  consistent  with  the potential  needed  to  explain
observations of the GD-1 stream.

\begin{figure*}[h!]
\epsscale{1.}
  \plotone{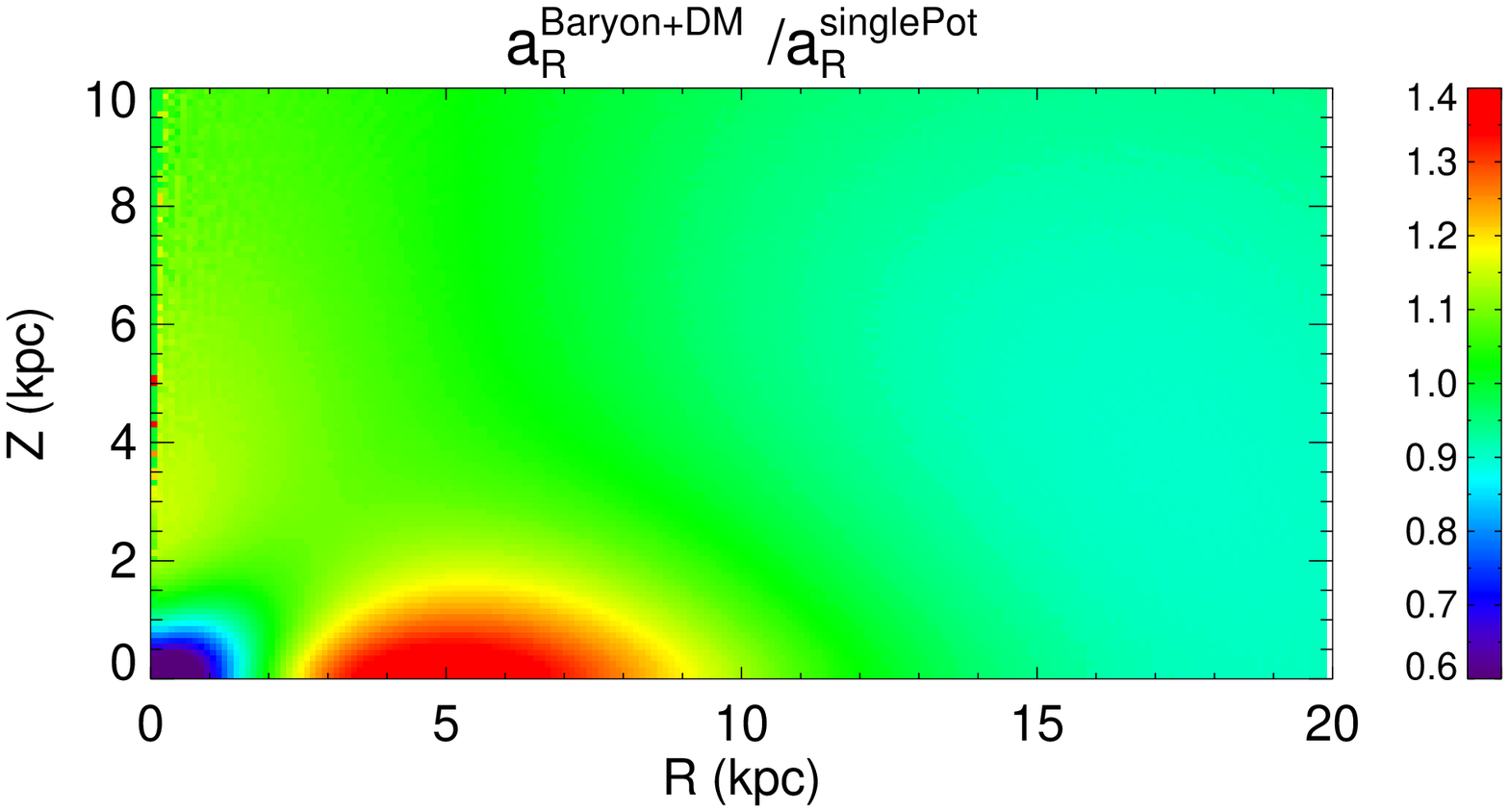}
  \plotone{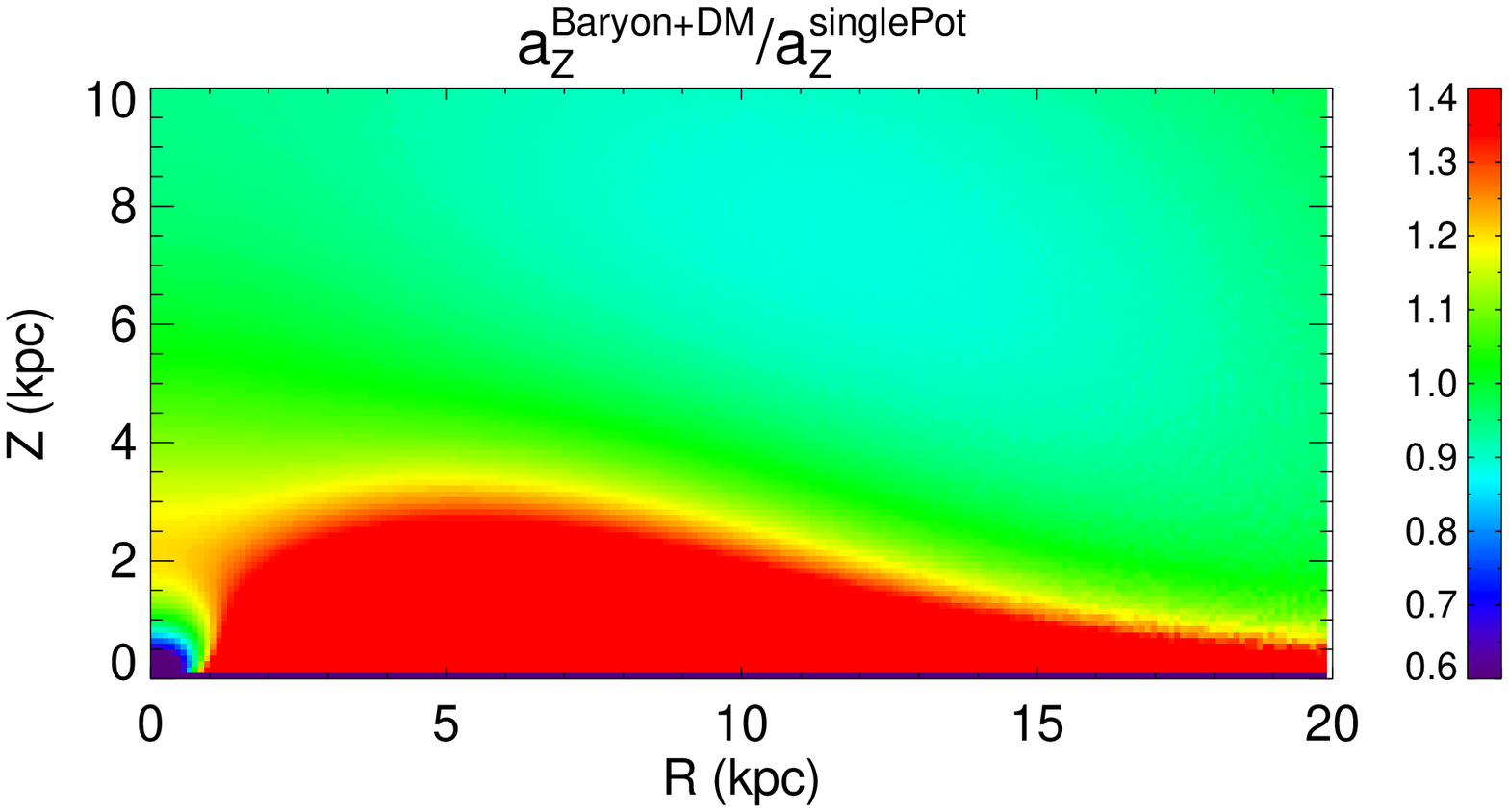}     
  \caption{These panels show the  ratio of the accelerations predicted
  by the K10 and BR13 models.  On top is ratio for $a_{R}$ and on
  the bottom is the ratio for $a_{Z}$.  The BR13 best-fit two-component
  model assumes $q_{DM}=0.7$; the  K10 single-component model based on
  Equation~\ref{eq:potDM} (see \S~\ref{sec:K2010}).}
\label{fig:rat2over1}
\end{figure*}

It is  certainly surprising that a single-component  potential can
provide  a  good description  of  acceleration  maps  for halo  stars,
especially given  the complex morphology  of the B13  baryon component
(see  the top  panels in  Figure~\ref{fig:BRfractions}).   We consider
this  in further detail  by comparing  our best-fit  two-component model
with a single-component model. The ratios  predicted for the $a_{R}$
and  $a_{Z}$  maps  are  shown   in  the  left  and  right  panels  of
Figure~\ref{fig:rat2over1} respectively.

As  can be seen  in both  panels of  Figure~\ref{fig:rat2over1}, above
$Z\sim$3-4 kpc, the  two models agree quite well  (shown in green and
matching to within 10\%).  This, in  fact, is the region where most of
constraints from  both the  K10 analysis and  our analysis  come from.
However, given the $Z<4$ kpc constraints from BR13, it is clear that a
single-component  model  does  not  have the  flexibility  to  explain
simultaneously both  the BR13  SEGUE G dwarfs  data and our  SDSS halo
stars data.  For example, at $R=8$ kpc and $Z=0$, the single-component
model  predicts  acceleration that  is  too small  by  a  factor of  6
compared  to the  best-fit  two-component model  and constraints  from
BR13.  We  conclude from  this that the  potential from K10  cannot be
extrapolated close to the Galactic plane.

The  best-fit model  adopted here  does not  suffer from  this problem
because close  to the plane it  is dominated by  the baryon component.
Nevertheless, recall that the contribution  of the dark matter halo to
the local  circular speed is a  factor of 2 smaller  than estimated by
BR13. 

\section{Testing MOND}\label{Sec:MOND}

Modified Newtonian  dynamics (MOND) is a proposed  alternative to dark
matter, which posits a breakdown of Newtonian dynamics in the limit of
small accelerations  \citep[][and references therein]{MOND2014}.  When
the   Newtonian   acceleration,  $a_N$,   is   much   larger  than   a
characteristic acceleration  scale, $a_o$, the acceleration  felt by a
test particle is Newtonian, $a=a_N$.   However, when $a_N << a_o$, the
acceleration felt by a test particle is much larger than the Newtonian
prediction,   $a  =   a_N  \,   \sqrt{a_o/a_N}$.   The  characteristic
acceleration scale is about $10^{-13}$ km/s$^2$, that is, very similar
to the range of accelerations felt  by halo stars in the volume probed
by the SDSS. In the acceleration range between these two extremes, the
acceleration is given by an interpolating function
\begin{equation}
                     a \, \mu(x) = a_N,
\end{equation} 
where  $x =  a/a_o$.   The interpolating  function $\mu(x)$  satisfies
$\mu(x) \approx 1$ for $x \gg 1$  , and $\mu(x) \approx x$ when $x \ll
1$.   For  example,  \citet{FB2005}  investigated  functions  such  as
$\mu_1(x) =  x/\sqrt{1+x^2}$ and  $\mu_2(x) = x/(1+x)$.   The physical
basis  of   MOND  theories   is  discussed  in   \citet{SMcG2002}  and
\citet{GRMOND}.

\begin{figure*}[h!]
\epsscale{1}
  \plottwo{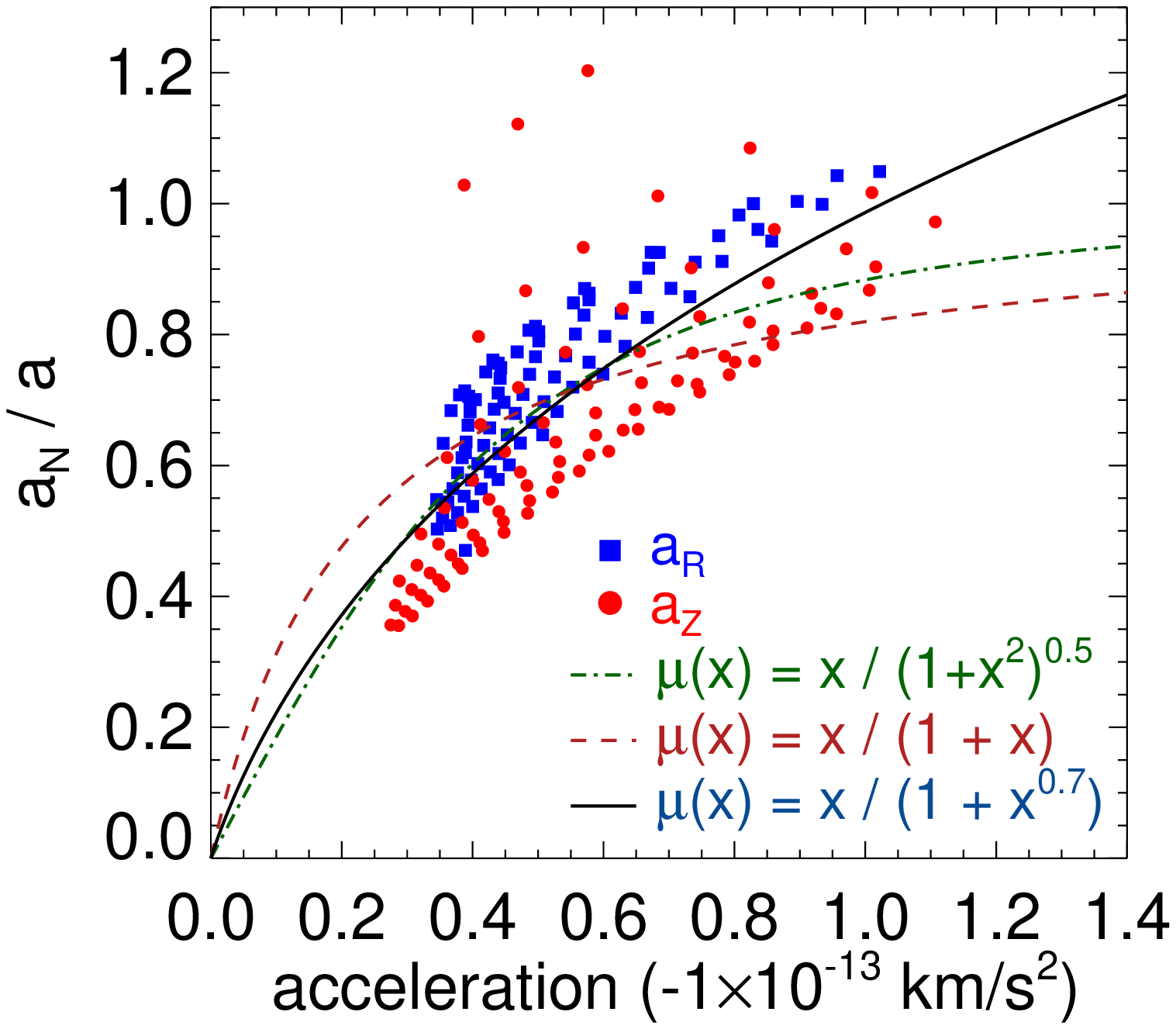}{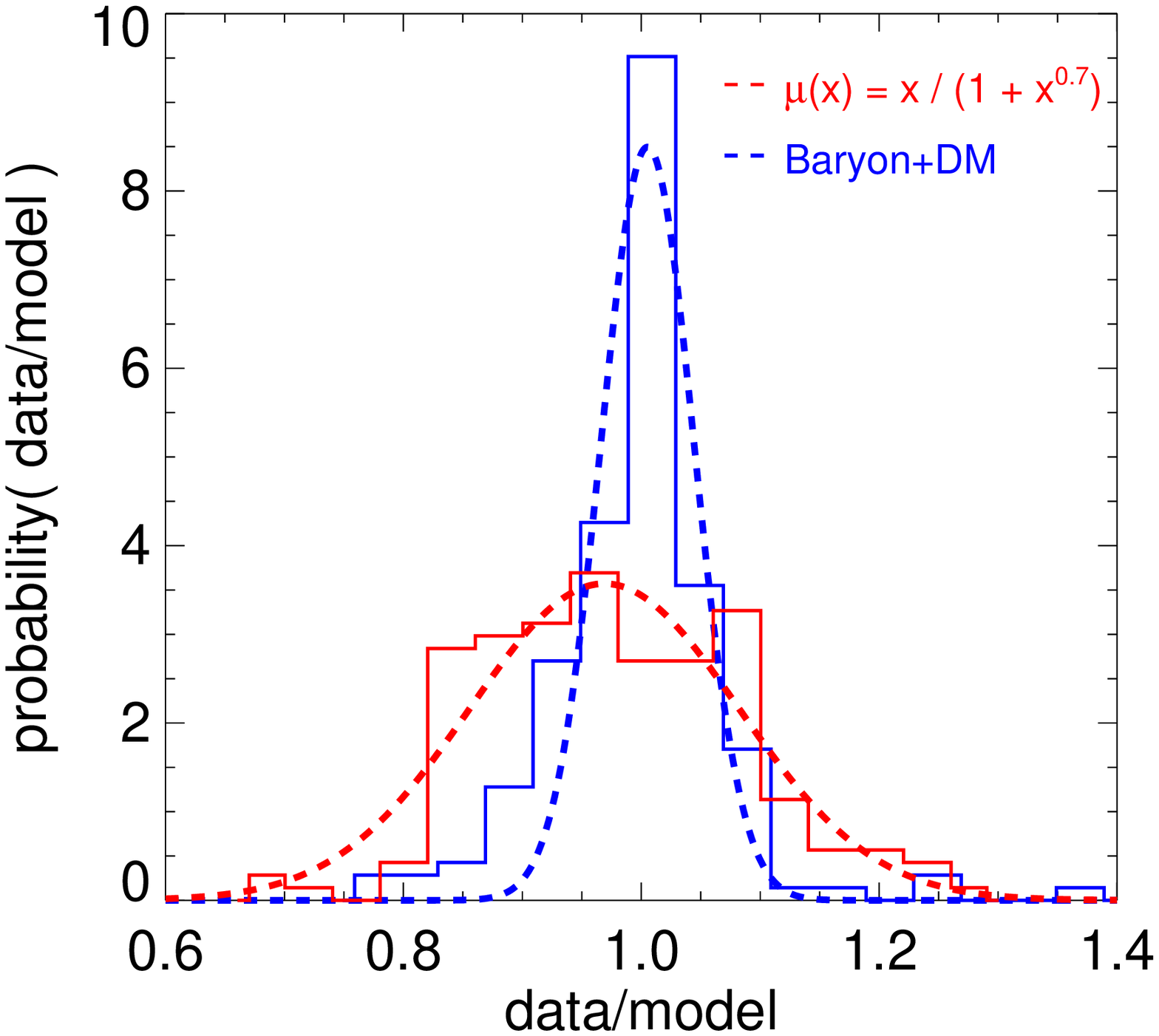}  
  \caption{A test of various MOND models.  The symbols in the left panel
           show the ratio of acceleration due to baryons from the BR13
           model  and the measured  acceleration for  halo stars  as a
           function  of the  former  (blue squares:  $a_R$, red  dots:
           $a_Z$).    Lines  show   MOND  predictions   for  different
           interpolating  functions,  $\mu(x)$,  with  $x=a/a_o$,  and
           different  values  of  characteristic  acceleration  scale,
           $a_o$ (dot-dashed:  $\mu(x)=x/\sqrt{1+x^2}$ and $a_o=0.53$;
           dashed:    $\mu(x)=x/(1+x)$    and    $a_o=0.22$;    solid:
           $\mu(x)=x/(1+x^{0.7})$ and $a_o=0.31$;  with $a_o$ in units
           of   10$^{-13}$   km/s$^2$).    The   right   panel   shows
           distributions of the data/model ratio, using both $a_R$ and
           $a_Z$,  for  the  best-fit  model with  dark  matter,  blue
           histogram, and the best-fit  MOND model (shown by the solid
           black  line in the  left panel),  red histogram.  The thick
           dashed lines are best-fit Gaussians with the widths of 0.04
           (dark matter model) and 0.11 (MOND model).}
  \label{fig:mond}
\end{figure*}

In addition to  sampling the relevant range of  acceleration, the SDSS
data  for  halo  stars  simultaneously constrains  two  components  of
acceleration and enables  a very powerful test of  the MOND model. The
left panel  in Figure~\ref{fig:mond}  shows the ratio  of acceleration
due to baryons  from the BR13 model and  the measured acceleration for
halo stars  as a function  of the measured acceleration.   MOND models
predict that  the two should be  correlated, and indeed  data for each
component ($a_R$ and $a_Z$) show remarkably small scatter about a mean
trend.   However, the two  trends are  significantly offset  from each
other  and  it   is  impossible  to  fit  them   both  with  a  single
interpolating  function. In  other words,  both the  magnitude  of the
measured  acceleration is  different  than predicted  by the  baryons,
\textit{and} the  direction of  the acceleration vector  is different.
Since MOND  modifies only the  former\footnote{Strictly speaking, this
  is true only for Milgrom's MOND;  it might be possible to avoid this
  problem via the  addition of a solenoidal vector  field to Newtonian
  acceleration;  see  Equation  19  in \citet{MOND2012}.},  it  cannot
explain the latter.

\begin{figure*}[h!]
\epsscale{1}
  \plotone{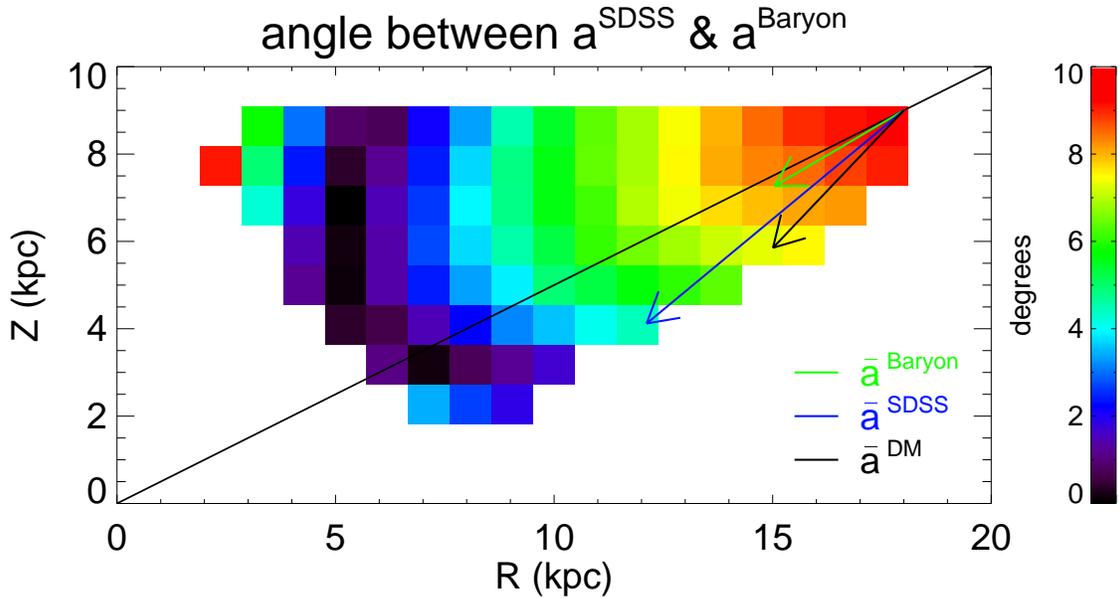}  
  \caption{The angle between the measured acceleration vector and the BR13
           baryon-based prediction.   The largest angles  are observed
           for $R>10$ kpc and  $Z>5$ kpc. The three vectors correspond
           to  the top  right  pixel,  and the  diagonal  line is  the
           direction towards the Galactic center from that pixel.  The
           vector  closest  to   the  diagonal  line  is  acceleration
           predicted  by  BR13 baryon  component  (the same  arbitrary
           length  scale is  used for  all three  vectors;  angles are
           correctly displayed).  It points approximately  towards the
           Galactic  center.  The   longest  vector  is  the  measured
           acceleration: it is stronger than the baryon prediction and
           it points  in a different direction (angle  between the two
           vectors  is 9.3  deg.).  MOND cannot  explain the  measured
           acceleration because it only  modifies the length of baryon
           prediction and  not its direction.  For the same  reason, a
           spherical  DM halo  cannot do  it either  -  its prediction
           always points directly towards the GC.  The third vector is
           a prediction by the  best-fit oblate dark matter halo.  The
           vector   sum  of  baryon   contribution  and   dark  matter
           contribution produces the measured acceleration.}
\label{fig:mondAngle}
\end{figure*}

For   further  illustration,   Figure~\ref{fig:mondAngle}   shows  the
variation of  the angle between  the measured acceleration  vector and
the acceleration vector predicted by the BR13 model for baryons within
the probed  volume; this angle is in  the range 0$^\circ$--10$^\circ$,
with a median value of  4.7$^\circ$.  The largest values are found for
the largest  $R$ and  $Z$; significantly detected  differences between
the direction of the measured acceleration vector and the direction of
the baryon-based  prediction are found  for $R>10$ kpc and  $Z>5$ kpc.
For example, at  the bin with $R=18.5$ kpc and  $Z=9.5$ kpc, the angle
between the measured acceleration vector and the direction towards the
Galactic   center  is   12.5$^\circ$,  and   the  angle   between  the
acceleration predicted  by the baryon model and  the direction towards
the  Galactic  center  is  3.2$^\circ$  -- this  is  a  difference  of
9.3$^\circ$ (see  vectors marked in the figure)!   This mismatch angle
cannot  be explained by  MOND; MOND  only modifies  the length  of the
baryon  prediction but  not its  direction.   For the  same reason,  a
spherical  dark matter  halo does  not work  either --  its prediction
always points  directly towards the GC.   The {\it vector  sum} of the
dark  matter contribution  for an  {\it  oblate halo}  and the  baryon
contribution produces a satisfactory model.

To quantitatively estimate the  disagreement between our data and MOND
models,  we  have   tried  three  different  interpolating  functions,
$\mu(x)=x/\sqrt{1+x^2}$,  $x/(1+x)$  and  $x/(1+x^{0.7})$,  where  the
index $0.7$  in the last one  was determined as a  free parameter. The
corresponding best-fit  values of the  characteristic acceleration are
$a_o=0.53$,  $0.22$  and  $0.31$,  respectively (in  units  of  10$^{-13}$
km/s$^2$). The scatter of data around the model prediction is smallest
for the third MOND model, 17\%.  This scatter is still more than three
times larger  than for  the best-fit model  with dark matter  from the
preceding section, which produces  a scatter of 4.8\%. The data/model
ratio   distributions    are   shown    in   the   right    panel   in
Figure~\ref{fig:mond}.  If  data  for  $a_R$  and  $a_Z$  are  treated
separately, then the scatter for  each can be decreased to about 10\%,
but the ratio  of best-fit values of $a_o$ (using  the third model) is
significantly  different   from  1  (0.39  for  $a_R$   and  0.27  for
$a_Z$). Because it is impossible to construct a single MOND model that
outperforms the  model with dark  matter, the latter  is statistically
preferred. 

Therefore,   thanks  to   precise   two-dimensional  measurements   of
acceleration  for halo  stars,  we can  reject  the MOND  model as  an
explanation of  the observed behavior.  The model  that incorporates a
dark   matter   halo   is   in   much  better   agreement   with   the
data. Nevertheless, we emphasize that these conclusions are critically
dependent on the accuracy of the baryon potential.

\section{Discussion and Conclusions}
\label{s:conclusion}

We  have demonstrated  that the  SDSS observations  of the  MW stellar
halo combined with the gravitational potential due to baryons derived by BR13, imply the
existence of  an invisible component  that contributes to  the overall
gravitational    potential.   At   large    galactocentric   distances
($\sim$20~kpc), the detection of  this component is highly significant
and robust  because the gravitational force experienced  by halo stars
is as  much as  three times  stronger than what  can be  attributed to
purely visible matter.

Our  results are  derived using  Jeans equations,  which  estimate the
gradient  of the gravitational potential  statistically from  the observed
spatial variation of stellar  counts and stellar kinematics (Equations
1 and 2).   The derivation of these equations  assumes a cylindrically
symmetric steady-state  system. Both  available MW data  and $N$--body
simulations indicate that these  assumptions cannot be fully satisfied
and  thus  the performance  of  the  Jeans  equations method  must  be
critically examined before drawing conclusions.

Using a  modern cosmologically derived  $N$--body simulation, designed
to mimic the  formation and evolution of a  MW--like galaxy, we showed
that the Jeans equations method can uncover true accelerations despite
deviations from  a steady-state  system with cylindrical  symmetry: we
recovered  true mean  per-bin accelerations  with random  errors below
20\% and a  bias below 10\%. Such a precision  is more than sufficient
to robustly  detect deviations  between the measured  acceleration and
the acceleration that can be attributed to baryonic material.

When  applied  to  SDSS  data, this  method  produced  two-dimensional
acceleration maps to heliocentric distances exceeding $\sim$10~kpc and
galactocentric  distances exceeding $\sim$20~kpc.  It was  possible to
probe  this  large  distance  range  thanks to  substantial  SDSS  sky
coverage    and    accurate    multi-color   photometry    to    faint
limits. Leveraging the baryonic  gravitational potential from BR13, we
showed  that the  gravitational  force experienced  by  halo stars  at
galactocentric  distances  of  $\sim$20~kpc  cannot be  explained,  in
a Newtonian  framework, by  only baryon  matter contributions.  At these
large galactocentric distances, the discrepancy is much larger than in
the Solar  neighborhood because the baryonic  material is concentrated  in the
Galactic disk, while the presumed dark matter is much more extended.

We attempted  to construct  a MOND model  in agreement with the data, but
found that our best-fit MOND  model is significantly outperformed by a
model  with dark  matter.  MOND's  inability  to explain  the data  is
related  to the  evidence for a  non-spherical dark  matter distribution
(which  comes from  the oblate  spatial distribution  of  halo stars).
{\it Specifically, the magnitude of the measured acceleration is different
  than  predicted  by  baryons,  and  the direction  of  the  measured
  acceleration vector is different,  too. Since MOND modifies only the
  former, it cannot explain  the latter}. Therefore, because of precise
two-dimensional  measurements  of the acceleration  of  halo stars,  MOND
models   can  be   rejected   irrespective  of   the details  assumed
in the interpolating    function   and    the    value   of the   characteristic
acceleration. Of course, these conclusions are critically dependent on
the accuracy of the baryon potential. 

The  large volume probed  by SDSS  halo stars  also provides  a strong
constraint  on the  shape of  the dark  matter halo  potential. Within
galactocentric  distances  of   $\sim$20~kpc,  the  dark  matter  halo
potential  is  well  described  by  an oblate  halo  with  axis  ratio
$q^\Phi_{DM}=0.7$$\pm$$0.1$;  this   corresponds  to  an   axis  ratio
$q^\rho_{DM}\sim0.4$$\pm$$0.1$    for   the   dark    matter   density
distribution.  The   quoted  uncertainties  attempt   to  account  for
systematic errors  but Gaussian behavior cannot  be guaranteed. The formal
random  errors  for  $q^\Phi_{DM}$  and $q^\rho_{DM}$  do  not  exceed
$\sim$0.05. The $R_{core}$ parameter in the logarithmic dark matter model,
and the  preference for logarithmic  over power-law model, are not as
well constrained as $q^\Phi_{DM}$.

The resulting best-fit two-component gravitational potential, based on
the baryonic  component from BR13,  and the dark  matter component  described by
eq.~\ref{eq:potDM}, is simultaneously consistent with relatively local
(within $\sim$3 kpc) measurements of disk stars, and with measurements
of  halo  stars  to  galactocentric distances  of  $\sim$20~kpc.   The
best-fit  potential   derived  here   is  also  consistent   with  the
gravitational  potential  required to  explain the spatial and  kinematic
behavior of the GD-1 tidal  stream.  {\it Given vastly different selection
  criteria,  spatial  distributions  and  kinematics  of  these  three
  populations, this consistency is indeed remarkable!} 

Nevertheless, it is  almost certain that the functional  form given by
eq.~\ref{eq:potDM}  {\it  cannot be  valid  throughout  the entire  MW
  halo}, as we discuss next. 

\subsection{The Range of Validity of Our Results \label{sec:validity}}

Our analysis is based on both  stellar  count and  kinematics data  from  the SDSS Galactic catalog, which extends to  no  more than  $\sim$20  kpc from  the
Galactic center.  This dataset excludes the vicinity immediately surrounding the Galactic 
center, as well as the region very close to the Galactic plane (closer than
$\sim$3 kpc). Therefore, our results should not  be extrapolated beyond this
limit.

There are already strong indications  that the stellar halo model from
J08, given  by Equation~\ref{eq:nuDH2}, cannot  be extrapolated beyond
a galactocentric  distance  of about  30  kpc.  \cite{Sesar2011}  found,
using  main  sequence   stars  detected  by  the  Canada-France-Hawaii
Telescope Legacy Survey (CFHTLS) in about 170 deg$^2$ of sky, that the
halo stellar number-density  profile becomes steeper at Galactocentric
distances greater than $\sim$28 kpc, with the power-law index changing
from $n= 2.62\pm0.04$ to  $n=3.8\pm0.1$.  They measured the oblateness
of the stellar halo  to be $q_H=0.70\pm$0.01 (statistical error only),
and they  detected no evidence  of the oblateness changing  across the
range of distances probed.  \cite{Deason2011} explored similar issues,
using a sample  of $\sim$20,000 BHB and blue  straggler stars detected
by  SDSS over  14,000 deg$^2$  of sky,  and obtained  almost identical
results to  those from \cite{Sesar2011}. Their best  fitting model has
an inner  power-law index  of $n=2.3$ and  an outer index  of $n=4.6$,
with the transition occurring at  $\sim$27 kpc, and a constant stellar
halo  flattening of $q_H=0.6$.  In addition,  the distributions  of RR
Lyrae stars from the SEKBO  survey \citep{Keller2008}, and of RR Lyrae
stars from SDSS Stripe 82 data \citep{Watkins2009,Sesar2010}, indicate
a steeper density profile beyond 30 kpc. These results are in general
agreement with the dual halo hypothesis advocated in \cite{Beers2halos},
and references therein (and also including opposing views). 

Similarly, the  finding by  B10 that the  velocity ellipsoid  for halo
stars  is invariant in  spherical coordinates  cannot be  valid beyond
about  30 kpc  from the  Galactic center.   \cite{Brown2010}  used the
Hypervelocity  Star survey  data to  measure the  halo radial-velocity
dispersion  out  to 75  kpc.   They  obtained  results in  statistical
agreement   with   similar   studies   by   \cite{Battaglia2005}   and
\cite{Xue2008},   which   they   summarized   as:  ``the   Milky   Way
radial-velocity  dispersion  drops from  $\sigma=110$  km s$^{-1}$  at
$R_{gc}=15$ kpc to $\sigma=85$  km s$^{-1}$ at $R_{gc}=80$ kpc'' (here
$R_{gc}$  is the  Galactocentric  spherical radius  $r_{sph}$ in  this
paper).  It is hard to  quantitatively and robustly estimate what this
gradient  implies for  the underlying  potential because  the velocity
dispersion in two orthogonal directions at distances beyond 30 kpc has
not been measured yet; these measurements will likely have to wait for
Gaia and LSST surveys (\citealt[][also see below]{IBJ}).  We note that
\citep{CVH2013}  constrained   the  dark  halo   potential  using  the
Sagittarius  dwarf tidal  stream at  a large  range  of galactocentric
radii.  They  found that the  potential is axisymmetric  and flattened
toward the disk plane within 10  kpc from the Galactic center, with $q
= 0.9$.  At larger radii, they argue for a triaxial shape in the outer
halo, consistent with the  \citet{Law2010} model, with deviations from
oblate halo starting at galactocentric distances of about 10 kpc.

\begin{figure*}[h!]
\epsscale{.32}
  \plotone{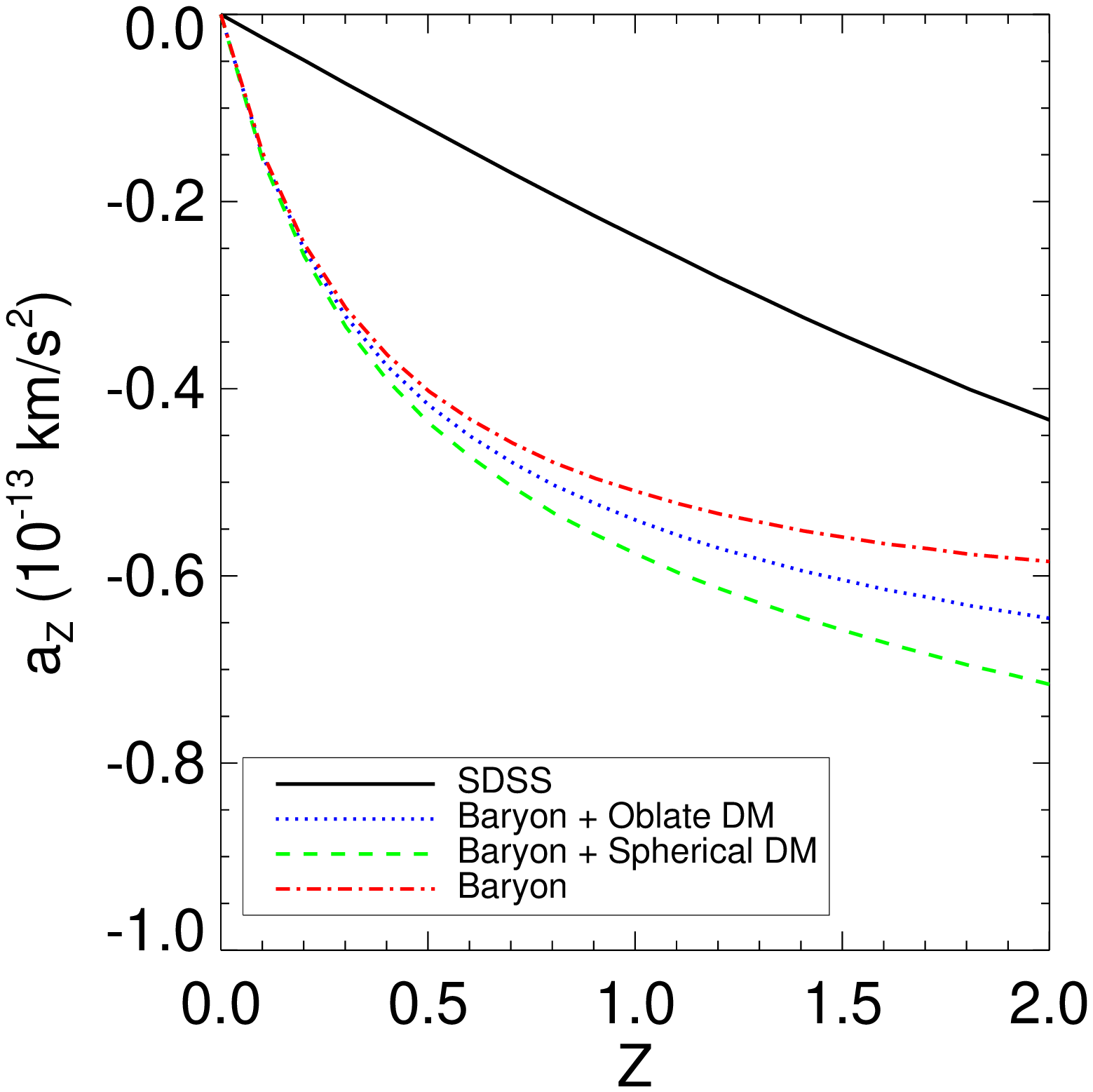}
  \plotone{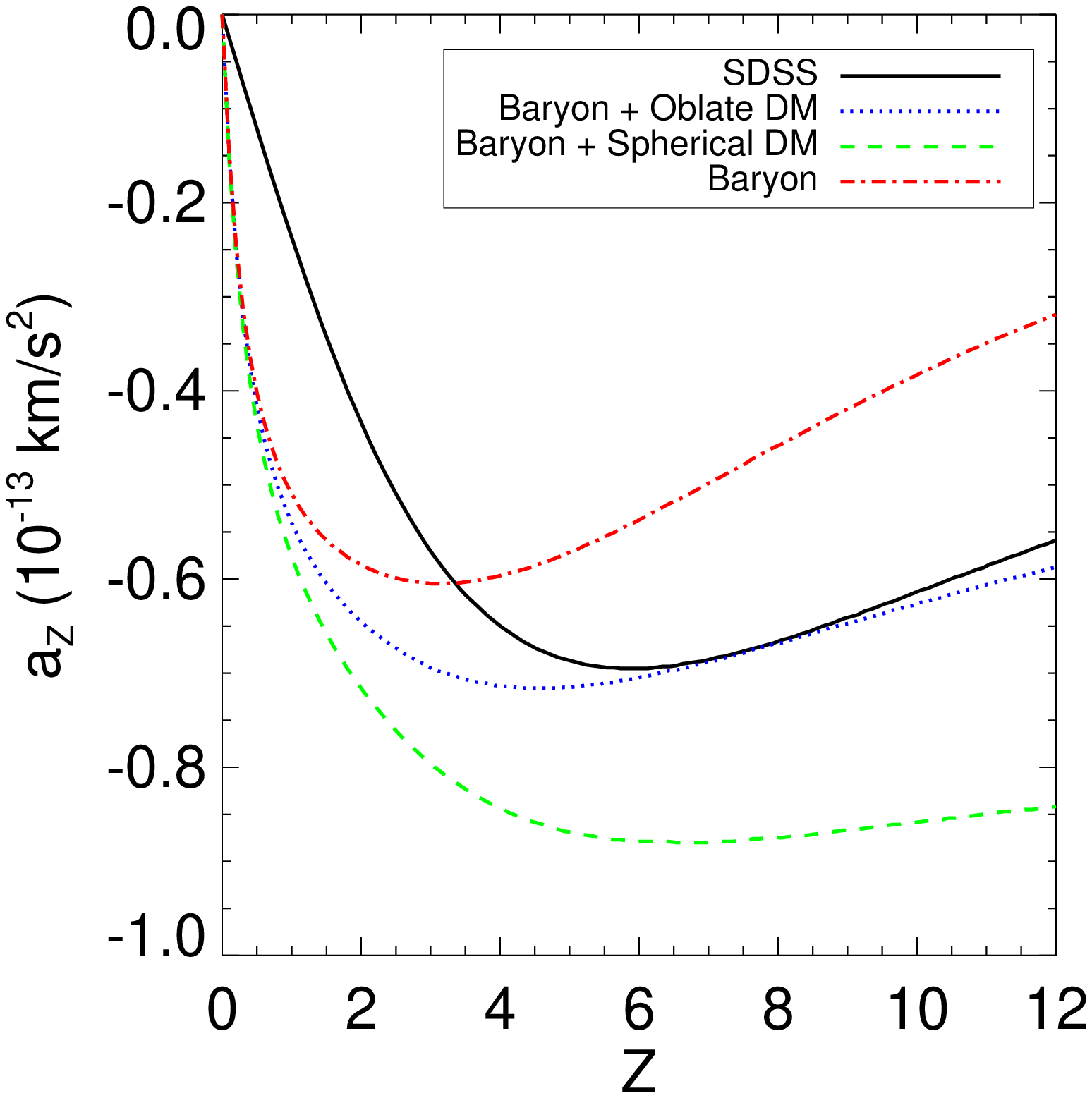}
  \plotone{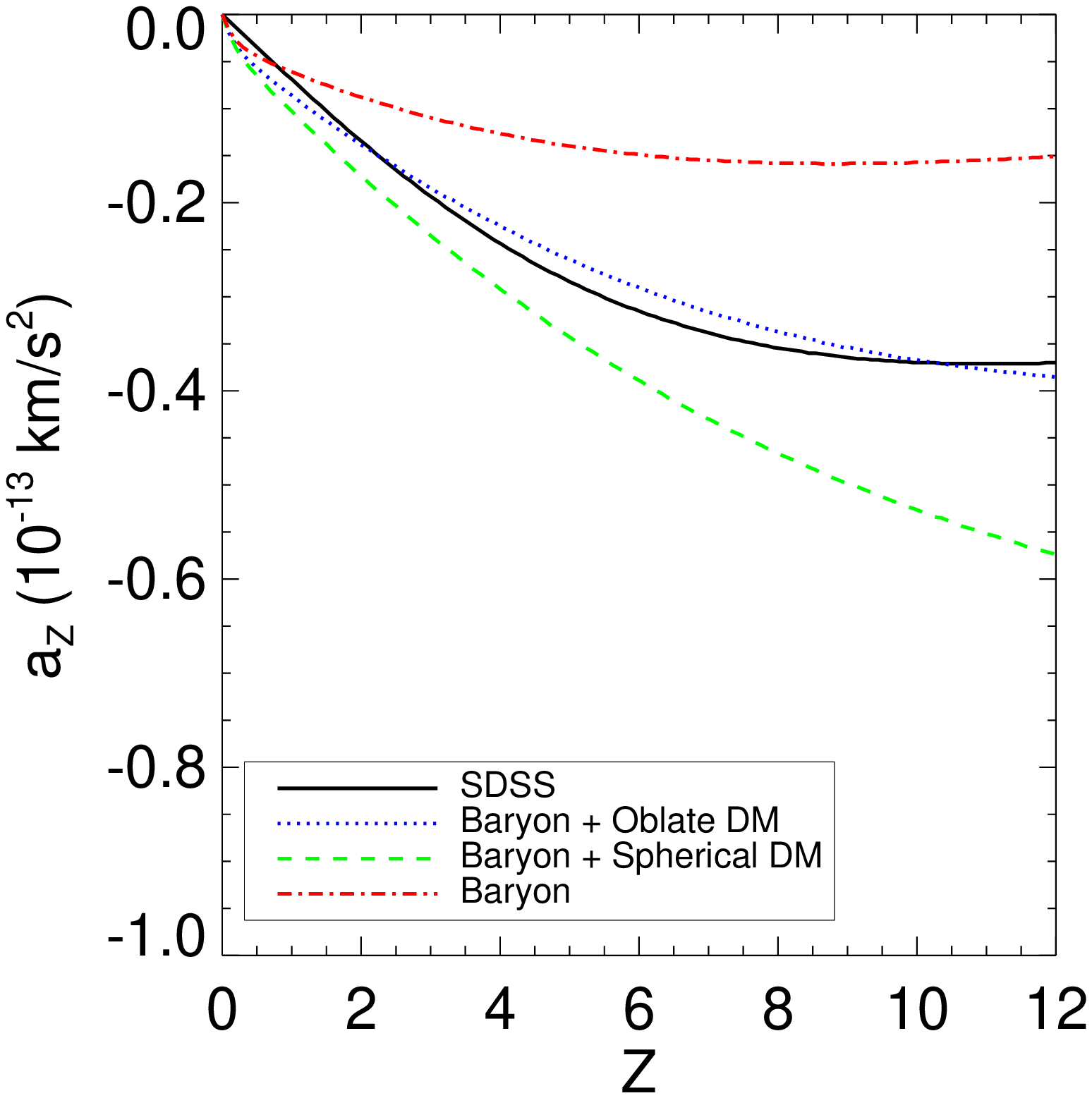}
  \caption{Comparisons of the data and the models for $a_Z$. In all panels, the
           black line is  data for halo stars, derived  from the J08 and B10
           results.  The green line  is the  2-component model  from BR13
           (which comes from data for disk stars within $Z<3$ kpc); it
           includes  the  baryon contribution,  shown  by the  red
           line,  and spherical  dark matter  model (not  shown).  The
           blue  line  is  the  sum  of the  baryon  contribution  and
           modified oblate  dark matter model. The  middle panel shows
           $a_Z(Z)$  for  $R=8$  kpc,  and  the  left  panel  shows  a
           zoomed-in version.   The right panel  corresponds to $R=15$
           kpc. The  modified best-fit  model (blue line)  agrees with
           data (middle  and right panels) for both  disk stars (green
           line,  $Z<2$ kpc)  and halo  stars (black line,  $Z>4$ kpc).
           However, at $R=8$ kpc and  within a few kpc from the plane,
           the acceleration of halo stars  in the $Z$ direction implied by
           J08 and B10 results is weaker than that experienced by disk
           stars (left panel).}
\label{fig:3panels}
\end{figure*}

The acceleration  of the halo  stars in  the $Z$  direction and  close  to the
Galactic   plane   is   weaker   than   that   experienced   by   disk
stars. Figure~\ref{fig:3panels} shows data  and models for $a_Z(Z)$ at
$R=8$ kpc and $R=15$ kpc.   While at $R=15$ kpc, the agreement between
halo and  disk star  accelerations is satisfactory,  at $R=8$  kpc and
within a few kpc from the plane, the implied acceleration of halo stars is
weaker  than   that  experienced  by  disk   stars,  with  discrepancy
increasing from about  a factor of 2 at $Z\sim1$  kpc to larger values
at  smaller $Z$  (note  that all  values  vanish at  $Z=0$ because  of
symmetry).  It  appears that either  the spatial distribution  of halo
stars from  J08, or kinematics  from B10 (or  both) has to  break down
close to  the plane. In  order to increase the implied  acceleration, more
curvature,  that is,  larger derivatives  of the spatial  distribution and
kinematics, are needed  close to the plane.  With  existing data it is
hard  to quantify what  is the  problem because  the fraction  of halo
stars is very  small close to the  plane and it is easy  to get sample
contamination (this is why B10  only analyzed halo stars that are more
than  $\sim$3 kpc from  the plane).  Furthermore, turn-off  halo stars
closer than  about 1 kpc are  saturated in SDSS imaging.  

We can postulate a minor modification of the $q_H$ parameter from 
eq.~\ref{eq:nuDH2}, from its best-fit value $q_H=0.64$ obtained by J08, 
that brings $a_Z$ acceleration component of halo stars in agreement with 
acceleration of  disk stars at $Z\sim1$  kpc. The third term on the right-hand side of 
eq.~\ref{eq:eq2} is proportional to $q_H^{-2}$ for small $Z$. If $q_H$ 
decreases from 0.64 at $Z > 4$ kpc to $q_H\sim0.45$ at $Z\sim1$  kpc, 
the resulting $a_Z$ for halo stars becomes similar to $a_Z$ for disk
stars. At the same time, the stellar number density given by 
eq.~\ref{eq:nuDH2} is insensitive to $q_H$ for $R=R_\odot$ and 
$Z=1$ kpc, and thus remains consistent with J08 and B10 analysis. 
The implication is that the halo is more oblate closer to the disk
midplane than far away from it, an idea that was already advocated
in the literature (e.g., \citealt{PSB1991}). 

It is likely that Gaia  will provide a  definitive resolution of this  puzzle. 
Gaia will  also likely resolve  the origin  of the divergence  of $a_R$  on the
symmetry  axis  ($R=0$). Because  values  for $\sigma_{\phi\phi}$  and
$\sigma_{\theta\theta}$ obtained by  B10 are different, $a_R$ computed
using   Equation~\ref{eq:eq1}   diverges   for   $R=0$   (for   $R=0$,
$\sigma_{RR} = \sigma_{\theta\theta}$). Proper motion accuracy below 
1 mas yr$^{-1}$ is required to map the behavior of $\sigma_{\phi\phi}$  and
$\sigma_{\theta\theta}$ close to the Galaxy's symmetry axis. 

Figure~\ref{fig:3panels}  also   illustrates  that  the   dark  matter
component  from the  original BR13  model produces  too much  force at
large $Z$.  This discrepancy  is resolved by replacing their spherical
dark  matter model  by the  new  oblate dark  matter model  introduced
here. While  this modification  results in a  satisfactory explanation
for the  measured halo star acceleration  maps, it fails  to produce a
sufficiently large dark matter halo contribution to the local circular
speed by about a factor of two. This failure suggests that the adopted
logarithmic  potential given by  Equation~\ref{eq:potDM} close  to the
Galactic  plane  does  not   have  sufficient  curvature  in  the  $Z$
direction.   In   other  words,  the   dark  matter  should   be  more
concentrated  toward the  plane  of  the disk  the  our adopted  model
predicts.

In  summary, these discrepancies  indicate that  both the  dark matter
distribution,  and the  spatial  distribution and  kinematics of  halo
stars, should be sensitive to the existence of a stellar disk, but the
current models do not capture this behavior.

In  addition  to the  data  limitations,  simulations  have their  own
caveats.  We have demonstrated  using an $N$--body simulation that the
Jeans equations method can  recover the true stellar acceleration with
a  bias below  10\%.   However,  the general validity  of  this conclusion  is
crucially dependent on the similarity between the simulated galaxy and
the MW.  We  showed that the simulated galaxy is similar  to the MW in
many important  ways, such as  the overall distribution of  halo stars
and their  kinematics. Nevertheless, there are  other untested aspects
that might  be important and  perhaps are biasing our  measurements of
the     dark     matter     halo     properties.      For     example,
\cite{Schlaufman09,Schlaufman2011}  have shown,  using  data from  the
SDSS SEGUE  spectroscopic survey,  that about 10\%  of the  halo stars
within $\sim$20 kpc  from the Sun cluster kinematically  on very small
spatial  scales  (typical radial-velocity  dispersion  is $\sim$20  km
s$^{-1}$).   It remains  to be  seen whether  simulated  galaxies also
include this effect,  and how it affects the  performance of the Jeans
equations method. 

Last but not least, in deriving our conclusions we assumed that the BR13
baryon potential, derived using a sample of disk G dwarfs, is perfect. 
They directly measured $a_Z$ (at $Z$=1.1 kpc and 4.5 $< R/{\rm kpc} < 9$) 
from the vertical dynamics and combined it with $[a_R(R)-a_R(R_\odot)]$ from 
the tangent-point rotation curve measurements. Their model produces 
$a_R(R_\odot)$ equivalent to local circular speed of $v_{circ}(R_\odot)$=220 km/s.
Since at this time $v_{circ}(R_\odot)$ is uncertain by possibly as much as 10\%,
a similar level of uncertainty is implied for their baryon potential. 

\subsection{Future Work}

It is possible to go beyond Jeans equations to use stellar kinematics to  probe the  full  phase  space distribution  of  stars \citep[e.g.,][]{Binney2013}.  
For example, as \citet{Valluri2012} recently demonstrated,  stellar orbits can be 
used to determine  not only  the shape  of the inner  halo, but  the global
shape of  the Galactic halo. The Valluri et al.~orbital spectral
analysis method provides a  strong complementary tool to the technique
presented here  for constraining the  potential of the Milky  Way halo
and its stellar distribution function. In addition, outcomes of methods such as 
``made-to-measure models'' (\citealt{Syer1996}; \citealt{Dehnen2009}),
direct modeling of the distribution function \citep{Piffl2014}, and modeling
of stellar tidal streams (K10; \citealt{Bonaca2014}) can be compared to 
constrain systematic errors of each method and improve understanding 
of the Milky Way mass distribution. 
 
The  full potential  of  these  methods will  be  reached by  upcoming
next-generation  surveys, such as  Gaia \citep{PerrymanGaia}  and LSST
\citep{LSSToverview}.   First,  Gaia   will  provide  measurements  of
distances and kinematics with a  similar faint flux limit as SDSS, but
with  much smaller errors  (for a  comparison of  SDSS, Gaia  and LSST
astrometric    and   photometric    errors,   see    figure    21   in
\citealt{IBJ}). In particular, Gaia data will be superior to currently
available data for quantifying the spatial distribution and kinematics
of halo stars close to the plane. 

LSST will obtain photometric  and kinematic measurements of comparable
accuracy  to those  of Gaia  at Gaia's  faint limit,  and  extend them
deeper by about  5~mag. With LSST, it will be  possible to extend this
study to about 10 times larger distance limit than possible today with
SDSS  data   \citep{IBJ}.  Most  notably,  it  will   be  possible  to
investigate whether the dark matter halo shows any trace of changes in
the  spatial  and kinematic  behavior  around  $\sim$30  kpc from  the
Galactic  center that  are  revealed by  halo  stars. Furthermore,  by
extending  the  limiting  distance,  the  impact of  baryons  will  be
diminished and the conclusions about dark matter behavior will be more
robust.

\section{Acknowledgments}
\label{s:acknowledgements}

We thank  Monica Valluri, Victor Debattista, and  Carlos Vera-Ciro for
illuminating  discussions. SL  and \v{Z}I  acknowledge support  by NSF
grants AST-0707901 \& AST-1008784 to the University of Washington.  SL
acknowledges support  from the Washington NASA  Space Grant Consortium
and Michigan  Society of Fellows.  \v{Z}I acknowledges  support by NSF
grant AST-0551161 to LSST for  design and development activity, by the
Croatian National Science Foundation grant O-1548-2009, and thanks the
Hungarian   Academy  of   Sciences   for  its   support  through   the
Distinguished  Guest  Professor  grant  No.   E-1109/6/2012.   JB  was
supported by NASA through Hubble Fellowship grant HST-HF-51285.01 from
the  Space  Telescope Science  Institute,  which  is  operated by  the
Association of  Universities for Research  in Astronomy, Incorporated,
under  NASA contract  NAS5-26555.   FG acknowledges  support from  NSF
grant AST-1108885.   Resources supporting  this work were  provided by
the NASA  High-End Computing (HEC)  Program through the  NASA Advanced
Supercomputing    (NAS)   Division    at    Ames   Research    Center.
\textit{Galfast} computations were performed  on Hybrid at the Physics
Department, University  of Split, financed by  the National Foundation
for  Science, Higher  Education and  Technological Development  of the
Republic of Croatia.


\pagebreak

\clearpage

\end{document}